\documentclass[pdflatex,sn-mathphys-num,onecolumn]{sn-jnl}% Math and Physical Sciences Numbered Reference Style
\usepackage[utf8]{inputenc}%
\usepackage[T1]{fontenc}%
\usepackage{CJKutf8}%
\usepackage{graphicx}%
\graphicspath{{figures/}}
\usepackage{multirow}%
\usepackage{makecell}%
\usepackage{amsmath,amssymb,amsfonts}%
\usepackage{amsthm}%
\usepackage[title]{appendix}%
\usepackage[table]{xcolor}%
\usepackage{textcomp}%
\usepackage{manyfoot}%
\usepackage{booktabs}%
\usepackage{algorithm}%
\usepackage{algorithmicx}%
\usepackage{algpseudocode}%
\usepackage{listings}%
\usepackage{tabularx}
\usepackage{adjustbox}
\usepackage{longtable}
\usepackage{siunitx}
\usepackage{pifont} % Package for symbols
\usepackage{subcaption} % For subfigures
\usepackage{placeins} % \FloatBarrier
\usepackage{xurl} % better line-breaking for URLs/DOIs in references
\usepackage{url}
\usepackage{doi}
\usepackage{silence}
\usepackage{lmodern}%improve readibility of small fonts
\providecommand{\burl}[1]{\url{#1}}
\makeatletter
\@ifundefined{endbibitem}{}{}
\@ifundefined{barticle}{}{}
\@ifundefined{botherref}{}{}
\@ifundefined{bbook}{}{}
\makeatother

\theoremstyle{thmstyleone}%
\theoremstyle{thmstyletwo}%
\theoremstyle{thmstylethree}%
\newcommand{\bb}{\textcolor{black}{\textbullet}\ }
\definecolor{hasAV}{HTML}{2E7D32}
\definecolor{noAV}{HTML}{C62828}
\definecolor{partialAV}{HTML}{F57F17}
\begin{document}
\title[Tools for antiviral drug design ]{Benchmarking open-source tools for in silico antiviral drug discovery}
\author[1]{\fnm{Daniel} C. \sur{Elton}}\email{dan@radvac.org}
\author[2]{\fnm{Preston} W. \sur{Estep}}\email{pwestep@radvac.org}


\affil[1,2]{\orgname{Rapid Deployment Vaccine Collaborative}, \orgaddress{\city{Waltham}, \state{MA}, \country{USA}}} %\street{Street},\postcode{100190}

%\affil[2]{\orgdiv{Department}, \orgname{Organization}, \orgaddress{\street{Street}, \city{City}, \postcode{10587}, \state{State}, \country{Country}}}

%----------------------------------------------------------------
\abstract{Antivirals are uniquely positioned to be deployed quickly during a new outbreak, especially when repurposed from approved drugs. Yet there are no FDA-approved antivirals for the majority of viral families with pandemic potential. Here we lay out the case for investing in technologies and techniques for antiviral drug discovery and designing antiviral combinations. We present a survey of open-source datasets and computational tools for in silico antiviral drug discovery, with a particular focus on the latest AI-based systems and docking tools. We then present our custom dataset of 43,005 viral protein-ligand binding measurements that we curated from BindingDb and other sources. Importantly, we found that 31\% of viral protein binding data in BindingDb required polyprotein sequences to be carefully split before the data were suitable for training or testing ML models. Using our custom dataset we fine-tuned the DrugFormDTA binding affinity prediction model (Khokhlov et al. 2025). We then benchmarked 15 open-source binding affinity prediction tools on a custom test set of 853 antiviral compounds spread across 16 different protein targets from 10 virus species. Models tested include Boltz-2, GNINA, FlowDock, Interformer, AutoDock-GPU, and others. We found that Boltz-2 and DrugFormDTA ranked highest overall among ML-based approaches, and GNINA did best among docking approaches, with notable variance across specific viral proteins. Fine-tuning DrugFormDTA on our custom cleaned antiviral dataset boosted performance from \textit{r}=0.5 to \textit{r}=0.7. As part of this work we also compiled a library of approved drugs and a comprehensive list of investigational and approved antiviral drugs that can be viewed at \href{https://antivirals-database.radvac.org}{antivirals-database.radvac.org}. Together, this work provides a foundation for future work towards new tools and platforms for rapid drug repurposing and rapid design of antiviral combinations. 
}
%\keywords{}
\maketitle
\tableofcontents
%---------------------------------------------------------------------------------
\section{Introduction}\label{sec1}

Speed matters during a pandemic. Using mathematical modeling researchers have estimated that if vaccines had been available 30 days earlier in Brazil, 31,656 deaths could have been averted in Rio de Janeiro alone.\cite{BarbosaLibotte2022,Amaku2021} Modeling also shows that the one week pause in the AstraZeneca vaccine which occurred in the EU in March 2021 killed hundreds of people in Italy alone.\cite{Faranda2021} Economist Maxwell Tabarrok has estimated that if the FDA's emergency authorization of Pfizer’s COVID-19 vaccine had been done four months earlier, then 130 to 350 thousand lives would have been saved.\footnote{See Tabarrok, 2022 \cite{Tabarrok2022InvisibleGraveyard}. Four months earlier would have been around August 10th, 2020. This would have been 20 days after Pfizer submitted positive Phase I/II data on the safety and immunogenicity of their vaccine to the journal \textit{Nature}. Decades of work with vaccines has shown that immunogenicity is highly correlated with efficacy.} 

%\cite{Albani2021} %NYC & Chicago modeling paper
%\cite{Wang2021} %"Austin and Round Rock, Texas" modeling paper

After the sequence of the SARS-CoV-2 virus became widely known on January 10th, 2020, Moderna designed an mRNA vaccine within hours and started manufacturing the vaccine five days later.\cite{Corbett2020} Unfortunately, another ten months and three weeks passed before Pfizer's mRNA vaccine was given emergency use authorization by the FDA. That initial EUA was for elderly and at-risk persons only, and it would be another four months until the vaccine was available to all adults in the United States. 

The Rapid Deployment Vaccine Collaborative (Radvac) (\href{https://radvac.org/}{radvac.org}) was founded in 2020 to develop a new pathway for vaccine production and dissemination. Radvac pursues open-source vaccine technology that enables the rapid design, synthesis, and distribution of vaccines in emergency situations. Radvac has developed and published twelve iterations of an experimental intranasal, multivalent, multi-epitope peptide vaccine. 

Radvac is skeptical that the government will adopt the regulatory and legal reforms required to ensure evidence-based vaccine authorizations and deployments during future pandemics. Thus, Radvac remains focused and committed to open-source vaccine technology and the development of platforms for radically decentralized vaccine manufacture and distribution. However, Radvac recognizes that vaccines have limitations even if the problems surrounding rapid manufacture and dissemination are solved. First and foremost, fear and skepticism of vaccines remain strong among the general public. Radvac recognizes that a large fraction of the population likely will be unwilling to take any vaccine that isn't extensively tested in a historically typical manner (preclinical and multi-phase clinical trials lasting between one and several years).\footnote{A large body of work suggests that vaccine hesitancy can be reduced by moving away from needles towards nasal and oral vaccination.}

Adaptive immunity takes about one to two weeks to be realized after vaccination, and this is unlikely to be sped up without considerable re-engineering of the human immune system. Most troubling, many pathogens have proven to be stubbornly hard to vaccinate against. The canonical example is HIV, a deadly pathogen for which there is no approved vaccine despite over four decades and many billions of dollars of research investment and multiple Phase III trials. Hepatitis C has also resisted thirty years of dedicated vaccine research. The development of a vaccine for RSV took 60 years and required a breakthrough in understanding the conformations of RSV's fusion protein. Developing vaccines for the Zika and Dengue viruses has proven difficult due to the potential for antibody-dependent enhancement.\footnote{Zika is very similar to Dengue virus, and Zika infection produces antibodies that have cross-reactive efficacy against Dengue. Unfortunately, laboratory studies show some Zika antibodies can promote Dengue infection by enabling the Dengue virus to infect macrophages. The same is true for Dengue antibodies - some promote Zika infection. Additionally, Dengue has four serotypes (DENV-1, DENV-2, DENV-3, DENV-4) that are about equally prevalent, and each serotype requires unique antibodies.} Many other current vaccines leave much to be desired as well. Most notably, vaccines for COVID-19 and flu save lives, but are largely ineffective at preventing illness and spread. 

In light of all this, Radvac has been researching technologies for the rapid identification and deployment of antivirals, with a particular focus on technologies that enable the rapid repurposing of existing drugs and generally-recognized-as-safe (GRAS) compounds. During the spread of an emerging pathogen, the ideal scenario is the identification of one or more compounds with pre-existing safety records and wide availability that can be deployed as antivirals essentially immediately (in hours or days, as opposed to many months or years for a new regulatory body approved drug). As shown in table \ref{tab:antiviral_landscape}, there are no FDA approved antiviral drugs for many families of viruses that have pandemic potential for humans. Radvac believes that this lack of effective antivirals constitutes a major gap in our collective defense against future pandemics. 

Table \ref{tab:antiviral-classes} provides an overview of ten classes of antiviral agents: neutralizing antibodies, virus-directed small
molecules, host-directed small molecules, recombinant soluble human receptors, antiviral CRISPR/Cas systems, interferons, antiviral peptides/peptoids, antiviral nucleic acid polymers, and sntisense oligonucleotides (ASOs) and small interfering RNAs (siRNAs). Radvac maintains an active interest in several of these classes, especially antiviral small molecules, antiviral peptides, antiviral nucleic acid polymers, and ASOs/SiRNAS. 

Antiviral peptides and antiviral nucleic acid polymers can be synthesized quickly, but delivery to cells remains challenging. In contrast, most small molecule drugs can be taken orally or nasally. As discussed, there is also a large repertoire of existing small molecule compound readily available for purchase and many that have existing safety records. Short oligonucleotide (ASOs and siRNAS, ranging from about 12 to 50 bases) can be designed against known pathogens with relative ease using standard base-pairing rules for nucleic acids. Recent advances in third generation oligonucleotide chemistries and in delivery technologies have allowed intranasal administration of antiviral oligonucleotides.\cite{Chokwassanasakulkit2024,Zhu2022}\footnote{Antiviral ASOs are an important topic and area of active research for Radvac, and Radvac is working on a separate publication focused on ASOs.} While there are many exciting new AI-enhanced tools for antibody design,\cite{Naughton2026,Pacesa2025,Stark2025} antibodies are still expensive and difficult to manufacture and deliver at scale. Antibodies also require cold storage and delivery via IV or injection, making rapid dissemination challenging. Finally, viruses can quickly evolve resistance to hyper-specialized antibodies, as was observed during the pandemic.\cite{Cox2022} Therefore, in this work we focus on small molecules, with a particular focus on virus-directed small molecules (as opposed to antivirals that target human proteins).

%---------------------------------------------------------------------- 
\subsection{Drug repurposing}
Drug repurposing has a long history.\cite{Ashburn2004} The antidepressant Bupropion was repurposed as a smoking cessation agent by GlaxoSmithKline in the late 1990s. Famously, the two most popular erectile dysfunction drugs on the market today both started as cardiovascular-related medications. However, there are not good economic incentives for companies to invest in repurposing efforts. The FDA currently only grants three years of marketing exclusivity for repurposed drugs, or seven years for drugs that treat rare diseases. While a company might get three years of marketing exclusivity, if the drug is off-patent (likely) then other companies can still produce and sell it at low cost in a generic form. Doctors will then naturally prescribe the cheaper generic off-label for the repurposed application. This is why there is little economic incentive for companies to invest in repurposing. In fact, it is nearly impossible in most cases to recoup the investment costs required for the preclinical research and large-scale clinical trials that the FDA demands. While finding a new chemical entity may be harder and more laborious, it is vastly more profitable as the end reward is 10-15 years of exclusive rights for both manufacture and marketing. Although this primary barrier is sufficient to kill most repurposing efforts, there is yet another substantial barrier that we discuss in greater detail below: repurposed drugs are far more likely to be effective when used in combinations, but if each drug in an effective combination is owned by different companies, the complexities of competition and cross-licensing agreements provide further disincentives for repurposing.

The lack of economic incentives for industry to focus on antiviral repurposing means that academic and non-profit research in this area is especially important. During the pandemic, many existing anti-inflammatory drugs were quickly and successfully utilized to treat severe COVID-19.\cite{PratsUribe2021}  Academic researchers also simultaneously explored repurposing existing drugs as antivirals. The effort yielded one definite repurposing success story -- baricitinib, a Janus kinase inhibitor first approved by the FDA for rheumatoid arthritis. Remdesivir is also frequently cited as a ``repurposing'' success -- while not previously approved, it had been heavily studied in prior clinical trials. Baricitinib treats COVID-19 via two mechanisms - the first being immunosuppression (helpful for preventing the notorious ``cytokine storm'') and the second being a mechanism that blocks SARS-CoV-2 from entering cells. During the pandemic the FDA issued an Emergency Use Authorization for baricitinib in combination with Remdesivir, reasoning that it was necessary to combine Baricitinib with an antiviral due to its immunosuppression. (The EUA was quite narrow, however, and only applied to patients requiring supplemental oxygen, mechanical ventilation, or ECMO.) 

Fraudulent trials, poor quality research, and a proliferation of pseudoscientific misinformation made the repurposing of hydroxychloroquine and ivermectin very popular, even after further research established they are not effective. EHR records from 2020 show that unsuccessful repurposing attempts were made with many existing drugs, the most common being hydroxychloroquine, azithromycin, lopinavir/ritonavir, and umifenovir.\cite{PratsUribe2021} In March 2020 around 50\% of patients hospitalized for COVID-19 in the United States received hydroxychloroquine.\cite{PratsUribe2021} Hydroxychloroquine turned out to be a false positive due to the use of Vero E6 and human hepatocyte cells in cell culture screens (where it blocks SARS-CoV-2 entry) rather than human lung cells (where it does not block entry).\cite{Dittmar2021} There were also many false positive hits in cell culture repurposing screens due to colloidal aggregation\cite{Glenn2024colloidalagg} and phospholipidosis.\cite{Tummino2021} We worry that these negative experiences with repurposing during the pandemic have overshadowed the success stories and dampened interest in researching repurposing as a pandemic-preparedness strategy. 

%---------------------------------------------------------------------------------------F----
\subsection{Broad-spectrum antivirals}
Broad spectrum antivirals (BSAs) are drugs that exhibit antiviral activity across several virus families. One of the most widely touted classes of BSAs are nucleoside analogs. The approved nucleoside analogs Ribavirin, Remdesivir, and Favipiravir all have established activity in cell culture across several viral families. 
There are also a handful of protease inhibitors which have activity against multiple viruses in cell culture.\cite{Li2021}
%Other approved nucleoside analogs include Sofosbuvir for HCV, molnupiravir for SARS-CoV-2, and brincidofovir for smallpox, but their utility as BSAs is unclear.  

Other BSAs are host-directed. For instance, the drugs camostat mesylate and nafamostat mesylate (both approved in Japan) inhibit TMPRSS2, an enzyme found on the surface of lung cells which is important for coronavirus and influenza virus cell entry. In vitro, camostat mesylate exhibits antiviral activity against SARS-CoV-2, MERS, HCoV-229E, and influenza A (H1N1, H3N2). In a Phase II trial for COVID-19, camostat mesylate showed a small but statistically non-significant reduction of illness duration.\cite{Kim2023}

Another example of a host-directed BSA is the immunosuppressant drug cyclosporine A, which is approved for psoriasis and a variety of other diseases.\cite{GarcaSerradilla2019} The primary effect of cyclosporine A in the body is to inhibit cyclophilin. The cyclosporine–cyclophilin complex then inhibits calcineurin which in turn reduces T cell activity. Many viruses rely on cyclophilin during replication (including HIV-1, HCV, HBV, and SARS-CoV-2). (In particular, cyclophilin modifies the backbone conformation of proline residues, assisting in the folding of certain proteins.) While cyclosporine A's potent immunosuppressive and immunomodulatory effects might naively seem like a liability in the treatment of viral illness, a Phase II trial of cyclosporine A for COVID-19 showed that cyclosporin A reduces the risk of the ``cytokine storm'' dramatically in hospitalized patients.\cite{Zidan2026} There have been several small unblinded trials of cyclosporin A for COVID-19.\cite{Elhabyan2025} The results are mixed, with benefits that are often not statistically significant, but overall the human trials data suggest potential weak effectiveness against SARS-CoV-2. 

Nitazoxanide, an FDA-approved antiparasitic medication, inhibits the replication of a broad range of RNA and DNA viruses in cell culture, including RSV, parainfluenza virus, coronaviruses, rotavirus, norovirus, hepatitis B, hepatitis C, dengue, yellow fever, Japanese encephalitis virus, and HIV.\cite{Rossignol2014} Nitazoxanide exhibits several different mechanisms of antiviral action that are relevant to differing degrees for different viruses.\cite{Li2021} Additionally, some work suggests that nitazoxanide upregulates interferon-stimulated genes, boosting innate antiviral response mechanisms.\cite{Jasenosky2019} Nitazoxanide is one of the best validated and most impressive BSAs since it has several clinical trials supporting its efficacy against several viral families. A Phase 2b/3 trial showed nitazoxanide reduced influenza symptom duration and viral shedding,\cite{Haffizulla2014} and other trials have demonstrated varying degrees of efficacy against rotavirus, norovirus, hepatitis B, hepatitis C, and SARS-CoV-2.\cite{Rossignol2006,Rossignol2014,Rossignol2022}

In 2022 researchers discovered that neomycin, an active ingredient in antibacterial ointments, triggers the innate immune system to produce a variety of antiviral factors in the nasal mucosa, mimicking the effects of interferon alpha.\cite{Mao2024} Intranasal neomycin is now considered an emerging BSA.\cite{Mao2024}  

BSAs are generally only weakly effective (they are ``jacks of all trades but masters of none''). For this reason some BSAs have struggled to obtain regulatory approvals for certain indications, and to become established choices in the marketplace of options available to clinicians. When regulators like the FDA review drug applications they only consider efficacy against a single disease within a certain population in aggregate, weighing benefits vs risk for that application in isolation and ignoring possible benefits for treating other diseases, including novel pathogens. Although the FDA is not legally required to demand that a new therapeutic is better than the existing standard of care, in recent decades the FDA has evolved toward requiring that drug developers demonstrate a definite improvement over existing treatments.\footnote{FDA expert Joseph Gulfo has convincingly argued that the FDA has overstepped their legal mandate by requiring that new drugs be better than existing on average, or requiring that they be lower cost. This leads a lack of options for doctors which causes many issues. Having multiple options is important since some patients can not tolerate certain drugs due to genetics or drug-interactions.\cite{Elton2024}} Unfortunately, this can make it very hard for BSAs that are weakly effective against many viruses to be approved by the FDA. For example, although the BSA Remdesivir showed weak efficacy against Ebola virus in preclinical work, its Phase III trial was terminated early due to its clear inferiority compared to monoclonal antibodies. Thus, when the pandemic began Remdesivir was not approved for any indication and only 5,000 doses were available globally.\cite{wang2024dayzero} If Remdesivir had been widely available, many lives could have been saved. 
\begin{table}[!ht]
\centering
\caption{Ten classes of antiviral agents.}
\label{tab:antiviral-classes}
\fontsize{7.5}{9.35}\selectfont
\setlength{\tabcolsep}{5pt} %6 is default
\begin{tabular}{p{2cm} p{6cm} p{6.5cm}}
\toprule
\textbf{Class} & \textbf{Advantages} & \textbf{Challenges} \\
\midrule

\textbf{Neutralizing antibodies}&
\bb High specificity and potency\newline
%\bb Multiple sourcing platforms (convalescent plasma, humanized mAbs, chicken IgY, bovine colostrum)\newline
\bb Fast-acting\newline
\bb Can be sourced from human convalescent plasma, humanized animals, llamas, chickens, bovine colostrum\newline
\bb New computational tools for custom design
%\bb IgY avoids complement activation and ADE\newline
&
\bb Expensive production requiring complex protein purification\newline
\bb Potential to trigger autoimmunity requires testing during Phase I\newline
\bb Generally narrow-spectrum, very prone to viral escape via surface-protein mutations\newline
\bb Require cold-chain storage and IV or subcutaneous delivery\newline
\bb Antibody-dependent enhancement risks\\

\midrule
\textbf{Virus-directed small molecules} &
\bb Oral consumption increases ease of distribution \& uptake\newline
%\bb Largest share of $\sim$100 approved antiviral drugs\newline
\bb Well-established infrastructure for manufacturing and distribution\newline
\bb Potentially broad-spectrum, especially for nucleoside analogs&
\bb Interaction with other medications\newline
\bb Hard to predict ADMET and side effects\newline
\bb Resistance emergence under monotherapy \newline
\bb Clinical translation attrition remains high\newline
\bb Synthesis can be difficult and expensive\\

\midrule
\textbf{Host-directed small molecules} &
\bb Much less susceptible to viral resistance mutations\newline
\bb Pro-apoptotic subclass selectively kills infected cells\newline
\bb More potential for broad-spectrum efficacy for many viruses &
\bb Interaction with other medications\newline
\bb Hard to predict ADMET and side effects\newline
\bb More side effects due to perturbation of host-cell physiology\newline
\bb Fewer approved precedents than virus-directed antivirals\newline
\bb Synthesis can be difficult and expensive\\

\midrule
\textbf{Recombinant soluble human receptors} &
\bb Possibly broad spectrum (carbohydrate receptors\cite{Ezzatpour2025})\newline
\bb More resistant to viral escape\newline &
\bb Limited to viruses with well-characterized receptor interactions\newline
\bb Expensive manufacturing requiring expensive purification and cold-chain delivery\newline
\bb Emerging technology with few clinical precedents (best example: HrsACE2 for COVID-19)\newline
\bb May perturb endogenous receptor signaling\\

\midrule
\textbf{CRISPR/Cas systems} &
\bb Programmable---guide RNAs \newline
\bb Potential for curative one-time gene therapy\newline
\bb Cas13b mismatch tolerance limits mutational escape&
\bb New technology in very early clinical stage\newline
\bb In vivo delivery to cells is a major barrier\newline
\bb Off-target editing risks\newline
\bb Cas proteins have immunogenicity risks\\

\midrule
\textbf{Interferons}&
\bb Broad-spectrum activity\newline
\bb Approved clinical track record (HCV, HBV)\newline
\bb Trigger innate immune system cascades\newline &
\bb Significant adverse effects (flu-like symptoms, cytopenias, neuropsychiatric toxicity)\newline
\bb Many viruses encode interferon evasion mechanisms\newline
\bb Low monotherapy efficacy for many viruses\newline
\bb Timing-dependent--late administration may worsen inflammation with little benefit\newline
\bb Requires IV or injection to deliver\\

\midrule
\textbf{Antiviral peptides and peptoids} &
\bb Can exhibit a diverse array of mechanisms of action (peptides can be entry, fusion, enzyme, and assembly inhibitors)\newline
\bb Relatively easy to design\newline 
\bb Easy, fast, and inexpensive to synthesize\newline &
\bb Difficulties with oral bioavailability and very short half-life\newline
\bb Typically require parenteral delivery\newline
\bb Higher manufacturing cost than small molecules\newline
\bb Poor clinical trial history\newline
\bb Potential immunogenicity risks\\

\midrule
\textbf{Nucleic acid polymers} &
\bb Exhibit novel mechanisms for directly inhibiting viral entry or replication \newline
\bb Computationally designable and potentially programmable from viral sequence data\newline 
\bb Easy and fast to synthesize\newline &
\bb High manufacturing complexity and cost\newline
\bb Novel technology with little late stage clinical validation\newline
\bb Intracellular delivery is a major barrier\newline
\bb Susceptible to nuclease degradation without chemical modification\newline
\bb Off-target gene silencing risks\\

\midrule
\textbf{Antisense oligonucleotides (ASOs) and small interfering RNAs (siRNAs)} &
\bb Programmable and rapidly designable from viral genome sequence\newline
\bb High specificity to conserved viral RNA regions; can be multiplexed to reduce escape\newline
\bb Potential for broad-spectrum activity by targeting conserved motifs or host proteins\newline
\bb Scalable synthetic manufacturing (no biologics expression systems required)\newline
\bb Precedent in approved RNA therapeutics supports regulatory pathway\newline &
\bb Limited clinical validation as antivirals to date compared to other modalities\newline
\bb In vivo delivery remains challenging\newline
\bb Susceptible to nuclease degradation without chemical modification\newline
\bb Risk of off-target gene silencing and immune activation\newline
\bb Typically require repeated dosing due to transient effects\\

\midrule
\textbf{Aptamers} &
\bb Short ssDNA/ssRNA that bind protein targets via 3D shape complementarity, like antibodies. 
&
\bb Hard to deliver into cells.\newline
\bb No clinical trials as antivirals in humans as of 2026.\\
\bottomrule
\end{tabular}
\end{table}
\FloatBarrier

%-------------------------------------------------------------------------------------------
\subsection{Antiviral combinations}
Many antivirals are only weakly active, and thus are currently left ``on the wayside'' by the pharmaceutical industry. However, several compounds that are weak in isolation can become very powerful in combination. Drugs can also be synergistic in combination -- achieving a greater effect than would be expected assuming they acted independently. Additionally, when antivirals that act on different targets are combined, it becomes much harder for viruses to evolve resistance to the treatment.

Combination therapies have become common in oncology, since cancers are good at evolving resistance to monotherapies. Similarly, combination antiviral therapies are standard for rapidly evolving viruses. For example, human immunodeficiency virus (HIV) is notorious for rapidly developing resistance to a range of therapies, and it is currently accepted that long-term control of HIV requires combinations of antiviral drugs. The FDA-approved HIV drug Triumeq contains three different antivirals that target two different steps in HIV's lifecycle. There are over twenty FDA-approved HIV drugs that utilize combinations. While combination antiviral therapies have become standard for HIV and hepatitis C (largely out of necessity), combination therapies remain little explored for other viruses and for pandemic preparedness in general -- although the success of Paxlovid against SARS-CoV-2 might be considered an exception. Paxlovid is a combination of two distinct, co-packaged drugs, nirmatrelvir and ritonavir, with ritonavir acting to prevent the rapid breakdown of nirmatrelvir in the liver. While Paxlovid has been remarkably effective, scientists are now finding SARS-CoV-2 strains that are resistant to nirmatrelvir, Paxlovid's active ingredient.\cite{Duan2023} 
 
Designing combination therapies naturally dovetails with repurposing, since repurposed compounds are naturally more likely to be weak antivirals rather than strong ones. During the pandemic, experimental repurposing screens showed several weak hits (micromolar potency), but no strong hits (nanomolar potency).\cite{Ghahremanpour2020,Hosseini2021} Additionally, a large body of pre-clinical work suggests that some GRAS herbal products such as mint and licorice root may have weak antiviral potency against SARS-CoV-2.\cite{Buck2023, Alikiaie2023} These weak antiviral compounds might be quite powerful if combined in the right way, but designing such combinations has not received much attention. In a 2025 commentary, Kainov et al. suggest that the best antiviral strategy involves combining two or three virus-directed antivirals and possibly one host-directed antiviral, with each drug acting on a different target.\cite{Kainov2025}  

There are several challenges to designing combinations. First, the side effect and toxicity profiles of each compound need to be well-understood. If side-effects or toxicity profiles overlap, the burden might be too great, possibly causing more harm than good. As an example, both hydroxychloroquine and azithromycin prolong the heart's QT interval by blocking the hERG potassium channel, and this overlap makes the combination dangerous. 

If side-effect data are not readily available, ML models trained on side-effect ``big data'' may be helpful to reduce risk, although the utility of such models for designing combination therapies has only been studied in a few distinct areas like cancer therapy and liver disease treatment. For instance, the MADRIGAL model is a custom transformer model that aims to predict the side effect profiles of drug combinations, in addition to potential efficacy.\cite{madrigalarxiv} 

The greatest challenge for designing drug combinations might be predicting drug-drug interactions. Drugs might work independently, synergistically, or antagonistically. While antagonism might seem unlikely, it is common enough to warrant serious concern. In a cell culture study of 73 combinations of 32 SARS-CoV-2 drug candidates, researchers found sixteen synergistic combinations and eight antagonistic combinations.\cite{Bobrowski2021} For example, they found hydroxychloroquine (HCQ) was antagonistic to remdesivir because HCQ impairs the intracellular activation of the remdesivir prodrug. While remdesivir and nirmatrelvir appeared synergistic in a mouse model,\cite{Gidari2023} a study of EHR records suggests that they are strongly antagonistic in humans, possibly because nirmatrelvir impairs the metabolism of remdesivir in humans.\cite{Choi2024}  

%When repurposing, it is often the case that the dose needed to achieve a significant therapeutic effect is much higher than the doses approved by the FDA or that were studied in clinical trials. Moving into such high dosing regimes generally increases side effect load and can lead to the uncovering of entirely new types of side effects. 
%A common issue with natural products is that the amount that actually reaches cells is very low due to poor absorption and/or high metabolism, limiting the intercellular concentration that can be achieved.   

To optimize combinations, brute-force screening in cell culture is difficult due to combinatorial explosion – if considering 100 compounds, the number of two-drug combinations is $100*99/2 = 4,950$ and the number of three drug combinations is $100*99*98/6 = 161,700$. Methods like pooled screening are currently being developed to work around this combinatorial explosion.\cite{Shyr2021} In general, combination therapies must be designed strategically, using knowledge of biological mechanisms and metabolic pathways. Large language models might be particularly useful here given their extensive in-built knowledge of biological pathways. One model worth noting here is TxGemma, a 28 billion parameter open-source LLM released by Google DeepMind in early 2025 which was trained specifically to predict and reason about drug efficacy and side effects.\cite{txgemma}

\begin{table}[htbp]
\centering
\caption{There are 29 viral families known to cause disease in humans. The taxonomy here follows the ICTV Master Species List as of 2025. The ICTV regularly makes major changes to the viral taxonomy. We define ``pandemic potential'' broadly to mean that a viral family/genus is capable of evolving species that cause disease that becomes globally endemic. The genus Alphacoronavirus is thought to be limited to causing mild illness, so often not considered to be a genera with ``pandemic potential''. Similarly, HIV is not believed to be able to transmit to more easily transmissible forms, so it is also sometimes omitted. We include both in our list. For a more detailed breakdown, see \href{https://antivirals-database.radvac.org/viral-families}{https://antivirals-database.radvac.org/viral-families}.}
\label{tab:antiviral_landscape}
\scriptsize
\renewcommand{\arraystretch}{0.95}
\begin{tabular*}{\textwidth}{@{}p{0.5cm}l@{\extracolsep{\fill}}l@{\hskip 5pt}l@{\hskip 5pt}l@{\hskip 4pt}l@{}}
\toprule
 & \textbf{Family} & \textbf{Key diseases} & \makecell{\textbf{FDA}\\[-2pt]\textbf{anti-}\\[-2pt]\textbf{virals?}} & \makecell{\textbf{FDA}\\[-2pt]\textbf{mAbs?}} & \makecell{\textbf{Pandemic}\\[-2pt]\textbf{poten-}\\[-2pt]\textbf{tial?}}\cite{Jochmans2023}\\
\midrule
% --- DNA viruses (8 families) ---
\multirow{8}{*}{\rotatebox[origin=c]{90}{\textbf{DNA}}}
% & Anelloviridae      & TTV (no confirmed disease)           & \textcolor{noAV}{No}     & \textcolor{noAV}{No}  & No \\
 & Parvoviridae       & Fifth disease, hydrops fetalis       & \textcolor{noAV}{No}     & \textcolor{noAV}{No}  & No \\
 & Polyomaviridae     & PML, BK nephropathy                  & \textcolor{noAV}{No}     & \textcolor{noAV}{No}  & No \\
 & Papillomaviridae   & Warts, cervical cancer               & \textcolor{noAV}{No}     & \textcolor{noAV}{No}  & No \\
 & Adenoviridae       & Pneumonia, gastroenteritis           & \textcolor{noAV}{No}$^a$ & \textcolor{noAV}{No} & \textcolor{noAV}{Yes} \\
 & Hepadnaviridae     & Hepatitis B                          & Yes                      & \textcolor{noAV}{No}  & No \\
 & Orthoherpesviridae & HSV, VZV, CMV, EBV                   & Yes                      & \textcolor{noAV}{No}  & No \\
 & Poxviridae         & Smallpox, mpox                       & Yes                      & \textcolor{noAV}{No}  & \textcolor{noAV}{Yes} \\
\midrule
% --- (+)ssRNA viruses (8 families) ---
\multirow{12}{*}{\rotatebox[origin=c]{90}{\textbf{(+)ssRNA}}}
 & Astroviridae       & Gastroenteritis                      & \textcolor{noAV}{No}    & \textcolor{noAV}{No}  & No \\
 & Picornaviridae     & Common cold, polio, Hep A            & \textcolor{noAV}{No}    & \textcolor{noAV}{No}  & \textcolor{noAV}{Yes} \\
 & Hepeviridae        & Hepatitis E                          & \textcolor{noAV}{No}    & \textcolor{noAV}{No} & No \\
 & Caliciviridae      & Norovirus gastroenteritis            & \textcolor{noAV}{No}$^a$& \textcolor{noAV}{No}  & No \\
 & Matonaviridae      & Rubella                              & \textcolor{noAV}{No}    & \textcolor{noAV}{No}  & No \\
 & Togaviridae        & Chikungunya, EEE                     & \textcolor{noAV}{No}    & \textcolor{noAV}{No}  & \textcolor{noAV}{Yes} \\
 & Flaviviridae       &                                      &                           &             &  \\
 &  \textit{ Genus Flavivirus} &Dengue, Zika, Yellow Fever, West Nile & \textcolor{noAV}{No}  & \textcolor{noAV}{No}  & \textcolor{noAV}{Yes} \\
 &  \textit{ Genus Hepacivirus}& Hepatitis C                 & Yes               & \textcolor{noAV}{No}  & No \\
 & Coronaviridae      &                                      &                          &                       &  \\
 &  \textit{ Genus Alphacoronavirus}& Common cold            & \textcolor{noAV}{No}     & \textcolor{noAV}{No}  & \textcolor{noAV}{Yes} \\
 &  \textit{ Genus Betacoronavirus} & COVID-19, SARS, MERS   & Yes                      & \textcolor{noAV}{No}$^\dagger$ & \textcolor{noAV}{Yes} \\
\midrule
% --- (-)ssRNA viruses (11 families) ---
\multirow{11}{*}{\rotatebox[origin=c]{90}{\textbf{(\textminus)ssRNA}}}
 & Bornaviridae       & Borna disease encephalitis             & \textcolor{noAV}{No}    & \textcolor{noAV}{No}  & No \\
 & Orthomyxoviridae   & Influenza                              & Yes                      & \textcolor{noAV}{No}  & \textcolor{noAV}{Yes} \\
 & Paramyxoviridae    & Measles, mumps, parainfluenza          & \textcolor{noAV}{No}   & \textcolor{noAV}{No}  & \textcolor{noAV}{Yes} \\
 & Pneumoviridae      & RSV, hMPV                              & \textcolor{noAV}{No}   & Yes (RSV)             & \textcolor{noAV}{Yes} \\
 & Rhabdoviridae      & Rabies                                 & \textcolor{noAV}{No }  & \textcolor{noAV}{No}  & No \\
 & Filoviridae        & Ebola, Marburg                         & \textcolor{noAV}{No}   & Yes (Ebola)           & \textcolor{noAV}{Yes} \\
 & Arenaviridae       & Lassa fever, LCMV                      & \textcolor{noAV}{No}$^a$& \textcolor{noAV}{No} & \textcolor{noAV}{Yes} \\
 & Hantaviridae       & HPS, HFRS                              & \textcolor{noAV}{No}   & \textcolor{noAV}{No}  & \textcolor{noAV}{Yes} \\
 & Nairoviridae       & Crimean-Congo hemorrhagic fever        & \textcolor{noAV}{No}   & \textcolor{noAV}{No}  & \textcolor{noAV}{Yes} \\
 & Phenuiviridae      & Rift Valley fever, SFTS                & \textcolor{noAV}{No}   & \textcolor{noAV}{No}  & \textcolor{noAV}{Yes} \\
 & Peribunyaviridae   & La Crosse encephalitis, Oropouche      & \textcolor{noAV}{No}   & \textcolor{noAV}{No}  & \textcolor{noAV}{Yes} \\
\midrule
% --- RT viruses ---
\multirow{1}{*}{\textbf{RT}}
 & Retroviridae       & HIV/AIDS, HTLV                         & Yes                    & Yes (HIV)             & \textcolor{noAV}{Yes} \\
\midrule
% --- dsRNA viruses ---
\multirow{1}{*}{{\tiny \textbf{dsRNA}}}
 & Sedoreoviridae    & Rotavirus gastroenteritis              & \textcolor{noAV}{No}   & \textcolor{noAV}{No}  & No \\
\midrule
\bottomrule
\end{tabular*}
\vspace{4pt}
\raggedright%\footnotesize
$^a$ Off-label treatment available (cidofovir for adenoviridae, ribavirin for Hepeviridae and Arenaviridae.)\\
$\dagger$ As of January, 2026 almost all emergency use authorizations for SARS-CoV-2 mAbs have been revoked; only pemivibart (Pemgarda) retains an EUA for pre-exposure prophylaxis in immunocompromised patients.
\end{table}

\begin{figure}[htbp]
\centering
\includegraphics[width=0.85\textwidth]{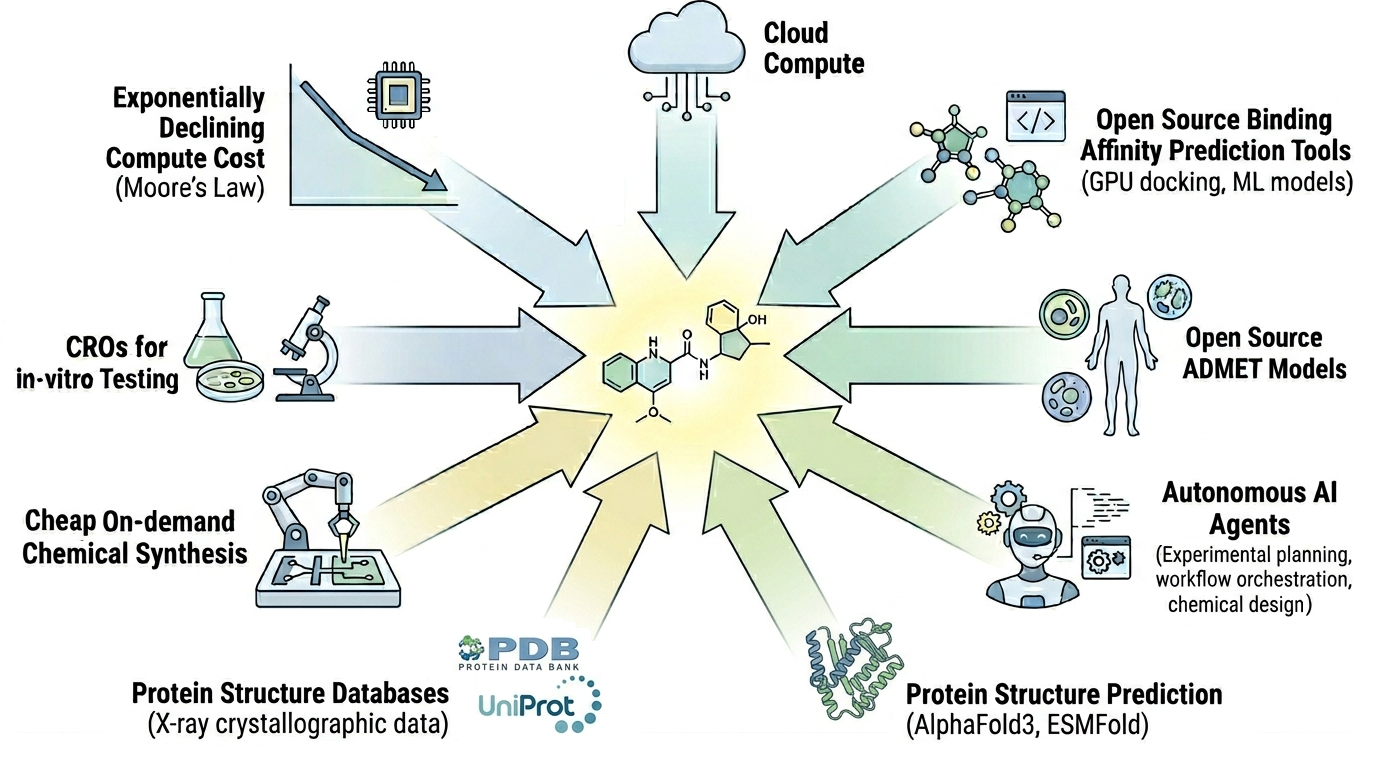}
\caption{Technologies and trends that are converging to accelerate early stage drug discovery and lower barriers to entry.}
\label{fig:converging_technologies}
\end{figure}

%-------------------------------------------------------------------------------------------
\section{Overview of drug-target affinity prediction}

For decades scientists have dreamed of being able to design drugs in silico, avoiding expensive and time-consuming laboratory testing. Now, a number of technologies and trends are converging to make in silico drug discovery a reality (see figure \ref{fig:converging_technologies}). 

Drug-target affinity (DTA) prediction attempts to predict the binding affinity of a chemical compound to a target protein. In the related task of compound-protein interaction (CPI) prediction, this is simplified to doing a binary classification – ie one seeks to classify compounds into ``interacts'' or ``does not interact''. 

Prediction methods fall into three major categories - machine learning (ML)-based, physics/docking based, and network based. This categorization is not completely clean, as ML methods are sometimes used to improve physics-based or network-based methods. 

\subsection{Overview of machine-learning based methods}
Using machine learning for DTA prediction is not new - it actually dates back to the 1980s under the banner of quantitative structure-activity relationship (QSAR) modeling. In early QSAR models, 2D molecular graphs were converted into fingerprint vectors that could be inputed into ML models. Some fingerprint vectors were constructed by hand using chemical knowledge, while others are algorithmically generated. Popular ML models were support-vector machines (SVMs), random forest models, and multilayer neural networks. In the mid 2010s QSAR modeling evolved away from using fingerprint vectors towards ML models that work with SMILES strings (like LSTMs) or with the molecular graph directly (like GNNs).\cite{Elton2019}

A key finding consistent across the literature is that machine learning models trained for DTA or CPI prediction do not generalize well to types of protein or ligand molecule that lie outside their training data. This is related to the fact that heuristics in drug discovery are generally brittle. If one imagines the relation between structure and affinity as a manifold the manifold exhibits many sharp ``activity cliffs''. An example of an activity cliff is when removing a single atom, such a single hydrogen, dramatically increases or decreases binding affinity. In some cases, removing a single hydrogen can drop the number of hydrogen bonds in the binding pocket from three to two, completely destabilizing a molecule and moving it from a strong binder to a non-binder. An analogy here is how sawing off one tiny section of a key makes it unable to turn a lock. The fact that chemical space is highly jagged means it is intrinsically hard to fit functions (ie with neural nets) that map structure to affinity. The brittleness of the problem extends to the protein side as well - in some cases a single amino acid substitution can cause binders to become non-binders and vice-versa. 

Since AI/ML models do not generalize, the best one can do is try to fit such models to as large and diverse a training set as possible. A key element of diversity is having both binders and non-binders in one's training dataset. A common pitfall is having too many binders (actives), and not enough non-binders. In such cases, models become biased towards predicting activity. A common intuition is that the ratio of binders to non-binders should be 50:50 in the training data. Indeed, Jiang et al.~showed that for CPI prediction their best validation performance was obtained when the ratio of binders to non-binders was 1:1.\cite{Jiang2022} Another pitfall is not being aware of clustering in datasets. A dataset may have thousands of molecules, but not be diverse since it consists of many slight variations on a few chemical scaffolds. When trained on such datasets, models end up learning very crude heuristics based on spurious (non-causal) features (heuristics like  ``if scaffold X is present, then molecule Y binds strongly''). Unfortunately, chemical datasets are full of biases leading to ML models learning spurious correlations. For instance, lead-discovery programs often are run by experimentally testing of dozens or hundreds of variations of a particular molecular scaffold and then mass-patenting all of the molecules that appear to have potential. Sets of molecules (and affinity data) published in patents often end up being scraped into BindingDb. Thus, the examples in BindingDb are often highly clustered.\footnote{Cleves and Jain say that human chemists have ``2D bias'' -- they think of molecular structures as 2D graphs. However, what is actually physically important is how pharmacophores (substructures important for binding) are positioned in 3D space. Two molecules may look very different in terms of their 2D graph but have a similar placement of pharmacophores.\cite{Cleves2007}}

%-------------------------------------------------------------------------------------------
\subsection{Overview of docking}
The main competitor to ML-based approaches is docking. Docking requires the 3D structure of the target protein and the 3D structure of the ligand, and use a physics-derived approximate energy scoring function to test different possible binding configurations. Docking methods perform a search over molecular binding configurations or ``poses''. Docking has been quite successful over the years, contributing to the development of HIV protease inhibitors and the neuraminidase inhibitors zanamivir (Relenza) and oseltamivir (Tamiflu). Docking also played an important role in the discovery of both nirmatrelvir (the active ingredient in Paxlovid) and Ensitrelvir, a SARS-CoV-2 Mpro inhibitor approved in Japan. Recently researchers have developed ways of increasing the accuracy of docking scoring functions by using machine-learning based scoring functions. 

%--------------------------------------------------------
\subsection{Overview of network-based approaches}
Network-based approaches utilize large knowledge graphs containing either protein-protein, drug-protein, drug-disease, or drug-drug interactions.\cite{deepDR,Zeng2020,Zhou2020,Wang2023,Huang2024,MorselliGysi2021} Network-based models for COVID-19 drug repurposing were very popular among academics during the pandemic, leading to dozens of publications, with approaches varying from simple network interpolations to methods that incorporate ML techniques like graph neural networks and large language models.\cite{Huang2024} 

There are two major limitations to network models which should be mentioned up front. The main limitation is that only drugs and compounds in the graph can be considered for repurposing. To be in the graph, a compound must have a (compound, disease) or (compound, protein target) association. Generally recognized as safe compounds and natural products that are not in the graph cannot be considered as candidates. Interestingly, these networks almost exclusively work with positive connections only (ie this drug binds to $X$ or treats $Y$) and rarely include negative connections (ie this drug does NOT bind $X$ or does NOT treat $Y$).

Another major limitation is that it may be difficult to incorporate a novel pathogen into a network model. To link a novel viral pathogen, you need to either link via virus-host interactions (if known) or by proximity to an existing viral disease already in the graph. Additionally, knowing a few viral-host interactions is likely not enough, you actually need many to clearly distinguish the novel pathogen from existing pathogens. Also, if the existing disease you link to has sparse connections, you may not get much out. On the other hand, if you can successfully link a new viral pathogen to the graph, you can do so without knowing anything about the virus's structure. Affinity prediction methods (discussed below) often require the crystal structure of a viral target protein, which may not be available for an emerging pathogen. 

Looking at some of the network-based repurposing publications from top journals, we first observed that network models tended to surface anti-inflammatory agents and rarely surfaced antivirals. While the method from network science pioneer Albert-L\'aszl\'o Barab\'asi's lab surfaced Ritonavir as a top repurposing candidate, it was an exception.\cite{MorselliGysi2021} No network-based approach we looked at recommended the broad-spectrum nucleoside analog antivirals Ribavirin (FDA approved for HCV) and Favipiravir (approved in Japan for influenza). While their utility for COVID-19 appears to be limited, Ribavirin and Favipiravir are obvious candidates for antiviral repurposing, and it would have been nice to see network models surfacing them. Favipiravir in particular has had a successful Phase II trial and was approved in Russia for COVID-19, although a Cochrane review finds the evidence for its use to be so low quality they are unable to recommend its use.\cite{Korula2022}

The network-based models we looked at all recommended a lot of drugs/supplements that we know now are false positives. For instance, Zeng et al.'s model listed ivermectin, sirolimus, and estradiol as top candidates for repurposing, compounds which subsequent studies have proven are ineffective and which can be dangerous to take.\cite{Zeng2020} Similarly, Wang et al.'s model gave questionable recommendations like testosterone, beta carotene, and others.\cite{Wang2023} The paper from Barab\'asi's lab ranked chloroquine 5th and ivermectin 16th, alongside a seemingly random selection of drugs.\cite{MorselliGysi2021} Among the top 100 drugs their model recommended for repurposing, only nine drugs had positive outcome in experiments with Vero E6 cell cultures, although three of those were in the top ten.\footnote{As an important side note, Vero E6 cells, which are derived from the African Green Monkey, were later shown to be a bad cell culture model for COVID-19 as they differ in significant ways from human airway cells, and the way SARS-CoV-2 enters E6 cells is different than how it enters into human cells. Many compounds (most famously ivermectin, chloroquine) showed promising antiviral activity in Vero E6 cells but failed in human respiratory cell experiments (e.g., Calu-3 cells).} 

After testing more compounds in human cells, Barab\'asi recommended three drugs for further study -- azelastine (an antihistamine), digoxin (a heart failure medication), and auranofin (a drug for rheumatoid arthritis). Azelastine has proven efficacy as an antiviral nasal spray, while as of 2026 digoxin and auranofin only have cell culture data supporting their use for SARS-CoV-2.

In general, it appears network-based modeling tends to always generate long lists of candidates which then have to be analyzed further. Furthermore, it appears these models always return a ranked list with ``top results'', even if there are actually no good candidates for repurposing.

%-------------------------------------------------------------------------------------
%-------------------------------------------------------------------------------------
%-------------------------------------------------------------------------------------
\section{Data sources}

When developing or testing any method for DTA or CPI prediction, the quality and diversity of data one has is extremely important. The relative success of different ML methods often comes down to the quantity and quality of data used rather than any specific algorithmic or architectural innovation. 

%-------------------------------------------------------------------------------------------
\subsection{Obtaining protein structures}
Care must be taken when selecting crystal structures from online sources like the Protein Databank (PDB). Structure files have been added continually to PDB since the 1980s and come in a wide variety of resolutions. Using a lower resolution file (e.g., $>3.0$ \AA) will severely degrade docking accuracy, so the highest possible resolution should always be sought out. It's also important to check if the viral protein is a dimer (or trimer). For example, SARS-CoV-2 MPro is a dimer - two proteins with the same sequence that are bound together. The result is a symmetric structure with two active sites. In the case of SARS-CoV-2 MPro all of the crystal structures are for the dimer, but for other proteins you sometimes see a mix of monomer and dimer structures. The dimer structure should always be used, since the interaction of the two monomers is likely to change the conformation of the binding site. 
 
Structures may contain missing parts, so the sequence data in the CIF file should always be checked to see if there are gaps. For instance, we found that all of the PDB structures for the Dengue-2 NS2B/NS3 protease had gaps around NS2B residues 77–84. The reason for this is that the NS2B/NS3 protease contains a highly flexible $\beta$-hairpin structure. When crystallized, the hairpins are disordered within the crystal lattice, causing X-rays to scatter randomly rather than creating a clear diffraction pattern. Therefore, that part of the structure cannot be resolved and appears as a gap in the structure file. 

Another thing to watch out for are heavy metal ions (such as Se, Au, or Ag) either incorporated into amino acids or bound to the protein. Selenium has a $K$ absorption edge that is perfectly tuned for synchrotron beam lines, resulting in an extremely high X-ray scattering cross-section. It is often incorporated into proteins by replacing the sulfur in methionine residues with selenium. The much higher cross section of selenium (Z=34) vs sulfur (Z=16) improves structure determination. Proteins are sometimes soaked in a gold solution to probe for druggable sites. Gold binds strongly to reactive regions of enzymes such as catalytic cysteines. Heavy atoms may also be added to stabilize dimeric proteins or to catalyze better crystal packing. We found that 25\% of SARS-CoV-2 MPro structures on PDB contain either missing amino acids or Se or Au metal ions, which emphasizes the importance of careful preprocessing. 

Protein binding sites are not rigid cavities -- most undergo conformational rearrangement upon ligand binding, a process called induced fit. This is not a big effect for proteins in general, but is known to be a large effect for some viral proteases. A study that looked at the bound and unbound structures of over 3,000 proteins found that the carbon backbone rarely moves but sidechains often move during binding.\cite{Guterres2020}

Thus, an important decision is whether to do docking using a crystal structure without a ligand (an ``\emph{apo}'' structure) or whether to use a crystal structure obtained with a ligand co-crystallized (a ``\emph{holo}'' structure). There are several benefits to using \emph{holo} structures. First, as mentioned, the \emph{holo} structure captures the type of conformation that is relevant for binding. Next, many ligands stabilize highly flexible parts of the structure, leading to less missing residues and improved structure resolution around the binding pocket. To give an example, the PDB structure 3U1J for Dengue-3 NS2B-NS3 resolves NS2B only to residue 69, leaving out the entire $\beta$-hairpin that helps form the binding pocket. By contrast, PDB structure 3U1I resolves all of the NS2B residues with no gaps since the protein is bound to a ligand. 

When selecting a structure file, look for a structure with a high resolution and a ligand bound to the active site ($<1.5$~\AA{} resolution is ideal).\cite{DrugFormDTA2025} In general, the following steps should be followed to process a protein structure:

\begin{enumerate}
    \item \textbf{CIF to PDB conversion.} CIF or mmCIF files need to be converted to PDB using Gemmi or OpenBabel.
    \item \textbf{Ligand removal.} Any ligands need to be removed. In the process, it is recommended that buffer molecules, and any non-standard residues are also removed. This is done by removing ``hetero-atom'' (HETATM) and ATOM records from the file. However, removing all HETATM records will also remove structural waters and metal ions, which may not be desired. Removing structural waters before docking is a common practice, but may not be a good idea. The HIV-1 protease, for instance, has important structural waters within the active site which mediate protein-ligand binding. SARS-CoV-2 PLPro has two crystallographic waters that might be important for binding.\cite{Rogers2023} Likewise, metal ions may be important if they are near the active site. 
    \item \textbf{Undue any effects from covalent bonds.} For structures containing covalent inhibitors (for example nirmatrelvir bound to Cys145), the residue the ligand was covalently bound to may have a modified side chain that needs to be fixed. 
    \item \textbf{Check metal ions.} Protein structures sometimes have Mg2+ and Zn2+ ions removed, since they interfere with crystallization.\cite{Rogers2023}
    \item \textbf{Hydrogen addition.} Hydrogens typically need to be added. They are typically missing from protein structures due to their very low X-ray cross section. This can be done using functions in RDKit.
    \item \textbf{Gasteiger charges.} Docking programs need partial charges for all atoms. They are usually estimated using the Gasteiger method, which is implemented in both RDKit and OpenBabel. The end result is file with partial charges called a PDBQT file.
\end{enumerate}

%------------------------------------------------------------
\begin{figure}[h]
\centering
\includegraphics[width=0.6\textwidth]{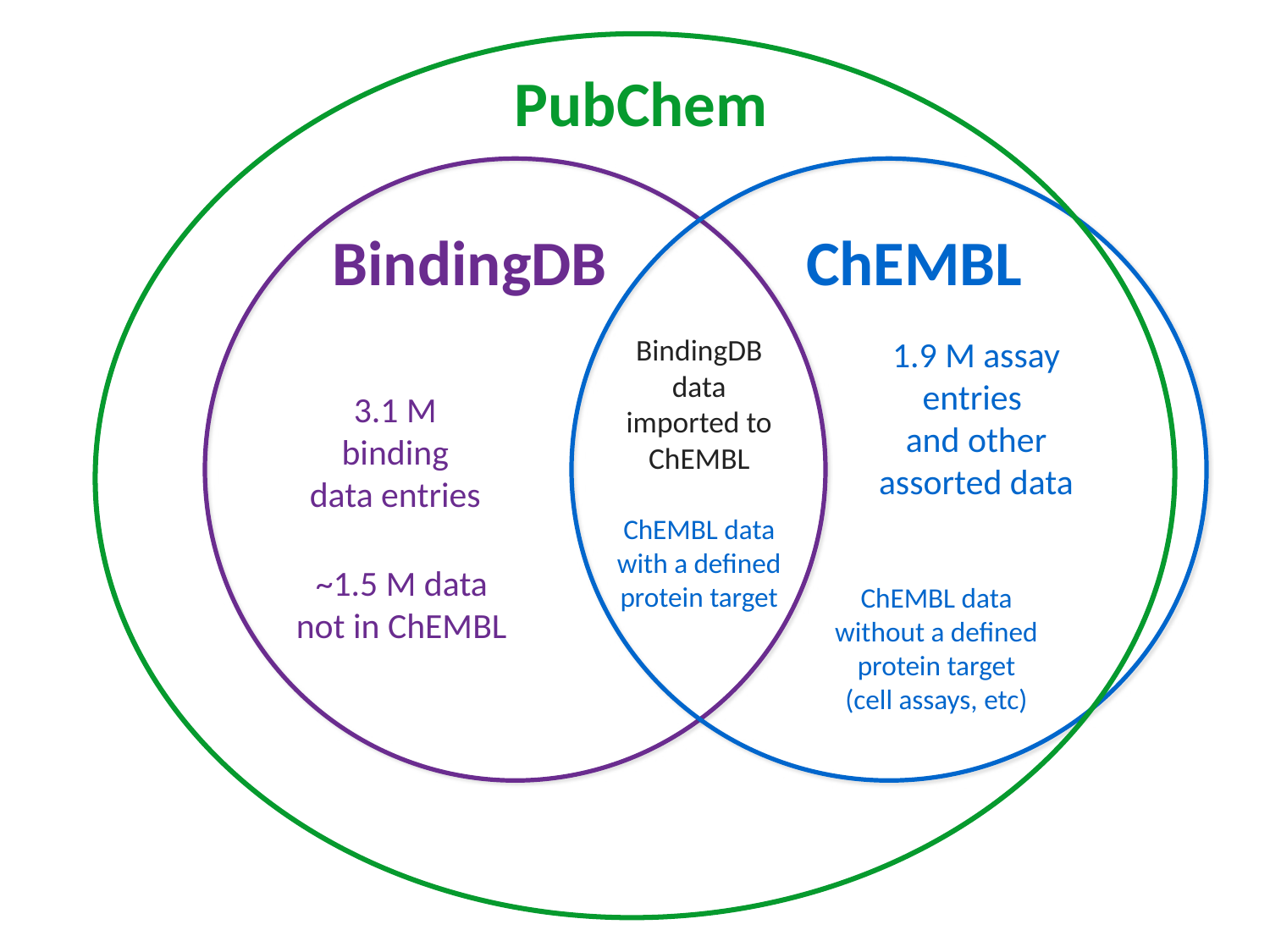}
\caption{Relation between BindingDb, ChEMBL, and PubChem}
\label{fig:bindingdb_chembl}
\end{figure}
\subsection{Protein-ligand binding affinity datasets}
The training data for ML models consists of experimental affinity measurements, which are typically IC$_{50}$, K$_i$, or K$_d$ values. For an overview of these metrics, see appendix \ref{sec:assay_metrics}. To facilitate combining IC$_{50}$ with K$_i$ data, it is common in ML papers to see the approximation that IC$_{50}$ $\approx$ K$_i$.\cite{2018_deepdta} A better approximation, however, is IC$_{50}$ $\approx$ 2K$_i$ (see eqn.~\ref{eqn:ic50}), although this is still a rather crude approximation. When constructing the Kiba binding affinity dataset (discussed below) researchers were able to use eqn.~\ref{eq:chengprusoff} to convert $IC_{50}$ to $K_i$.\cite{Tang2014} Unfortunately this is impossible to do BindingDb or ChEMBL data generally, since it requires knowing the substrate concentration for the assay and estimating the Michaelis–Menten parameter, which depends on temperature, pH, and salinity. That sort of detailed information is typically inaccessible. 

IC$_{50}$ values are often said to be more assay-dependent than K$_i$/K$_d$ values, with the implication being that IC$_{50}$ data is noisier and less preferable to K$_i$/K$_d$ data. However, the results of Landrum \& Riniker call that into question. They found an $R^2$ correlation of only 0.31 between IC$_{50}$ assays randomly sampled from ChEMBL.\cite{Landrum2024} Surprisingly, they found an even lower correlation between $K_i$ values of $R^2$ = 0.13. After removing $K_i$ for the same protein-ligand pair that differed by a factor of $10^3$ (obvious units transcription errors) and removing 239 ``contaminated'' assays (all for the same protein), they got $R^2 = 0.65$ for $K_i$. This was higher than the $R^2 = 0.60$ obtained for IC$_{50}$ after a similar level of curation, but only slightly.

\begin{figure}[h]
\centering
\includegraphics[width=0.85\textwidth]{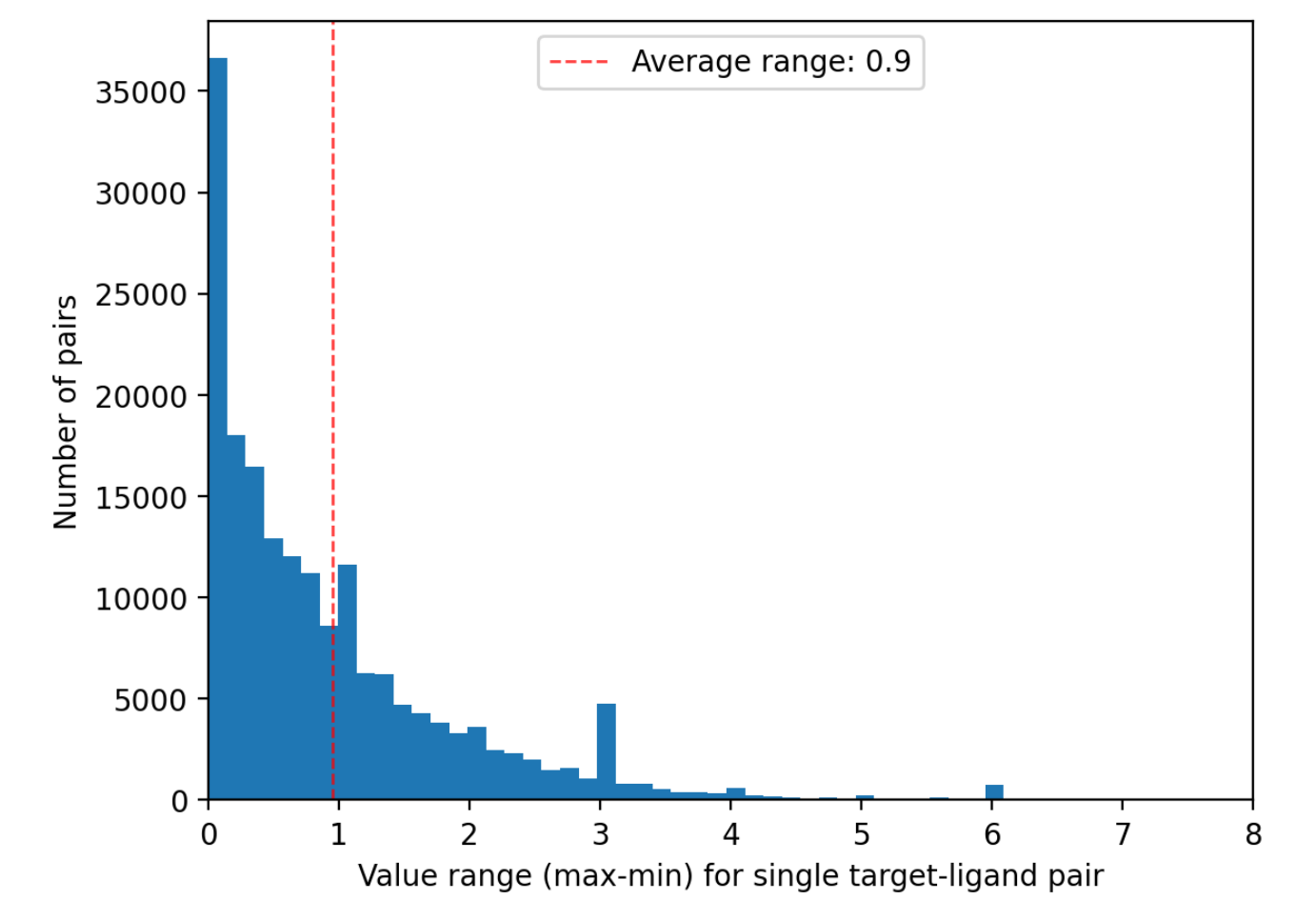}
\caption{Histogram showing the value ranges for different protein-ligand pairs on BindingDb, emphasizing the extreme variabiliy of this experimental data. The x-axis has a logarithmic scale, so each unit corresponds to a 10x difference. Figure taken from ref.~\cite{DrugFormDTA2025}.}
\label{fig:bindingdb_max_min}
\end{figure}
\paragraph{\href{https://www.ebi.ac.uk/chembl/}{ChEMBL}}
The ChEMBL database consists of curated binding affinity data from peer-reviewed scientific articles. ChEMBL also includes cell-based assay results, ADMET data, and other assorted data. Metadata on the assay type may be missing, and to get details on the assay requires fishing around in the literature. Martin et al.~(2023) found that 13.72\% of all virus-related CHEMBL assays did not state the cell type used.\cite{Martin2023SMACC} The authors also express a concern that some of the antiviral screening assays results in ChEMBL may actually be results from counter-assays. (A counter-assay is an assay done with a particular component missing to detect false positives.) 

\paragraph{\href{https://www.bindingdb.org}{BindingDb}}
BindingDb launched in 2004 and was first published in 2007.\cite{Liu2007} It is by far the largest binding affinity dataset online, and as of December 2025 contains 3.5 M binding affinity entries for 1.3 M small molecules across 11,367 protein targets. The BindingDb team pulls from ChEMBL monthly, so researchers using BindingDb don't have to worry much about missing ChEMBL data. Currently, about 50\% of BindingDb is data pulled from ChEMBL, with 25\% collected by the BindingDb team from US patents and 25\% from journal articles.\cite{2015_bindingdb_in_2015} Recently, almost all new BindingDb data comes from patents. A tiny fraction of BindingDb data is submitted directly through a form on their website. Where possible there are links to UniProt IDs, but BindingDb does not contain structure data (sequence information is provided, however). BindingDb data is extracted with the help of ``semi-automated'' pipeline, and it is not clear how much careful manual curation there is. BindingDb data requires extensive preprocessing before it can be used to train a binding affinity prediction model. Outlier values and outlier proteins (ie very large ones) must be removed to avoid training instabilities. Khokhlov et al.~found that 10\% of BindingDb entries were missing binding affinity values or had $EC_{50}$ values, suggesting a tissue-culture assay, not a binding assay.\cite{DrugFormDTA2025} A large fraction of the data has a threshold form (ie ``$>$ 100 nM''). To convert this data into a usable form, Khokhlov et al.~ shift the data towards the threshold, by multiplying or dividing the threshold by $\sqrt{10} = 3.16$. For a justification of this method, see the supplementary information section of their paper.\cite{DrugFormDTA2025} Also, for machine learning, it makes sense to aggregate assays that test the same (protein,ligand) pair, which reduces the size of the dataset by about 660,933.\cite{DrugFormDTA2025}

\paragraph{\href{https://pubchem.ncbi.nlm.nih.gov/}{PubChem}}
PubChem does not do any data curation themselves, but aggregate from other sources, including from BindingDb and ChEMBL. The relation between these three is shown in figure \ref{fig:bindingdb_chembl}. Metadata providing details about assays is lost during ingestion by PubChem.

\paragraph{\href{http://www.pdbbind.org.cn}{PDBbind}}
Unlike BindingDb and ChEMBL the PDBbind is limited to binding assays where experimental 3D crystal structures for the protein-ligand complex have been deposited in the Protein Data Bank (PDB).\cite{wang2004pdbbind,Wang2005} Thus, it is much smaller than the aforementioned datasets. PDBbind data curated after version 2020 is only accessible to paid users. The 2020 version of PDBbind contains about 19,500 assay results.  

\paragraph{\href{http://lsd.docking.org/}{Large-Scale Docking data}}
The large-scale docking platform, maintained at lsd.docking.org, was started in 2024 as a place for researchers to share the results of their docking runs. As of 2026, there is data for 11 protein targets. As far as viral targets, there are 1.1 billion docking results for SARS-CoV-2 MPro and 686 million docking results for SARS-CoV-2 NSP3\_Mac1. While training ML models on docking data is becoming popular, it remains controversial.\cite{DrugFormDTA2025}

\paragraph{Others}
Binding MOAD (15,223 assay results) overlaps heavily with PDBbind. DUDE-E is a library of ``decoy'' compounds which presents a challenging test environment for affinity prediction methods.\cite{Mysinger2012} LIT-PCBA (2022) is a curated and ``debiased'' dataset of PubChem assays, sorting compounds into ``active'' and ``inactive'' for a variety of important human receptors.\cite{TranNguyen2020} A 2025 ``audit'' of LIT-PCBA uncovered significant train-test leakage.\cite{Huang2025} PLAS-5k is a supplemental dataset of 5,000 protein-ligand complex structures and binding affinities derived from molecular dynamics simulations using the MMPBSA (Molecular Mechanics Poisson-Boltzmann Surface Area) method.\cite{Korlepara2022}

%------------------------------------------------------- 
\subsection{Protein-ligand structure datasets}

\paragraph{\href{https://github.com/plinder-org/plinder}{PLINDER}}
PLINDER contains no binding affinity data; instead, it provides curated 3D structural data for 449,383 protein-ligand systems extracted from the entire PDB, each annotated with over 500 features including domain classifications, ligand properties, and experimental quality metrics.\cite{Durairaj2024plinder} This dataset has been used to train or fine-tune ``deep learning docking'' methods like DiffDock and FlowDock which predict the docked complex. The PLINDER team worked very hard to do a train-test split with very low overlap.  

\paragraph{\href{https://github.com/plinder-org/plinder}{Others}}
BioLip2 is a curated set of protein-ligand complexes.\cite{Zhang2023}

%------------------------------------------------------------------------------
\subsection{Common binding affinity evaluation datasets}
The community of researchers working on computational binding affinity  prediction use several common test sets, also called validation, evaluation, or benchmark sets. Unfortunately, the field is plagued with the problem of varying degrees of overlap between model training sets and these standard test sets, making rigorous comparison of methods nearly impossible. Train-test set overlap can take the form of overlap in proteins, ligands, or specific (protein,ligand) pairs, and all three are problematic. In order to remedy this situation, standard train, validation, and test splits for BindingDb, Davis, and Kiba have been released by the Therapeutic Data Commons, but it doesn't appear they have been widely adopted. Another major issue that is worth emphasizing here is that the popular Davis and Kiba test sets are non-diverse, focusing on kinase inhibitors only.

%------------------------------------------------- 
\paragraph{The Comparative Assessment of Scoring Function (CASF) Benchmark}
The CASF challenge was run in 2016 with a hidden benchmark dataset containing high quality crystal structures for 185 protein-ligand complexes together with experimental affinities.\cite{Su2018} A set of decoys (non-binders) was also included. Later, the benchmark data was made public, and in 2025 Graber et al.~showed that the data has a lot of overlap with PDBbind.\cite{Graber2025} In response, Graber et al.~developed ``PDBbind CleanSplit'', a train-test set pair that has minimal overlap.\cite{Graber2025} Interestingly, they found that an older method from 2020 called graphDelta came in second when trained and tested on this split, while newer supposedly better methods from 2023 and 2024 did significantly worse, a very telling finding which indicates the newer models were not actually better in terms of generalization. They argue that their own method, called Pafnucy, generalizes better than existing methods.

%-------------------------------------------------- 
\paragraph{The ``Davis'' Dataset}
The Davis dataset focuses on kinase proteins, providing experimentally-derived $K_d$ values for each protein-ligand pair.\cite{Davis2011} It contains data on 30,056 bindings across 442 proteins and 72 compounds. Note that several of these proteins are actually mutant variants of a single ``base'' protein. In our investigation we found that none of the (protein, molecule) pairs in the Davis dataset overlap with BindingDb, but 259/442 of the proteins in the Davis dataset are in BindingDb.

%------------------------------------------------- 
\paragraph{The ``Kiba'' Dataset}
The Kiba dataset is another collection of kinase protein data that is popularly used as a test set.\cite{Tang2014} Data is derived from ChEMBL and STITCH, and IC$_{50}$, $K_d$, and $K_i$ values are unified into a single score. It contains 246,088 binding scores across 467 kinase targets for 52,498 compounds. We analyzed for overlap with BindingDb and found that it is not a truly independent set. In our own investigation we found 2,268 (protein, molecule) pairs that overlap with BindingDb (1.9\% of total). Furthermore, 93.9\% of the proteins exist in BindingDb.

%------------------------------------------------------------------------------
\subsection{Antiviral datasets}

\paragraph{\href{https://drugvirus.info}{DrugVirus.info 2.0}} 
This dataset, released in 2022, contains ``safe in man'' antivirals that are suspected to be broad-spectrum since they bind to enzymatic targets common among viruses.\cite{Ianevski2022} The version 2.0 update contains data on 255 approved, investigational, and experimental antivirals covering 86 human viruses from 25 viral families. It is based on manual curation of over 2,000 PubMed articles to collect SI, IC$_{50}$/EC$_{50}$, and CC$_{50}$/TC$_{50}$ values. 

\paragraph{\href{https://www.antiviraldb.com}{AntiviralDB}}
This dataset, released in 2025, contains 137 approved antiviral regimens and 697 experimental agents with laboratory-confirmed in vitro and/or in vivo activities (IC$_{50}$, EC$_{50}$, CC$_{50}$) against human viral infections, covering 35 human viruses from 17 viral families.\cite{Huang2025avdb}

\paragraph{\href{https://smacc.mml.unc.edu}{Small Molecule Antiviral Compound Collection (SMACC)}}
This dataset, released in 2023, is derived from the ChEMBL bioactivity database. It contains approximately 30,000 assays for 13 viruses. The dataset is split into a ``phenotypically curated'' dataset (11,123 entries, 6,000 compounds) containing readouts on virus replication inhibition in cellular assays, and a ``target-based'' dataset containing readouts from protein-ligand binding assays. Overall the data is 80.7\% IC$_{50}$, 7.5\% EC$_{50}$, 1.9\% EC$_{90}$, and 9.1\% K$_i$ or K$_d$ values. 

\paragraph{\href{https://smacc.mml.unc.edu}{Heli-SMACC}} 
This dataset extends SMACC with 20,432 bioactivity entries for viral, human, and bacterial helicases. The vast majority of activities are for human helicases ($>$90\%), not viral helicases.\cite{Martin2024} 

\paragraph{\href{https://go.drugbank.com/categories/DBCAT000066}{DrugBank}} 
Drugbank provides a free dataset of drugs for academic use, containing 225 approved antivirals.\cite{Knox2024} Additional data on clinical trial results is available under a paid license.  

\paragraph{\href{https://idrblab.org/ttd/}{Therapeutic Target Database}}
This dataset provides structure and activity data on targets, drugs, and drug-drug and drug-target interactions.\cite{Zhou2024TTD}

\paragraph{Wikipedia}
Wikipedia lists approximately 144 antiviral drugs in their antiviral drug category and 21 antiviral combinations in the Combination Antiviral Drugs category.

%------------------------------------------------------------------------------
\subsection{Antiviral combination datasets}
The DrugVirus.info website referenced earlier contains a dataset with data on 540 antiviral combinations.\cite{Ianevski2022} Fatehi et al. (2024) created the CombTVir dataset, which is just a filtered version of the DrugVirus.info dataset with biologics (mAbs, peptides, convalescent plasma) removed (after filtering, one obtains 372 small-molecule combinations).\cite{Majidifar2024}  Some of the same researchers who created DrugVirus.info released a larger combinations dataset in 2020 called antiviralcombi.info. That dataset contained data on 985 antiviral combinations across 68 viruses that was obtained from PubMed, clinicaltrials.gov, DrugBank, DrugCentral, the Chinese Clinical Trials Register, and EU Clinical Trials Register. The antiviralcombi.info website was taken down in 2021.

Jin et al.~(2021) assembled several data sources to create a dataset for training and testing their ComboNet deep learning model.\cite{Jin2021} The datasets they utilized were 88 SARS-CoV-2 combinations from NCATS, 71 SARS-CoV-2 combinations from Bobrowski et al., 20 SARS-CoV-2 combinations from Riva et al., and 114 HIV combinations from Tan et al.\cite{Jin2021} 

%-------------------------------------------------------------------------------------------
%--------------------------------------------------------------------------------------------
\section{Docking and physics-based modeling tools}

\begin{figure}[h]
\centering
\includegraphics[width=0.6\textwidth]{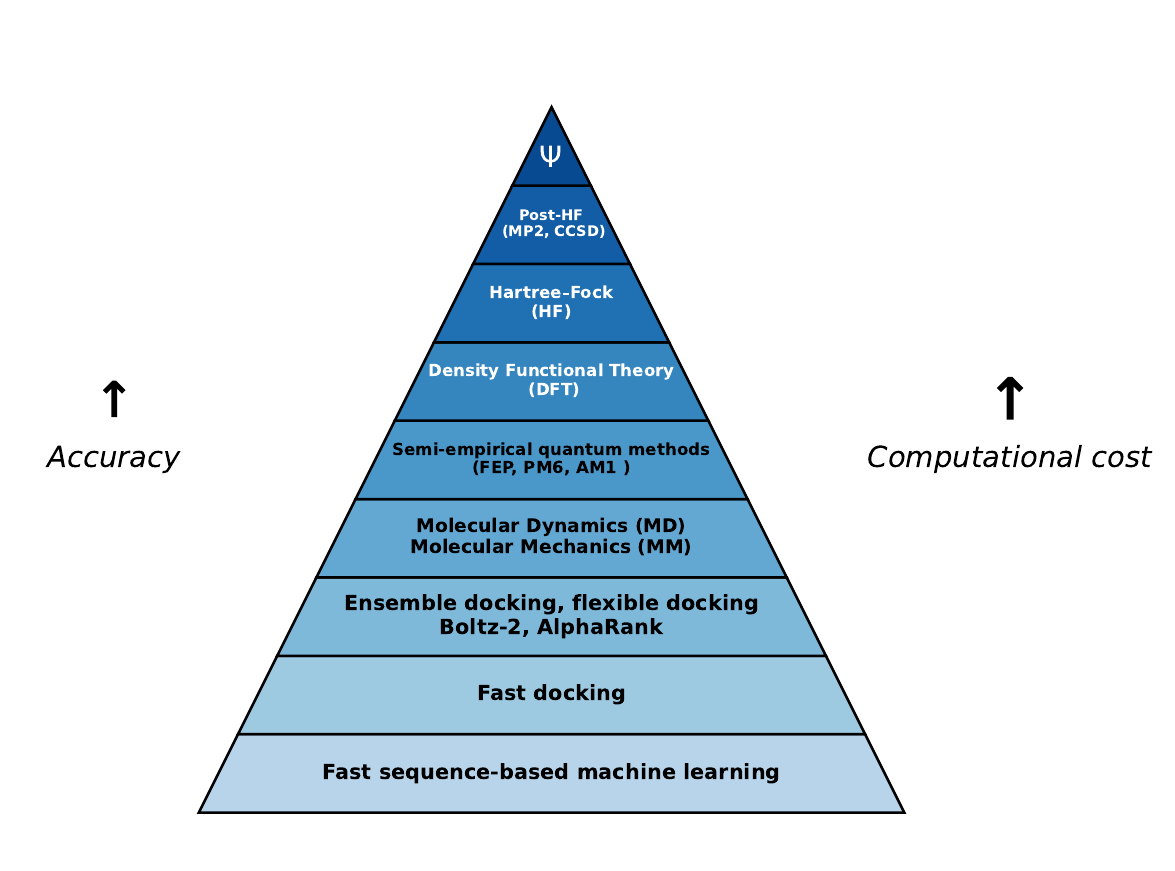}
\caption{The hierarchy of quantum chemistry methods, with additional classical physics and docking methods added to the pyramid.}
\label{fig:quantum_chemistry_pyramid}
\end{figure}

\subsection{The hierarchy of methods}

The fundamental physics of drug-protein interactions is fairly well understood, and since the early 2000s crystal structures for target proteins have been available with high enough resolution to facilitate \text{in silico} simulation. The difficulty lies in finding approximations that make prediction computationally feasible without giving up too much in predictive accuracy. Within quantum chemistry, there is a well known series of approximations that are typically displayed on either a ladder or pyramid diagram. In figure \ref{fig:quantum_chemistry_pyramid} we have expanded that pyramid to include semi-classical and classical physics-based techniques.  

Quantum methods are rarely used in drug-discovery due to their high computational cost. Some academic labs use the PM7 method with the COSMO solvent model as a final step in their in silico screening pipeline.\cite{Klamt1993} SQM2.20 (2024) is a semi-empirical quantum mechanics inspired scoring function which greatly outperforms classical docking scoring functions and achieves an accuracy close to that of density functional theory (DFT).\cite{Pecina2024} DFT calculations of binding free energy take hours or days on large-scale compute clusters, while SQM2.20 runs in about 20 minutes on a single CPU core. Unfortunately, the code for SQM2.20 has not been open-sourced.

It is worth spending a moment to consider the state of molecular dynamics simulation in drug discovery. In 2008 D.E. Shaw Research developed Anton, a custom supercomputer for molecular dynamics simulation that uses application-specific integrated circuits (ASICs) to accelerate molecular dynamics. This supercomputer eventually enabled the first millisecond-scale all-atom protein simulations. As of 2023, D.E. Shaw Research had six drugs in human clinical trials — two developed independently from concept to trial, and four developed with Relay Therapeutics. Their chief lead is zovegalisib (RLY-2608), which is in Phase 3 trials for HR+/HER2- metastatic breast cancer. The truth is that D. E. Shaw is the only publicly-known success story for a pure molecular dynamics approach to drug discovery. It is also a bit of an unusual company, having continually received funding from hedge fund billionaire David Shaw through a period of ten+ years ``in the wilderness'' without any drugs in clinical trials. 

Docking methods can be broken into blind docking methods and guided/constrained methods. Blind docking involves docking over the entire protein, while constrained docking involves docking to a particular active site or binding pocket within a cropped box, often no larger than 30~\AA on each side. Naturally, blind docking is far more computationally expensive, (between 10 to 100 more expensive depending on the protein). Docking programs can also be broken into rigid ligand methods (only rotations considered) and methods with a flexible ligand. Making the ligand flexible increases computational cost by 20x to 50x or more depending on the ligand. Protein flexibility is generally not considered explicitly within docking. To take into account protein flexibility, molecular dynamics simulations are sometimes run to sample conformational space, and then docking is done on frozen snapshots of the protein. Finally, docking methods differ on the scoring (energy) function. 

In the next few subsections we discuss some docking programs of note. While recent systematic benchmarks are lacking, a 2018 update on the 2016 CASF benchmark found that Glide and AutoDock Vina were among the top two performers,\cite{Su2018} so we start with those: 

%-------------------------------------
\subsection{Glide \& the Schr\"{o}dinger Suite}
Glide appears to be the most commercially successful docking program, and has been under continual development by Schr\"{o}dinger Research since 2001.\cite{Friesner2004} Today, Glide is part of the ``Schr\"{o}dinger Suite'', a suite of about twenty software tools. The suite is a self-contained software ecosystem that even contains its own custom version of Python. Glide is available in four forms - XP, WS, SP, and HTVS. Glide WS (2024) is the most accurate (since it includes WaterMap water energetics mapping) but takes about 10 minutes to run per ligand on multicore cloud CPUs. Glide XP is their standard model, and is quoted as taking 5-10 minutes to run per ligand on ``cloud CPUs''. The cheaper Glide SP is quoted as taking 20 seconds per ligand, and Glide HTVS takes about two seconds per ligand. Two programs in the suite are provided to clean the protein structure (ie add hydrogens, add/remove waters) and predict the ligand's 3D structure - these are the ``Preparation Wizard'' and the ``LigPrep tool''. Another tool in the suite, QikProp, can be used to rapidly estimate ADME-related properties.

%------------------------------------------ 
\subsection{AutoDock, AutoDock Vina, Autodock GNINA}
AutoDock started development in 1989, leading to the popular AutoDock 4.0.0 - 4.2.6 series (2009-2014) which later became AutoDock-GPU. AutoDock Vina (2010) is considered by some to be the main successor to AutoDock 4, and is a popular docking program that remains a common baseline for comparison with other techniques.\cite{trott2010autodock} Both AutoDock and AutoDock Vina are maintained by Scripps Research, and the latest major release appears to be AutoDock Vina 2.1 (2021).\cite{Eberhardt2021} AutoDock-GPU (2021) is Scripps' successor to AutoDock 4.2.6 (2014) and received its last major release in November 2024.\cite{SantosMartins2021} A GPU-accelerated version of AutoDock Vina 2.1 was released in 2024 outside the auspices of Scripp's.\cite{Tang2024} 

The Vina scoring function approximates the standard chemical potential via a weighted sum of steric interactions, hydrophobic effects, loss of torsional entropy, and hydrogen bonding, ignoring the computationally expensive electrostatic components found in other physics-based models. As with most other docking techniques, Vina's accuracy is limited by the method's complete ignorance of protein conformational states and the approximate energy calculation which ignores many-body interactions like Van der Waals interactions.\cite{eberhardt2021autodock} 

Autodock GNINA (Grid-based Neural Network Interaction Assessment) attempts to remedy those deficiencies by replacing the standard energy function with a 3D Convolutional Neural Network (CNN).\cite{mcnutt2021gnina} It is built on the 2013 ``Smina'' fork of Vina.\cite{Koes2013} Their method discretizes the protein-ligand complex into a voxelized grid—analogous to a 3D image, where atoms are treated as channels (analogous to the RGB channels in color image file). Somewhat surprisingly, this representation allows CNN to learn spatial features relevant for binding.\cite{Ragoza2017} GNINA was trained on the PDBbind database.\cite{wang2004pdbbind,Wang2005} Benchmarking work on the CASF and DUD-E datasets demonstrate that GNINA's CNN scoring consistently outperforms Vina.\cite{mcnutt2021gnina}

%---------------------------------------------
\subsection{GOLD}
GOLD from the Cambridge Crystallographic Data Center (CCDC) is considered a top docking program, and a 2016 benchmark showed GOLD slightly outperforming Glide.\cite{Wang2016} A 2023 benchmark on the COX-1 and COX-2 proteins found GOLD ranked \#1 among several methods, dramatically outperforming Glide on those particular targets.\cite{Shamsian_2023} Unfortunately, it is difficult to get a free academic license for GOLD, and it is generally considered a commercial program sold by CCDC Limited or through OpenEye, a third-party vendor.

%---------------------------------------------
\subsection{Rosetta \& RosettaVS}
Rosetta is a popular package for molecular dynamics simulation. Binding free energy can be calculated using the Generalized Born and Surface Area solvation method, often abbreviated MM-GBSA. The method requires 1 to 10 nanoseconds of simulation, and explicit solvation with water molecules is recommended. This is obviously very computationally intensive, but can be done at scale on a GPU cluster. Rosetta Virtual Screening (RosettaVS), released in 2024 by Zhou et al., is a major addition to Rosetta which includes the Rosetta GALigandDock docking program and a varity of scoring functions including a force-field based scoring (RosettaGenFF-VS) and machine-learning based scoring.\cite{Zhou2024} The RosettaGenFF-VS scoring equation for the binding free energy $G$ is: 
\begin{equation}
    \begin{aligned}
        \Delta G &= \Delta H - T \Delta S \\ 
        \Delta H &= E_{\text{complex}} - E_{\text{protein}} - E_{\text{ligand}} \\  
        \Delta S &\approx 0.4 N_{\text{rotor}}
    \end{aligned}
\end{equation}
where $N_{\text{rotor}}$ is the number of rotational bonds in the ligand. The assumption is that essentially all rotational motions of the molecule cease after binding. Rosetta also has a ``detailed calculation'' mode for estimating $\Delta G$ which does a more precise estimate of the entropy change using Monte Carlo sampling. 

%-------------------------------------
\subsection{DOCK}
The DOCK software (\href{http://docking.org}{docking.org}) from UCSF is one of the most popular docking programs that is available for free academic use. The first version was released in 1982, and it has been updated regularly since then. There are two DOCK codebases -- the DOCK 3.x series (currently 3.8), which runs on a mix of Fortran and C, and the Dock 6.x series (currently 6.13), which is written in C++.\cite{Coleman2013,Allen2015} The two codes expose different docking techniques to the user and are seen as complementary. Generally speaking, DOCK 3.8 is good for fast rigid docking while 6.13 is better for slower flexible docking and the ``anchor-and-grow'' technique. 

%-------------------------------------
\subsection{DiffDock and FlowDock}
DiffDock (2022) represents a fundamentally new approach to docking and is a product of the Regina Barzilay and Tommi Jaakkola labs at MIT.\cite{Corso2022}  DiffDock uses ``geometric deep learning'' inspired by diffusion-based generative modeling, a technique first developed for generating images. Diffusion models start with random noise and iteratively refine the noise to minimize a learned loss function. In this case, the optimization occurs not in pixel space but in the space of docking poses, hard-coding the degrees of freedom for a docking ligand as an ``implicit bias''.\cite{Corso2022} DiffDock, by design, does not output an affinity score, but only a ``confidence level'' for the final docked structure, which is unrelated to affinity. Therefore, an additional scoring module must be added to get an affinity. We tried using DiffDock on our test set, and then feeding the docked structures into Autodock GNINA for affinity prediction. The resulting correlation with experimental values was almost zero ($r \approx 0.02$). Upon investigation, we found that the ligand was too far from the protein, which may have been due to the DiffDock program not properly transforming the coordinates back into the original space (this is a known issue with the current codebase). For the SARS-CoV-2 main protease specifically, we found that 84\% of time, DiffDock had not placed the ligand at the active site and therefore failed to find the right binding pocket. The SARS-CoV-2 main protease has a large surface area with many small cavities which make blind docking difficult. As an experiment, we used Claude Code to implement ``pocket guided'' docking with DiffDock - essentially telling DiffDock to try docking the ligand at the SARS-CoV-2 active site. Across 178 ligands, the correlation was $r = 0.240$. According to Claude Code's analysis, DiffDock does not do a good job minimizing steric clashes between atoms. In some cases, direct atom-atom clashes caused positive binding energy, resulting in negative $K_i$, which is nonsensical. DiffDock is already obsolete as subsequent work has improved and extended the DiffDock technique, including DiffDock-L (2024), DiffDock-CB, DiffDock-Pocket, and FlowDock (2025).\cite{Morehead2025} To minimize work, we chose to focus on FlowDock. DiffDock-CB may also be very competitive, but is more complex to implement as it requires running and managing a technique called Confidence Bootstrapping which is similar to an active learning loop.

%-------------------------------------
\subsection{Interformer}
Interformer (2024) is a docking system from Tencent's AI Lab that incorporates three machine learning models.\cite{Interformer2024} A ``interaction-aware mixture density network'' model is used to model electrostatic interactions between protein and ligand atoms. A ``pose-scoring'' model is used during sampling. Finally, there is a transformer-based model specifically designed to predict the binding affinity of the final pose. The models were all trained on PDBBind crystal structure data. On the PoseBusters benchmark, Interformer achieved an 84.09\% docking success rate (RMSD $< 2$ \AA), outperforming AutoDock Vina, GNINA, and DiffDock. Interformer was applied to internal drug discovery projects to find binders for SARS-CoV-2 Mpro and LSD1 (a target for cancer treatment). The authors report identifying compounds with experimental IC$_{50}$ values of 16 nM (Mpro) and 0.7 nM (LSD1), which are quite impressive figures.

%-------------------------------------
\subsection{Protenix-Dock}
Protenix-Dock was released in February 2025 \href{https://github.com/bytedance/Protenix-Dock}{on Github} by Chinese tech giant ByteDance, the makers of TikTok. Unlike the main Protenix model (discussed below), which is a machine-learning based system, Protenix-Dock is a classical rigid docking program that uses empirical scoring functions rather than deep learning. 

 %--------------------------------------- 
\subsection{Other docking software}
Older docking programs that appear to now be obsolete include Slide (1998 - 2008) and ParaDockS (2011). rDock (1998-2014) is a docking program that started as a closed-source commercial product developed by Vernalis Research and then shifted to open-source development in 2012 under a partnership with the University of York.\cite{RuizCarmona2014} Development of rDock ceased in 2014, but some additional development continued from 2019-2022 under a fork called RxDock. PLANTS (2006 - 2009) is an open-source docking program that uses ant colony optimization.\cite{Korb2006} A new version of PLANTS called VirtualFlow ANTS (2021) has been made available in VirtualFlow (the upgrades over the 2009 version appear to be minor, however).\cite{Gorgulla2021, Gorgulla2020} DINK (2013) is a fork of AutoDock that has been tailored for processing large ligands.\cite{Dhanik2013} FlexAID (2015) is a docking program that was worse than AutoDock Vina and rDock during testing but was better for highly flexible proteins.\cite{Gaudreault2015} QuickVina 2 (2015) is a variant of Vina that speeds up docking using heuristics and may be useful as an ultrafast screening tool.\cite{Alhossary2015} Vina-Carb (2016) is a specialized version of Vina which is tailored for handling carbohydrate molecules like oligosaccharides and glycosides, which are highly flexible molecules that can exhibit complex torsional motions.\cite{Nivedha2016} Vinardo (2016) is an upgraded scoring function for the ``Smina'' fork.\cite{Quiroga2016} VinaXB (2016) is a modified version of AutoDock Vina that incorporates an empirical halogen bond scoring functions.\cite{Koebel2016} GalaxyPepDock (2015), MOLS 2.0 (2015), and AutoDock CrankPep (2019) are specialized codes for peptide docking.\cite{Lee2015, Paul_2016, Zhang2019} Not surprisingly, AutoDock CrankPep outperforms Glide, GOLD, and other programs for docking cyclic peptides.\cite{Zhao2024} SEED (2018) is a specialized docking code for docking small fragments.\cite{Marchand2018}

KarmaDock (2023) is a deep-learning guided docking method designed for speed and scale over accuracy.\cite{Zhang2023} In KarmaDock, an $E(n)$-equivariant GNN encodes both the protein pocket and the ligand into ``node embeddings'' and then a Mixture Density Network is used for scoring. SwissDock (2011 - 2024) is the name for a series of docking codes developed by the Molecular Modelling Group of the University of Lausanne and the SIB Swiss Institute of Bioinformatics.\cite{Bugnon_2024} The source code for SwissDock has not been made available, but they have made available a web portal (swissdock.ch) where users can run docking jobs. HADDOCK3 (``High Ambiguity Driven DOCKing 3'') (2025) is a docking framework that is described as enabling ``integrative modelling of biomolecular complexes''. Rather than performing purely ``blind'' sampling, HADDOCK can incorporate prior information (e.g., known or suspected pocket/interface residues, constraints from cryo-EM data, etc) into constraints that bias sampling toward plausible binding sites. HADDOCK was designed with an eye towards docking larger structures like antibodies, RNAs, proteins, and peptides.\cite{Giulini2025} Giulini et al.~(2025) describe using HADDOCK3 to refine and dock antibodies.\cite{Giulini2025} 

EnzyDock is a docking program designed for enzymes that uses the CHARMM forcefield and techniques from simulated annealing, and is thus more computationally expensive.\cite{Das2019} A 2021 work showed that EnzyDock performed the best among six different docking programs when it came to predicting binding modes to SARS-CoV-2, slightly outperforming Glide and DOCK and greatly outperforming AutoDock and AutoDock Vina.\cite{Zev2021}

In addition to Glide and GOLD (discussed above), there are many other commercial docking programs including ICM-Pro from MolSoft LLC, FlexX and Chemical Space Docking from BioSolveIT, FRED and OEDocking from OpenEye Scientific, Surflex-Dock and StarDrop from Optibrium, and MOE-dock from Chemical Computing Group. 

%---------------------------------------------------------------------------------
\subsection{Consensus docking}
In recent years, a large number of papers have presented frameworks for running multiple open-source docking programs and aggregating the results, a process called ``consensus docking''. To give a highly non-exhaustive list, we found DockingPie (2022),\cite{Rosignoli2022} MILCDock (2022),\cite{Morris2022} MetaDock (2023),\cite{Kamal2023} ESSENSE-Dock (2023),\cite{Nelen2023} CobDock (2024),\cite{Ugurlu2024} and DOCKM8 (2024).\cite{Lacour2024} Of these and others, DOCKM8 was the most comprehensive in terms of the number of sub-programs and scoring functions it implements, so we focused on DOCKM8 in our testing. Consensus docking is also easy to implement oneself - all one has to do is write a wrapper script (or ``harness'') to run several docking codes, and then aggregate the results. Simply averaging output scores doesn't work, because different docking programs use different output scales. Approaches include taking the mean of the ranks, the geometric mean of the ranks, or averaging over z-score (standard deviations from the mean output score).

%---------------------------------------------------------------------------------
\subsection{Drugit}
\textit{Drugit} (2025) is a platform that gameifies drug design. It is built on top of Rosetta and includes features for physics-based energy minimization. \textit{Drugit} builds off the success of \textit{Foldit} (2008 - present), which turned protein folding into a game.\cite{Koepnick2019} In the third CACHE challenge, the \textit{Drugit} team did well, advancing to the second round and discovering a SARS-CoV-2 NSP3-MAC1 inhibitor that was experimentally confirmed with a $K_D$ of $\approx$ 10 $\mu$M.\cite{Herasymenko2026} The CACHE-3 organizers said the result was ``sobering'' but noted that many of the users of \textit{Foldit}/\textit{Drugit} during the challenge were professional chemists.

%------------------------------------------------------------------------------------
%------------------------------------------------------------------------------------
\section{Machine learning based tools}

\begin{table}[h]
\caption{Some recent machine learning based architectures for drug-target affinity prediction}\label{tab2:DTAmethods}
\begin{tabular*}{\textwidth}{@{\extracolsep\fill}lcccccc}
\toprule%
Name             & year & molecular encoder & protein encoder & Davis $r^2$ & KIBA $r^2$ & reference\\
\midrule
DeepDTA          & 2018 & CNN             & CNN             & 0.631     & 0.673     & \cite{2018_deepdta}\\
DeepGLSTM        & 2021 & GCN             & Bi-LSTM         & 0.679     & 0.789     & \cite{DeepGLSTM}\\
ELECTRA-DTA      & 2022 & CNN-SEB         & CNN-SEB         & 0.671     & 0.727     & \cite{ELECTRA-DTA}\\
MGraphDTA        & 2022 & GNN             & CNN             & 0.710     & 0.801     & \cite{MGraphDTA}\\
DGDTA            & 2023 & GAT             & Bi-LSTM         & 0.707    & 0.809      & \cite{DGDTA}\\
MSGNN-DTA        & 2023 & GNN             & GNN             & 0.719     & 0.818     & \cite{MSGNN-DTA}\\
MMFA-DTA         & 2024 & GAT             & GAT+CNN         & 0.688     & 0.777     & \cite{MMFA-DTA}\\
\botrule
\end{tabular*}
\footnotetext{A small fraction of the binding affinity models that have been published in recent years. CBB = Convolutional Neural Network, CNN-SEB = NN, Squeeze and Excitation Block, GNN = Graph Neural Network, GCN = Graph Convolutional Network, GAT = Graph Attention Network, Bi-LSTM = Bidirectional Long Short-Term Memory.} 
\end{table}

A very large number of machine learning based methodologies using ``deep'' (multi-layer) networks have been proposed for DTA prediction since the 2018 DeepDTA paper.\cite{2018_deepdta} A semi-random subset of recent models is shown in table \ref{tab2:DTAmethods}. There have been dozens of other models published in recent years including PADME, \cite{PADME}, DGraphDTA,\cite{DGraphDTA}, GraphDTA,\cite{GraphDTA}, TransformerCPI,\cite{TransformerCPI} MATT-DTI,\cite{MATT_DTI} DeepDTAF,\cite{DeepDTAF} MolTrans,\cite{MolTrans} BridgeDPI,\cite{BridgeDPI} STAMP-DPI,\cite{STAMP-DTI} WGNN-DTA,\cite{WGNN-DTA}, DATA-DTA,\cite{DataDTA} GRA-DTA,\cite{GRA-DTA} GTAMP-DTA,\cite{GTAMP-DTA} TEFDTA,\cite{TEFDTA} DeepDTAGen, \cite{DeepDTAGen} DrugFormDTA,\cite{DrugFormDTA2025} FusionDTI,\cite{FusionDTI} InceptionDTA,\cite{InceptionDTA}, OnionNet,\cite{Zheng2019} PocketDTA,\cite{Zhao2024PocketDTA} and DrugCLIP.\cite{Jia2026}

%The ``DeepPurpose'' library \cite{DeepPurpose} was specifically designed with repurposing in mind. It is a toolkit with multiple ML models included with different architectures. 

What differentiates these different approaches are the particular neural network architectures they use and the ways they encode molecules and proteins into feature vectors that neural networks can operate on. Early models encoded molecules as SMILES strings and encoded proteins as amino acid strings. The invention of the graph neural network molecular graph structure to be encoded more directly into a feature vector, making the graph structure more transparent than SMILES (where it is implicit).\cite{Elton2019} Some of the latest models use pretrained encoders like ChemFormer for molecules or the Evolutionary-Scale Modeling (ESM) transformer for proteins to generate feature vectors. Both ChemFormer and ESM have been trained on massive datasets and have learned how to extract useful features. Other models like DGraphDTA (2020) and WGNN-DTA (2022) use a contact map to encode information about a protein’s 3D structure.\cite{DGraphDTA, WGNN-DTA} Contact maps are not always readily available, so both methods have ways of estimating the contact map -- DGraphDTA uses the ``PconsC4'' method to estimate the binary contact map from an AA sequence, while WGNN-DTA uses an ESM-based method to predict a weighted contact map.

AEV-PLIG is a method that requires a PDB file giving the protein structure along with a separate SDF file of the ligand in its bound pose with coordinates in the same coordinate system. The name of the model comes from the use of ``Atomic Environment Vectors (AEVs) and ``Protein–Ligand Interaction Graphs'' (PLIGs). AEV-PLIG uses a graph neural network to encode the protein and ligand structure jointly.\cite{Valsson2025}  

Many of these papers have only been tested on small non-diverse datasets like Davis and Kiba, and thus have questionable generalizability. In next two sections we discuss three models that appear to be the best available as of early 2026 - Boltz-2, Protenix, and DrugFormDTA.

%------------------------------------------------------------------------------------
\subsection{AlphaFold, Boltz, and Boltz-2}
DeepMind's original AlphaFold system debuted in 2018, winning that year's Critical Assessment of Structure Prediction (CASP) challenge by a small margin. The big advance, however, came in 2021 with the release of AlphaFold2.\cite{Jumper2021} AlphaFold2 was the first method able to predict protein structures close to the resolution typically achieved in X-ray crystallography experiments (ie about 1 \AA~RMSD), at least for a subset of proteins. While AlphaFold2 was a step-change advance, it is important to emphasize that AlphaFold2 is still inaccurate for some classes of proteins and fails completely when it comes to disordered proteins. 

AlphaFold2 was followed by AlphaFold3 in May 2024.\cite{Abramson2024} In addition to improving protein prediction, AlphaFold3 extended the AlphaFold2 system to enable the prediction of biomolecular complexes. On the PoseBusters v1 dataset, containing 428 protein–ligand complexes, AlphaFold3 outperformed AutoDock Vina and RoseTTAFold All-Atom, two popular docking methods (the task was the prediction of protein-ligand structure, and prediction was deemed successful if the ``pocket-aligned ligand RMSD'' was less than 2.0 \AA).\cite{Abramson2024} While AlphaFold3 does not predict binding affinity directly, the protein-ligand structures it predicts can be fed directly into free energy estimation software like docking scoring function or quantum chemistry software with little or no further optimization required (the optimization of protein-ligand binding pose is the computationally expensive part). Thus, the utility of AlphaFold3 for drug discovery is clear. Indeed, AlphaFold3 was developed jointly between DeepMind and Isomorphic Labs, a drug-discovery startup at Google. Isomorphic labs has entered into multi-billion dollar partnerships with Eli Lilly and Novartis.

Given the commercial implications of the AlphaFold3 advance, it is not surprising that Google did not release the code or model weights upon publication of AlphaFold3 in May 2024. Instead, they made available a web-based interface that allowed people to run the model a few times for free for non-commercial use, and then pay for further non-commercial use. This sparked backlash from the academic community. Eventually, Google released AlphaFold3 with an open-source, non-commercial license in November 2024.

That episode spurred the development of Boltz-1, the first iteration of which was published a preprint on November 20, 2024.\cite{Boltz1} Loosely speaking, Boltz-1  can be thought of as a fully open-source version of AlphaFold3 with an additional ``affinity prediction'' branch added on. Boltz-2 was announced in June 2025.\cite{Passaro2025}

AlphaFold is a very complex system, consisting of extensive sequence preprocessing and six interacting modules that operate in a recursive fashion. A full exposition of how AlphaFold works is well beyond the scope of this paper. However, we will highlight a few key aspects that can be explained briefly: 

The AlphaFold and Boltz family of models use multiple sequence alignment (MSA) to identify evolutionarily conserved regions. Regions of the genome that are highly conserved are likely to be important for structural integrity and/or function, and regions that co-mutate together are likely spatially close. An attention-based network processes sequences (up to 150-200), identifying key regions. The ``evoformer'' module generates a 2D ``contact probability map'' while enforcing key aspects of geometric consistency. Alphafold2 runs the entire network multiple times in a process called ``recycling''. During recycling the predicted structure is fed back in, leading to successive refinement of the structure. 

These systems also use a process called ``diffusion'' to stochastically explore structural space. Diffusion does not happen randomly - there are many hard-coded constraints regarding bond lengths, bond angles, and chirality (all amino acids must be left-handed -- simply mirror imaging an amino acid is not allowed). Boltz and Boltz-2 introduce ``Boltz steering'', a feature that applies simple physics-based potentials (e.g., repulsive, stereochemical) to the coordinates during the diffusion process. While the use of Boltz steering increases runtime by about 2x, it also increases accuracy.  

Despite the introduction of Boltz steering, AlphaFold3 is still better at protein structure prediction. This may be due to Google's greater use of ``distillation''. During distillation, the model is run on sequences for which there is no crystal structure. Some structures deemed to be very good quality are kept and used for fine-tuning. The exact way in which this is done at Google has not been described publicly. Boltz-2 was fine-tuned with distillation from AlphaFold2 outputs where AlphaFold2 had high-confidence. Boltz-2 was also trained on data from molecular dynamics trajectories in an attempt to get the model to understand multiple conformations better. 

While not as good at structure prediction, Boltz and Boltz-2 provide two major new features that increase its utility vs AlphaFold3. First, they allow the user to input conditioning constraints. For instance, the user may specify certain distance constraints between amino-acids or that binding must be confined to a narrow region (eg, the ``binding pocket''). The next major innovation is that the Boltz model applies an ``affinity module'' to predict binding affinity. 

Boltz-2 was trained on 1.2 million (protein,ligand) binding values (K$_i$, K$_d$, IC$_{50}$, EC$_{50}$, etc) from a variety of datasets. Using so much data was unprecedented at the time (some of the works cited in table \ref{tab2:DTAmethods} are only trained on $\approx$ 10,000 values). Since K$_i$ and K$_d$ data are combined with IC$_{50}$ and others during training, the resulting output is not an affinity value, but a generalized ``measure of affinity''.  

The Boltz-2 team found their model outperformed quantum free-energy perturbation (FEP) methods on three different test sets. However, one of their test set was limited to kinases, which are a very data rich set of proteins. External validation studies show that Boltz-2's performance is highly variable with weak generalizeability, as is the case with ML models in general. On the PL-REX dataset consisting of a diverse set of 10 proteins, Boltz-2's free-energy scoring function barely outperforms docking scoring functions from commercial docking programs like GlideSP and Gold ChemPLP, while requiring a longer runtime. On the Ultralarge Virtual Screening Hit dataset, Boltz-2 performs better at classifying hits than docking methods (AUCROC $\approx$ 0.75 vs $\approx$ 0.58-0.65 for docking methods).\cite{Sindt2025,Bret2026} However, Boltz-2's scoring function showed an insensitivity to binding pose, suggesting it is not modeling physics but is doing a form of statistical interpolation between training data points. They also found Boltz-2 is insensitive to the mutation of binding site residues.\cite{Bret2026} The authors had trouble making sense of these findings in light of Boltz-2's overall performance.\cite{Bret2026} While Boltz-2 is capable of some generalization, these results indicate that Boltz-2, like other machine learning based methods, is poor when it comes to generalizing to novel proteins. 

On a blog post on LinkedIn, Xi Chen discusses a particular ligand for both PLK1 and BRD4 where he found the values had been swapped in ChEMBL, so the binding data for both proteins was incorrect. The incorrect data was part of Boltz-2's training data, however Chen found that Boltz-2's predictions did not repeat the error. Chen uses this fact to argue that Boltz-2 is learning physics. However, some of Chen's other results contradict this notion. He reports that Boltz-2 does poorly when buried waters are present in the structure. He also reports how Boltz-2 tends to predict affinities within a particular range, in other words, it does not predict outliers well. Both results suggest statistical interpolation, often using spurious correlations, not the learning of physics. 

On the OpenADMET challenge, which included SARS-CoV-2 MPro data, Boltz-2 performed ``very poorly, with a mean absolute error worst among any method''. However, it's important to note that several of the other methods received significant fine-tuning for the task at hand. 

On a set of small molecules (<500 Da) known as ``molecular glues'',  Boltz-2 did poorly vs a FEP method, and for 4/6 protein targets Boltz-2 had Pearson's $R=0$ or negative, suggesting extreme failure to predict.\cite{Lukauskis2025} Essentially, the researchers found that Boltz-2 was worthless for molecular glues. This is probably due to molecular glues not appearing much in Boltz-2's training data. The authors note that the FEP method was much more computationally expensive (\$28-\$128 in cloud compute cost per compound vs a cost of \$0.06 for Boltz-2). 

In totality, the benchmarks so far suggest that Boltz-2, like other ML methods, should be used with great caution when applied outside of distribution of examples in the training data.

%-------------------------------------
\subsection{DrugFormDTA}
DrugFormDTA was released by Khokhlov et al.~in 2025.\cite{DrugFormDTA2025} It uses SMILES to represent molecules and an amino acid sequence to represent proteins. They leverage Chemformer to transform the SMILES into embeddings. Chemformer is a bi-directional autoregressive transformer model pretrained on over 100 million molecules. To encode proteins into embeddings, they leverage ESM-2 (Evolutionary Scale Modeling-2), using the 150 M parameter model. ESM-2 is a 30-layer transformer trained on 27 M amino acid sequences. Work shows that (surprisingly), ESM-2 learns implicit structural information during training, and an extension of ESM-2 has been developed that can predict contact maps.

An important feature of the DrugFormDTA model is they do not assume IC$_{50} \approx $ K$_i$, and they have separate outputs to predict both IC$_{50}$ and K$_i$. During training, if one value is not present, the model's prediction is used instead. 

There are a couple reasons why DrugFormDTA was not developed in an optimal manner for use with viral proteins. The authors of DrugFormDTA filtered BindingDb to mammalian proteins and human virus proteins (getting rid of bacterial proteins and also non-human virus proteins (0.8\% of rows)). They also weighted human proteins higher by a factor of $\sqrt{10} = 3.16$. They also removed any protein target with more than 2,500 amino acids, which includes 59\% of viral polyproteins in the dataset. That restriction dropped 91\% of the viral polyprotein types and 31\% of the total virus data rows.
\begin{figure}[h]
\centering
\includegraphics[width=0.8\textwidth]{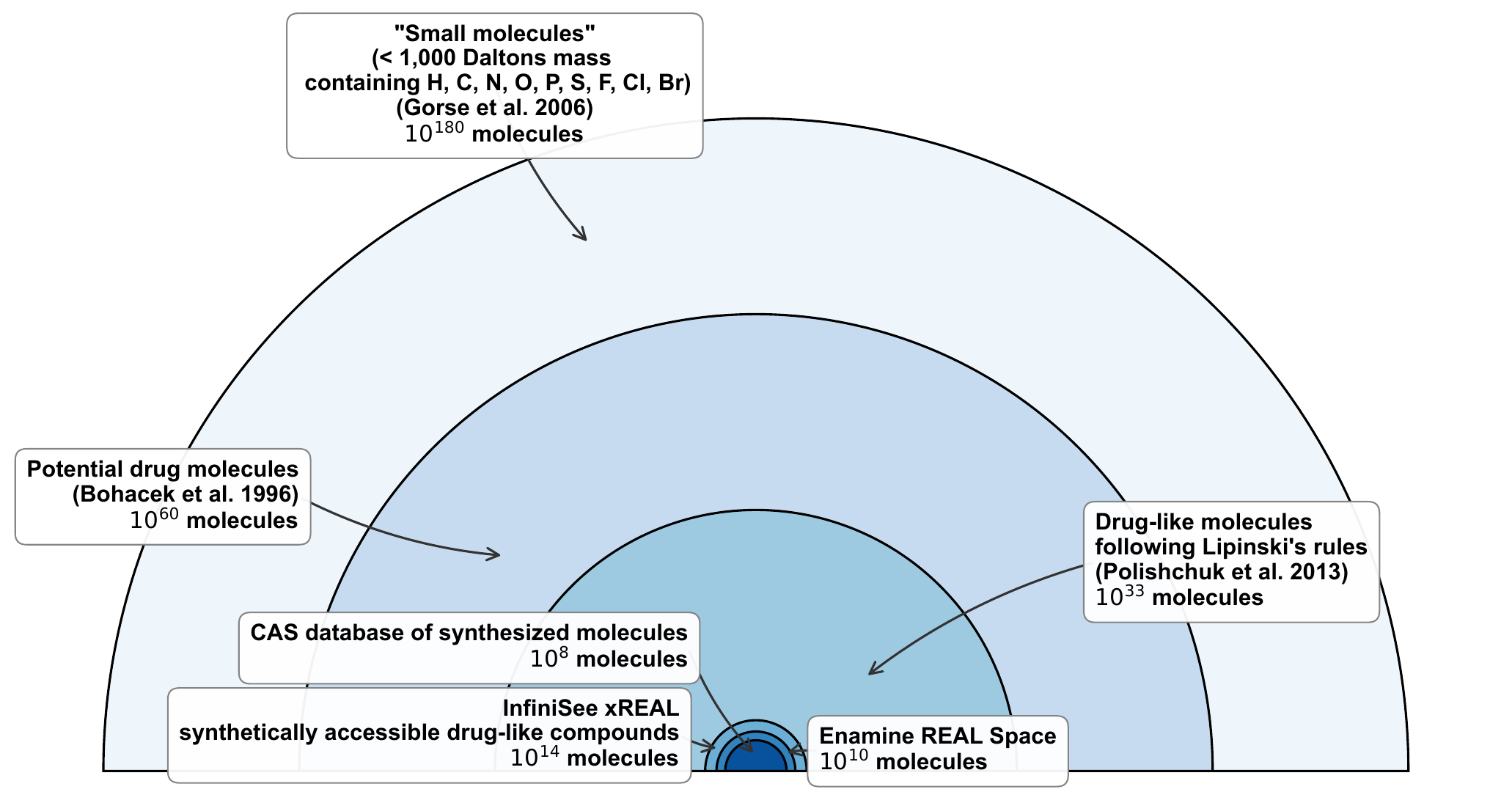}
\caption{Some estimates of the size of small molecule chemical space and the number of potential drug molecules.}
\label{fig:chemical_space_nested_euler}
\end{figure}

%----------------------------------------
\subsection{Protenix and AlphaRank}
Protenix is an open-source system modeled after AlphaFold3 which was released by Chinese company ByteDance in January 2025. In February 2026, ByteDance released Protenix-v1, a 368M parameter model that they claim is the first open-source structure prediction model to outperform AlphaFold3 across diverse benchmarks while adhering to the same training data cutoff, model scale, and inference budget.\cite{Protenix2026} Protenix-v1 is released under an Apache 2.0 license, making it available for both academic and commercial use. The Protenix ecosystem also includes Protenix-Mini, a lightweight variant that reduces inference cost, and PXDesign, a protein design suite. AlphaRank is a module that was developed by fine-tuning the Protenix ``pairformer'' module on protein-ligand binding data from ChEMBL and BindingDb. At a high-level, the AlphaRank system is very similar to Boltz-2. Some initial AlphaRank results were published in a 2025 ICML GenBio Workshop paper that compares AlphaRank to FEP, but no further papers or code have been released as of March 2026.\cite{hong2025how}

%--------------------------------------------------------
\subsection{Direct ML approaches}
Direct ML approaches take in a molecule's graph structure (in a string format like SMILES) along with the full amino acid sequence for a virus, and then output a score between 0 and 1 (0=ineffective, 1=effective). (The task is to directly predict effectiveness, not protein binding.) The attractiveness of this approach is that effectiveness includes important considerations like absorption, distribution, metabolism, excretion, and toxicity (ADMET) properties. However, there are some major difficulties with this approach. The main issue is that the models can only be trained on a limited number of molecules for which effectiveness data is available plus possibly a roughly equal number of ``decoys''. Given the diversity of chemical structures and the diversity of viral amino acid sequences, a few thousand examples is not really enough to train a generalizable model. We were particularly interested in following the approach Cant\"{u}rk et al.~used to develop a ML model for COVID-19 drug repurposing which appears in a 2020 preprint.\cite{Canturk2020} However, Cant\"{u}rk et al.~did not include decoys in training and only tested their model on a small number of approved antivirals they held-out for testing, so it is hard to evaluate the utility of their approach. In our view, the drugs their model recommended for repurposing did not make much sense. 

%---------------------------------------------------------------------------------
%--------------------------------------------------------------------------------- 
\section{Strategies for navigating chemical space}

Much of the challenge in in silico screening boils down to finding efficient ways to search through chemical space. The size of chemical space is very vast, even when constrained to drug-like small molecules (see figure \ref{fig:chemical_space_nested_euler}) In this section we non-comprehensively touch on some considerations and techniques. 

%-------------------------------------------------------------
\subsection{Ultra-large combinatorial chemical spaces}
There are now large pre-defined chemical spaces of molecules that are predicted to be synthetically accessible and are advertised as available on-demand. In general, these spaces are generated by applying validated reaction steps to large sets of chemical building blocks, yielding combinatorial libraries that now reach into the trillions of molecules. Table~\ref{tab:ulcs} summarizes the major commercially available ultra-large combinatorial spaces. For many of those spaces, the complete list of molecular structures can be downloaded. Another option is to generate synthesizable molecular structures on-the-fly using the SynFlowNet generative model.\cite{cretu2025synflownet} 

\begin{table}[htbp]
\centering
\caption{Commercially available ultra-large combinatorial chemical spaces. Shipping and synthetic feasibility data from \url{https://www.alipheron.com/spaces}. The non-commercial ZINC15 and SAVI-Space-2024 databases are provided as references. (The synthetic accessibility of ZINC15 is 100\% since all compounds are synthesized and available.)}
\label{tab:ulcs}
\begin{tabular}{llrcp{1.8cm}}
\toprule
Vendor & Space & \shortstack{Compounds\\(billions)} & \shortstack{Shipping\\(weeks)} & \shortstack{Synthetic\\feasibility (\%)} \\
\midrule
Enamine        & REAL Space      & 83      & 3--4  & $>$80 \\
Chemspace      & Freedom Space   & 296     & 5--6  & $>$80 \\
eMolecules\cite{Zagribelnyy2026}  & Synple-eXplore  & 5,320 & 3--4  & $>$85 \\
PharmaBlock    & Sky Space       & 56.8    & 4--6  & $>$85 \\
WuXi AppTec    & GalaXi          & 28.6    & 4--8  & 60--80 \\
Life Chemicals & LifeCheMyriads  & 50      & 6--8  & $>$80 \\
Molecule.One   & D2B-SpaceM1     & 1.5     & 2--6  & $>$85 \\
XtalPi         & VAST            & 9.2     & 2--3  & $>$80 \\
\href{https://www.otavachemicals.com/}{Otava}  & ChemInfinita    & 800     & 5--8  & 55--85 \\
\bottomrule
docking.org    & ZINC15          & 0.23    &  n/a  & 100\% \\
NCI            & SAVI-Space-2024 & 7.5     &  n/a  & high   \\
\end{tabular}
\end{table}

%-------------------------------------------------------------
\subsection{Similarity search}
The most popular type of similarity search that one sees in academic papers use fingerprint algorithms like MACCS keys or Morgan/ECFP fingerprints. These methods encode information about the 2D molecular graph into a long binary vector. In drug design, however, what matters is not the 2D graph but the 3D structure and in particular the location of pharmacophores in 3D space. Two molecules can have very different looking 2D graphs but have similar 3D shape. The Ultrafast Shape Recognition with CREDO Atom Types (USRCAT) fingerprint encodes information about the 3D shape.\cite{Schreyer2012} Each atom is classified into four categories: hydrophobic, aromatic, H-Bond Donor (e.g., the O in -OH, N in -NH), and H-bond acceptor (eg lone O and N).

%-------------------------------------------------------------
\subsection{The Lipinski rules and other heuristics}
\begin{table}[htbp]
\centering
\begin{tabular}{llp{2cm}p{2cm}}
\toprule
Criteria Set & Rule & FDA approved antiviral drugs (n=73) & FDA approved drugs with oral delivery route (n=716) \\
\midrule
\multirow{5}{*}{Lipinski (1997) \cite{Lipinski1997}} & Molecular weight $<$ 500 Da & 56.2 & 83.7 \\
 & Hydrogen bond donors $\leq$ 5 & 100.0 & 97.6 \\
 & Hydrogen bond acceptors $\leq$ 10 & 89.0 & 93.3 \\
 & LogP between $-2$ and 5 & 83.6 & 82.1 \\
 & \textbf{All Lipinski rules} & \textbf{54.8} & \textbf{71.5} \\
 \midrule
 \multirow{3}{*}{Egan (2000) \cite{Egan2000}} & TPSA $\leq$ 132 \AA$^2$ & 52.1 & 86.0 \\
 & LogP $\leq$ 5.9 & 87.7 & 93.6 \\
 & \textbf{All Egan criteria} & \textbf{49.3} & \textbf{81.4} \\
\midrule
\multirow{3}{*}{Veber (2002) \cite{Veber2002} } & Rotatable bonds $\leq$ 10 & 72.6 & 92.6 \\
 & TPSA $\leq$ 140 \AA$^2$ or HBD+HBA $\leq$ 12 & 74.0 & 91.9 \\
 & \textbf{All Veber criteria} & \textbf{60.3} & \textbf{87.3} \\
%\midrule
%\textbf{Combined} & \textbf{All 6 criteria} & \textbf{47.9} & \textbf{69.4} \\
\bottomrule
\end{tabular}
\caption{Testing the Lipinski and Veber rules on 73 drugs approved by the FDA as orally-administered small molecule antivirals. IV drugs (like Remdesivir, Foscarnet, Cidofovir), topical antivirals, and nasal antivirals were excluded. If FDA approved antiviral drugs with other delivery routes are included, all of the pass rates decrease. They are compared with 716 FDA approved drugs which are marked with "ORAL" as the delivery route in the FDA data dump.}
\label{tab:lipinski_comparison}
\end{table}
The Lipinski rules, also known as the ``rule of five'', are a set of criteria developed by Christopher Lipinski at Pfizer and published in 1997 which have had an outsized influence on drug discovery.\cite{Lipinski1997} They are often applied to narrow chemical space towards compounds likely to have high permeability through the intestinal lining and through cellular membranes. For instance, they were used by Du et al. in their machine learning based search for SARS-CoV-2 MPro/3CLpro inhibitors.\cite{Du2024}

The Lipinski rules are as follows: 

\begin{itemize}
\item Molecular mass must be less than 500 Daltons.
\item No more than 5 hydrogen bond donors (including hydroxyl, amino, etc)
\item No more than 10 hydrogen bond acceptors (all nitrogen or oxygen atoms)
\item Octanol-water partition coefficient not greater than 5 or fat-water partition coefficient (logP) between -2 and
5.
\end{itemize}

The term ``rule of 5'' comes from the fact that the number 5 appears multiple times. The Lipinkski criteria were followed by the Egan criteria (2000) and Veber criterion, published by Daniel Veber and colleagues from GlaxoSmithKline in 2002:\cite{Veber2002} 
\begin{itemize}
\item No more than 10 rotatable bonds. 
\item The polar surface area (PSA) should be $\leq$ 140 \AA$^2$ or, equivalently, the number of hydrogen bond donors plus acceptors should be $\leq$ 12.
\end{itemize}

By studying rat bioavailability data for 1,100 compounds, Veber argues that the first Lipinski rule is too strict and not a fundamental limiter. Molecules with mass $>$ 500 Daltons can have high bioavailability if the number of rotatable bonds is constrained. Rotatable bonds increase solubility, but too many can decrease membrane permeability, since it becomes harder for the molecule to adopt a compact, low-polarity conformation. Too many rotatable bonds also make protein binding difficult -- the molecule can be thought of as being ``floppy''. 

%To be effective, drugs must be able to permeate the cell membrane or (more likely) be transported across by one of numerous transporters. Thus, some have selected that drugs should mimic metabolic compounds (ie the many mostly intermediary compounds that exist in human metabolic processses.) Hagan found that 90\% of marketed drugs have a Tanimoto similarity of more than 0.5 to the closest human metabolite.\cite{OHagan2014}
We test all three rules on approved antiviral compounds in table \ref{tab:lipinski_comparison}. Only about 50\% of FDA approved antivirals pass the Lipinski and Egan criteria, and only 60.3\% pass the Veber criteria. Among FDA-approved drugs overall, 71.5\% pass the Lipinski criteria, 81.4\% pass the Egan criteria, and 87.3\% pass the Veber criteria. As drug discovery continues to evolve, all three of these criteria appear to be becoming less relevant over time. 

%-------------------------------------------------------
\subsection{PAINS \& ``bad apples''}
Pan-assay interference compounds (PAINS) are compounds that cause false positives in screening, usually due to profligate chemical reactivity. Typically these are compounds that easily bind to wide variety of proteins, either covalently or electrostatically. PAINS may bind extensively to an enzyme without actually inhibiting the active site. Some enzymes require metal ions to operate, and some PAINS compounds chelate those metal ions, removing them from the assay (this shuts down the enzyme's activity, giving the appearance of binding). Other PAINS interfere with assays by causing protein aggregation, by catalyzing the formation of colloidal structures, or by directly interacting with fluorescent probes. Not surprisingly, many PAINS are cytotoxic, while others, like curcumin, are not (curcumin is especially notorious since it exhibits multiple PAINS mechanisms). In the drug-discovery literature the term ``PAINS'' often specifically to compounds exhibiting a set of structural patterns that were identified by Baell \& Holloway in 2010.\cite{Baell2010} The RDKit library contains a catalog of about 480 PAINS substructure patterns across three classes. It is common in early-phase screening to remove compounds that have PAINS substructures. The BioAssay-Data Associative Promiscuity Pattern Learning Engine (``Badapple'') version 2.0 (2025) is the latest-and-greatest open-source tool for eliminating PAINS, using substructure matching and interpretable machine learning.\cite{Ringer2025}

%-----------------------------------------
\subsection{Previous large-scale screening initiatives}
In this section we describe two large-scale screening initiatives in depth to give a sense of how they are typically done. 
\begin{figure}[ht]
\centering
\includegraphics[width=\textwidth]{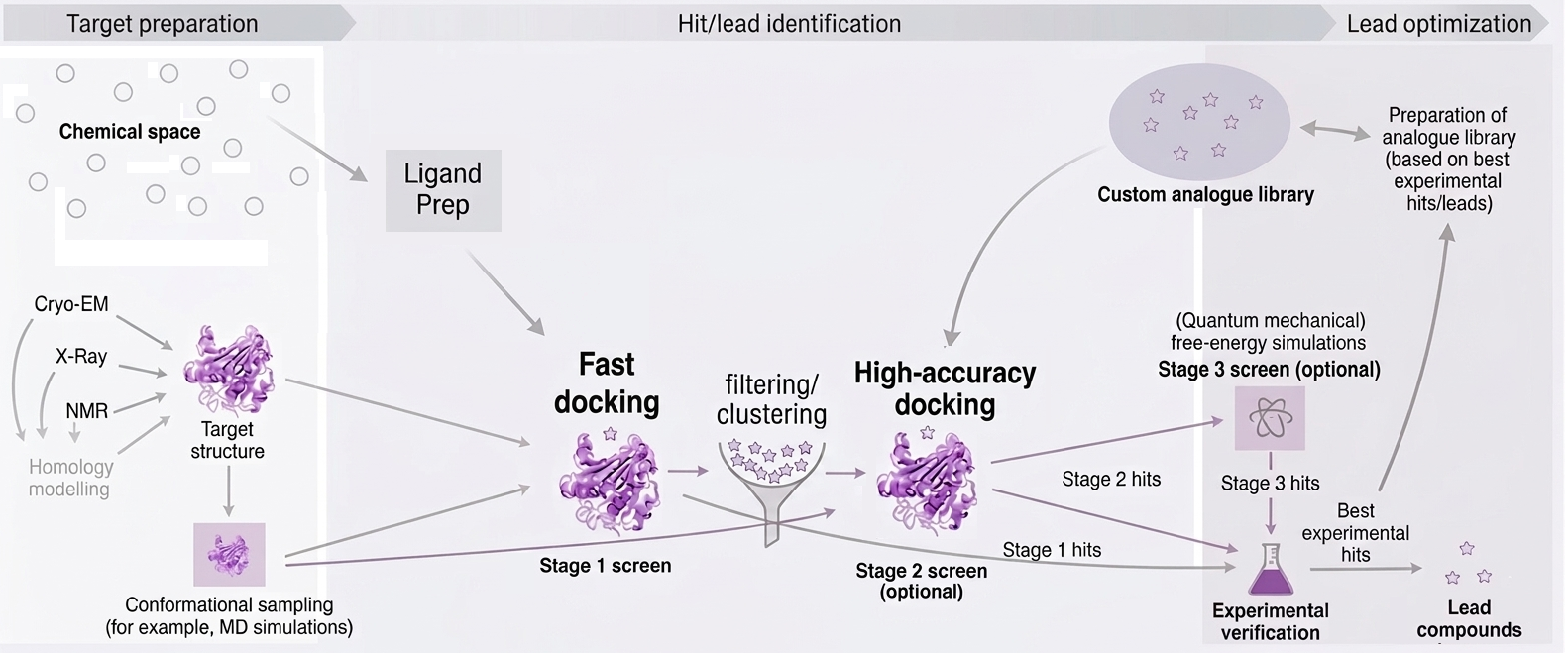}
\caption{Overview of the typical flow in early stage in silico screening, adapted in part from Gorgulla, et al.\cite{Gorgulla2020}. }
\label{fig:flowchart}
\end{figure}

Schuller et al.~(2023) did Glide docking using the BioAscent library of 125,000 compounds against the SARS-CoV-2 NSP3-macrodomain, focusing the docking on the ADP-ribose pocket. The ADP-ribose pocket is a catalysis site which removes mono-ADP-ribose ``tags'' from proteins, reversing the actions of the host interferon response.\cite{Schuller2023} 2,000 compounds were selected for experimental assaying using homogeneous time-resolved fluorescence assay (HTRF), of which 1,786 could be obtained from their chemical supplier, BioAscent. The most potent inhibitors obtained had IC$_{50}$ of 18$\mu$M and 28$\mu$M. However, further crystallographic screens did not show these compounds binding to the protein. A follow-up search by looking at compounds with ``analogous scaffolds'' yielded compound with IC$_{50}$s of 4.9 $\mu$M and 3.1$\mu$M, among others. Interestingly, the authors also did experimental screening of 1600 FDA-approved drugs. They ran into several known assay interference compounds (PAINS) like biotin, which they ignored and did not mark for follow-up. The most potent compounds found via the approved drug screen were the selenium-containing compound ebselen (IC$_{50}$ = 3.7~$\mu$M)) and the mercury-containing compound thimerosal (IC$_{50}$ = 62 µM). A very high Hill slope on ebselen indicated promiscuous binding on multiple sites or possible assay interference. Overall, the two most promising FDA-approved drugs were two antibiotics: aztreonam (IC$_{50}$ = 29 µM) and ceftazidime (IC$_{50}$ = 37 µM). A follow-up crystallographic screen showed aztreonam in a groove next to the ADP-ribose pocket reaction pocket, not in the ADP-ribose pocket. Since they focused their docking on the pocket, this unexpected binding mode would not have been discovered via docking.

Fraser et al.~(2025) did large-scale screening against the TMPRSS2 protein, which is a serine protease important for SARS-CoV-2 entry.\cite{Fraser2025} There was no crystal structure available, so they used a model developed in RosettaCM. More specifically, 1,500 protein structures were generated, and they were tested against known serine protease inhibitors. The best-performing model was docked using DOCK with 1.5 million compounds from the ZINC15 library. The compounds were filtered by the following criteria - deep penetration into subsite S1, no large molecular strain, no buried unsatisfied H-bond donors, and less than three unsatisfied H-bond acceptors, and the presence of an H-bond with Asp326. This led to about 1,500 compounds. Using clustering they selected 108 for experimental testing. One hit ($< 100 \mu M$) was obtained which had an IC$_{50}$ of 22 nM. Once a crystal structure for TMPRSS2 became available, the group did a second docking campaign against 200 million compounds from ZINC-22 focused on the experimentally-determined binding pocket. 59 compounds were tested, leading to two hits with IC$_{50}$s of 12 and 9.4 $\mu$M.

%billion compound docking against SARS-CoV-2 targets \cite{Rogers2023}

\begin{table*}[htbp]
\centering
\caption{Examples of large-scale screening runs, mostly against SARS-CoV-2 targets. The definition of a ``hit'' varies across papers. For instance, looking at the first three rows, Fink et al.\ defined a hit as IC$_{50}$ < 300 $\mu$M in their first campaign, IC$_{50}$ < 200 $\mu$M in their second campaign, and IC$_{50}$ < 150 $\mu$M for their third (covalent) campaign.\cite{Fink2023} Generally, ``hit'' was defined quite liberally, ie $< 200 \mu$M rather than $< 1 \mu$M.}
\label{tab:bruteforce_vs}
\resizebox{\textwidth}{!}{%
\begin{tabular}{llllrlrrll}
\toprule
Method & \# Mols & Target               & \# Tested & Hit rate & Most potent hit & \shortstack{CPU\\core hrs} & \shortstack{\# CPU\\cores} & Year & Ref. \\
\midrule
DOCK3.7     & 335 M & SARS-CoV-2 MPro      & 97   & 6\%    & IC$_{50}$=20 $\mu$M  &         &      & 2023 & \cite{Fink2023} \\
DOCK3.7     & 862 M & SARS-CoV-2 MPro      & 146  & 11.6\% & IC$_{50}$=29 $\mu$M  & 481,000 & 1000 & 2023 & \cite{Fink2023} \\
DOCKovalent & 6.5 M & SARS-CoV-2 MPro      & 194  & 6\%    & IC$_{50}$=97 $\mu$M  &         &      & 2023 & \cite{Fink2023} \\
DOCK3.7     & 235 M & SARS-CoV-2 MPro      & 100  & 3.0\%  & IC$_{50}$=26 $\mu$M  &         &      & 2022 &  \\
DOCK3.7     & 400 M & SARS-CoV-2 NSP3 mac1 & 124  & 40.3\% & IC$_{50}$=42 $\mu$M  & 139,497 & 1000 & 2023 & \cite{Gahbauer2023} \\
Glide SP    & 0.25M & SARS-CoV-2 NSP3 mac1 & 1786 &        & IC$_{50}$=18 $\mu$M  &         &      & 2023 & \cite{Schuller2023} \\
DOCK3.7     & 1.1 B & SARS-CoV-2 NSP14     & 72   & 2.7\%  & IC$_{50}$=6 $\mu$M   & 121,018 & 1000 & 2023 & \cite{Singh2023} \\
DOCK3.7     & 437 M & EP4R                 & 71   & 14.1\% & IC$_{50}$=850 nM     & 273,692 & 1500 & 2023 & \\
DOCK        & 200 M & TMPRSS2              & 108  & 0.9\%  & IC$_{50}$=121 nM     &         &      & 2025 & \\
DOCK        & 200 M & TMPRSS2              & 59   & 3.4\%  & $K_i$=9.4 $\mu$M     &         &      & 2025 & \\
VirtualFlow & 1.3 B & KEAP1                & 590  & 11.7\% & $K_d$=114 nM         & 5.4 M   & 8000 & 2020 & \\
Chem Space Docking & 858 M & ROCK1         & 69   & 39\%   & $K_i$=38 nM          &         &      & 2022 & \cite{Beroza2022} \\
%AutoDock-GPU & 1.4 B & SARS-CoV-2 (5 targets) &      &        &                   &         &      & 2023 & \cite{Rogers2023} \\
DOCK3.7     & 20 M  & SARS-CoV-2 NSP3 mac1 & 60   & 33\%   & IC$_{50}$=183 $\mu$M &         & 500  & 2021 & \cite{Schuller2021} \\
\bottomrule
\end{tabular}
}
\end{table*}
 %------------------------------------------------------------
%\begin{figure}[h]
%\centering
%\includegraphics[width=0.5\textwidth]{CACHE-challenge-2.png}
%\caption{Figure from \url{https://cache-challenge.org/challenges/finding-ligands-targeting-the-conserved-rna-binding-site-of-sars-cov-2-nsp13}}
%\label{fig:cache_challenge_2}
%\end{figure}

%---------------------------------------
\subsection{Learning from the CACHE challenges}
The CACHE (Critical Assessment of Computational Hit-finding Experiments) challenges, first launched in December 2021, are a series of challenges which have solicited teams to predict compounds that will be experimentally active against a given target. These challenges rigorously test methods by choosing targets that do not already have a lot of known actives and by experimentally testing predictions, rather than relying on previously existing experimental data. Most of these challenges have two rounds of experiments. 

Some of the CACHE challenges provide the teams with fragment screening data. Fragment screening is a technique where protein crystals are soaked or grown in high concentrations of small molecule ``fragments'' (typically MW 110-250 Da).\cite{Badger2011} The proteins are soaked/grown in solutions with 3-5 fragments, so a screen of 3000 fragments can be done in 1,000 X-ray crystallography runs. During the pandemic there were several large-scale fragment screens against the SARS-CoV-2 NSP3 macrodomain and NSP1.\cite{Schuller2021,Lennartz2025}

%--------------------------------------- 
\paragraph{CACHE challenge \#1}
The first CACHE challenge targeted the WD-40 repeat (WDR) domain of the Leucine-rich repeat serine/threonine-protein kinase-2 (LRRK2) protein. The top performing team managed to win with only VINA and GNINA docking combined with careful selection and filtering of molecules.\cite{Dunn2024} (They used VINA and GNINA separately and then combined results.) Another team docked molecular fragments to find pharmacophores. They then did pharmacophore screening over 179 million compounds from the ZINC20, MCULE, and MCULE Ultimate datasets using the PHARMIT PHARMER search method. The hits were screened using fast VINA docking and the 3,572 with the highest score were selected. In parallel, they docked 7 million molecules from Molport's library using GNINA. The top 1,000 were then docked again, this time using multiple conformations generated from MD simulations. They top 1,000 were docked with both VINA and GNINA and the best score from either was used for final ranking. The final set was clustered using FP2 fingerprint vectors to weed out molecules that were similar and increase the diversity of their final set, which had 109 molecules.

%--------------------------------------- 
\paragraph{CACHE challenge \#2}
The 2nd CACHE Challenge targeted the RNA Site of the SARS-CoV‑2 helicase (NSP3).\cite{Herasymenko2025} Teams started with the results of a fragment screen, which found some fragments bind right near the RNA binding pocket. The 2nd place team started with the \textit{Foldit} platform, where users are tasked with constructing molecules that fit in the binding pocket. They also looked at bound fragments in a published crystal structure fragment screen (published on PDB) and grew those fragments into potential small molecule inhibitors. They then did a fingerprint similarity search to find similar molecules, followed by flexible docking with \textit{RosettaLigand} on all of the candidates accrued so far. Deep learning scoring was done last. The first placed team adopted a traditional strategy. From the existing crystal structure fragment screen, the team developed a set of pharmacophores they thought were important. They then generated a large dataset of molecules with ``pharmacophore constraints'' and ran docking on those molecules using Schr\"{o}dinger's Glide software. During their docking study they tested binding to variations on the protein structure obtained using molecular dynamics simulation. The outputs were then refined using another scoring function (\textit{HYDE} from BioSolvIT). The top 300 candidates were visually inspected and studied by medicinal chemists to finalize their selection. Another high-performing workflow leveraged machine learning first, then ``high-throughput docking'', then ``deep learning docking'', and finally consensus scoring. The challenge organizers created a list of six top workflows based on aggregated performance on both the first and second rounds. A high diversity of approaches are visible in the top six workflow, so it is hard to make any firm conclusions regarding workflow. However, while one of the two top teams did not use ML, it appears that hybrid (ML + docking) approaches perform well, generally. Additionally, all six of the top workflows at had at least four filtering stages, and five out of six had at least six stages. Among the teams in general, 4/23 had only two filtering stages and 1/23 had only three. It appears that having four or more stages is important to be competitive. In the third CACHE challenge (discussed below) only two out of 23 teams had only two stages and only 1 had 3 stages, and again none of those were identified as top approaches. 

%------------------------------- 
\paragraph{CACHE challenge \#3}
The third CACHE challenge targeted the SARS-CoV-2 NSP3 macrodomain, which is an enzyme that removes ADP-ribose modifications from proteins, counteracting part of the innate immune system's response.\cite{Herasymenko2026} Three of the top seven teams used molecular dynamics and perturbation methods to approximate the free-energy. Based on lessons learned from prior efforts, the organizers of the challenge suggested that all participants filter out ``chemical liabilities'' (PAINS) and they required participants to filter out carboxylic acids, which have poor cell membrane permeability. 

%------------------------------- 
\paragraph{Takeaways from the CACHE challenges}
Across the four CACHE challenges where results have been published so far, multi-stage pipelines having three or more stages almost always outperformed pipelines with only one or two stages. Fragment-based approaches have appeared in 40\% of the approaches that successfully led to an experimental hit, and machine learning has played a role in $\approx 70$\%. Two commercial docking programs were heavily utilized by many teams - Glide from Schr\"{o}dinger and ICM-pro from MolSoft. The most popular open-source docking programs were AutoDock Vina and GNINA (some others that appear are PLANTS, Smina, rDock, and DiffDock). Free energy perturbation (FEP) and MMPBSA, two molecular dynamics based methods, became more popular in CACHE \#3, suggesting MD methods are making inroads. The molecular dynamics toolchains used were NAMD+FEP+CHARMM, GROMACS+PyAutoFEP, AmberTools, and GROMACS+gmx\_MMPBSA. One high-level takeaway is that a wide variety of approaches are competitive - there is no clear ``best approach''. This suggests room for further methods development and optimization. Another takeaway, remarked upon by the CACHE organizers in CACHE \#3, is that the judgment of medicinal chemistry experts still appears to matter, especially when it comes to final selection.

%-----------------------------------------
\subsection{Screening orchestration software}
There are now open-source tools for orchestrating screening campaigns on cloud CPUs or high performance computing (HPC) clusters. VirtualFlow is open-source screening workflow orchestration software developed at Harvard Medical School and used by the commercial consultancy \href{https://www.virtualdiscovery.net/}{virtualdiscovery.net}.\cite{Gorgulla2020} It is designed specifically for operation on computing clusters that run \textit{slurm} job handling software. A similar piece of open-source software is WARPDock (2023), which is designed for Oracle Cloud Infrastructure.\cite{McDougal2023} In the past few years there has also been a proliferation of commercial platforms that ``glue'' together open-source tools, allowing users to easily run screening campaigns in the cloud using open-source tools (for instance \href{https://neurosnap.ai/small-molecule}{Neurosnap}, \href{https://www.tamarind.bio/}{Tamarind Bio}, \href{https://www.deepmirror.ai}{DeepMirror}, \href{https://rowansci.com}{Rowan Scientific}, \href{https://www.litefold.in}{LiteFold}, and \href{https://boltz.bio/}{Boltz.bio}, among others). These platforms sit alongside similar platforms for protein design such as \href{https://latentlabs.com/}{Latent Labs}, \href{https://www.ariax.bio}{Ariax}, \href{https://www.cradle.bio}{Cradle.bio}, and \href{https://proteinbase.com}{ProteinBase}. 

%---------------------------------------------------------------------------
%---------------------------------------------------------------------------
%---------------------------------------------------------------------------
\section{Our contribution - testing open-source tools on novel antiviral dataset}
\subsection{Building a proper antiviral-ligand binding dataset}
\begin{table}[h]
\centering
\begin{tabular}{lrrrrrrrp{1.75cm}}
\toprule
\textbf{Virus Type} & \textbf{Total} & \textbf{\# Proteins} & \textbf{\# Ligands} & \textbf{Ki} & \textbf{Kd} & \textbf{Ki+Kd} & \textbf{IC50} & \textbf{\% Polyprotein targets} \\
\midrule
HIV & 31,018 & 118 & 19,743 & 5,776 & 401 & 6,138 & 25,011 & 0.0\% \\
SARS-CoV-2 & 14,479 & 5 & 11,545 & 385 & 273 & 658 & 13,833 & 99.3\% \\
HCV & 7,240 & 33 & 5,064 & 1,295 & 78 & 1,373 & 5,870 & 35.9\% \\
Influenza & 4,471 & 26 & 2,350 & 190 & 132 & 322 & 4,154 & 0.0\% \\
Herpesvirus & 2,669 & 16 & 2,264 & 41 & 1 & 42 & 2,638 & 0.0\% \\
SARS-CoV-1 & 2,268 & 6 & 1,693 & 337 & 98 & 435 & 1,853 & 89.6\% \\
HCoV-229E & 606 & 1 & 266 & 5 & 0 & 5 & 601 & 100.0\% \\
Dengue & 587 & 5 & 430 & 134 & 12 & 146 & 448 & 77.0\% \\
Rhinovirus & 466 & 1 & 395 & 167 & 0 & 167 & 299 & 0.0\% \\
HBV & 403 & 3 & 260 & 0 & 2 & 2 & 401 & 0.0\% \\
West Nile & 347 & 3 & 324 & 48 & 10 & 58 & 315 & 89.0\% \\
SV40 & 173 & 1 & 173 & 0 & 0 & 0 & 173 & 0.0\% \\
Other$^*$ (26 others) & 1,223 & 38 & 1,043 & 74 & 64 & 138 & 1,083 & 12.0\% \\
\midrule
\textbf{Total} & \textbf{65,950} & \textbf{238} & \textbf{44,337} & \textbf{8,452} & \textbf{1,071} & \textbf{9,481} & \textbf{56,679} & \textbf{31.1\%} \\
\bottomrule
\end{tabular}
\caption{Overview of the human virus protein data in BindingDb. \small{$^*$Other viruses = HTLV (123), GBV-B (121), Coxsackie (112), HPV (104), IBV (101), Vaccinia (97), MERS-CoV (97), Ab-MLV (80), Norovirus (73), ASV (67), RSV (66), HCoV-OC43 (37), WMSV (35), HPIV (27), Poliovirus (20), AEV (15), MMTV (13), RSV-avian (8), Zika (7), FCoV (6), Reovirus (4), HCoV-EMC (4), BVDV (1), MVV (1), FuSV (1).}}
\label{tab:affinity_summary}
\end{table}

We wanted a custom dataset for fine-tuning and testing ML models specifically for viral proteins. To start, we analyzed what viral data was in BindingDb already. Table \ref{tab:affinity_summary} presents an overview of the virus data in BindingDb as of November 2025. On the BindingDb website there is also a separate SARS-CoV-2 dataset, which to our surprise we found contains several thousand rows not in BindingDb, so it was included in our larger test set. 

%------------------------------------------------
\subsection{Polyprotein processing}
Many positive-sense single-stranded RNA viruses translate their genome as one or two large polyproteins that must be cleaved into individual functional proteins by viral proteases. Such viruses include coronaviruses, hepatitis C virus, and picornaviruses. Using polyproteins allows the virus to encode proteins more efficiently and also helps ensure stoichiometry of the individual proteins. In coronaviruses, the genome encodes two overlapping open reading frames (ORF1a and ORF1ab) that are translated into polyproteins \textit{Pp1a} and \textit{Pp1ab}. Many other viruses use polyproteins as well. The \textit{GAG} and \textit{GAG-POL} genes encode the \textit{Gag} and \text{Gag-Pol} polyproteins in HIV, which are cleaved by an aspartic protease to yield structural proteins (matrix, capsid, nucleocapsid) and enzymes (protease, reverse transcriptase, integrase). HCV encodes a single $\sim$3,000 amino acid polyprotein that is processed by both host peptidases and viral proteases (NS2 and NS3/NS4A) into 10 mature proteins.

In BindingDb, many targets are listed as polyproteins or as polyprotein slices rather than the actual protein target used in the assay. We found 30\% of the viral data has a polyprotein target, and this rises to 99.3\% for SARS-CoV-2. Polyprotein slices were easy to handle by creating a custom mapping function. For instance, the slice ``Replicase polyprotein 1ab [3264-3569]'' in BindingDb corresponds to SARS-CoV-2 main protease. 

We were able to map some of the polyproteins to a target using ligand names which sometimes contain the target name (ie. ``CLpro inhibitor 6a''). However, in most cases the only way to tell is to look at the attached reference which is usually either a scientific article or patent. We downloaded the abstracts for the papers and patents via the PubMed API and USPTO API and fed them into Anthropic's low-cost Haiku model. Claude Code then ``manually'' checked 60 entries and found 59/60 were correct. The first author also reviewed the 60 entries, performing a quick sanity check. The one ``incorrectly classified'' entry that surfaced was actually a mistake in BindingDb, not an issue with the LLM.\footnote{BindingDb had ``Human rhinovirus B'' as the organism, but the paper abstract (PMID 15203137) states that the protein was a Hepatitis A virus (HAV) 3C proteinase. Thus, BindingDb either got the virus name or the paper reference wrong. Upon further investigation, we determined that BindingDb had gotten the virus name wrong.}

There still remained some targets that are marked ``unknown'', which were entries without a literature reference or patent reference. Some of those rows contained a ChEMBL ID. While it might be possible to track down literature via ChEMBL ID and then figure out the target from the literature, we decided not to undertake that work. 

\subsection{Fine-tuning and testing the DrugFormDTA model}
\begin{table}[htbp]
\centering
\caption{Curation steps for viral protein-ligand data}
\label{tab:pipeline-attrition}
\begin{tabular}{lr}
\toprule
Stage & Records \\
\midrule
Raw BindingDb viral protein data & 65,950  \\
After removing inexact values (ie ``> 1,000'') & 56,881 \\
After removing polyproteins we couldn't map to a specific protein & 51,127 \\
After removing proteins missing a reference sequence & 50,413 \\
After aggregation (mean of duplicate measurements) & 42,379   \\
After adding new Heli-SMACC (409) and SMACC (NCATS) (238) external data & 43,026 \\
After removing invalid SMILES (21) & \textbf{43,005} \\
\bottomrule
\end{tabular}
\end{table}

\begin{table}[htbp]
\centering
\caption{Small hold-out test set for testing the fine-tuned DrugFormDTA model.}
\label{tab:initial_test_set}
\begin{tabular}{llr}
\toprule
Virus & Protein Target & Ligands \\
\midrule
Dengue & NS2B-NS3 Protease & 19 \\
HCV & HCV\_NS3\_NS4A & 30 \\
MERS-CoV & PLpro & 8 \\
SARS-CoV-2 & Mpro & 40 \\
West\_Nile & NS2B-NS3 Protease & 30 \\
Zika & NS2B-NS3 Protease & 14 \\
\midrule
\textbf{Total} & & \textbf{141} \\
\bottomrule
\end{tabular}
\end{table}

Table \ref{tab:pipeline-attrition} summarizes the major steps in the data curation. BindingDb data was combined with the SMACC and Heli-SMACC data. The original Heli-SMACC dataset had 20,431 rows. However 18,970 rows were for human helicases and 234 were for bacterial helicases, leaving 1,227 rows spread across 11 viruses. 

\begin{figure}[ht]
\centering
\makebox[\textwidth][c]{%
\begin{subfigure}[t]{0.49\textwidth}
\includegraphics[width=\textwidth]{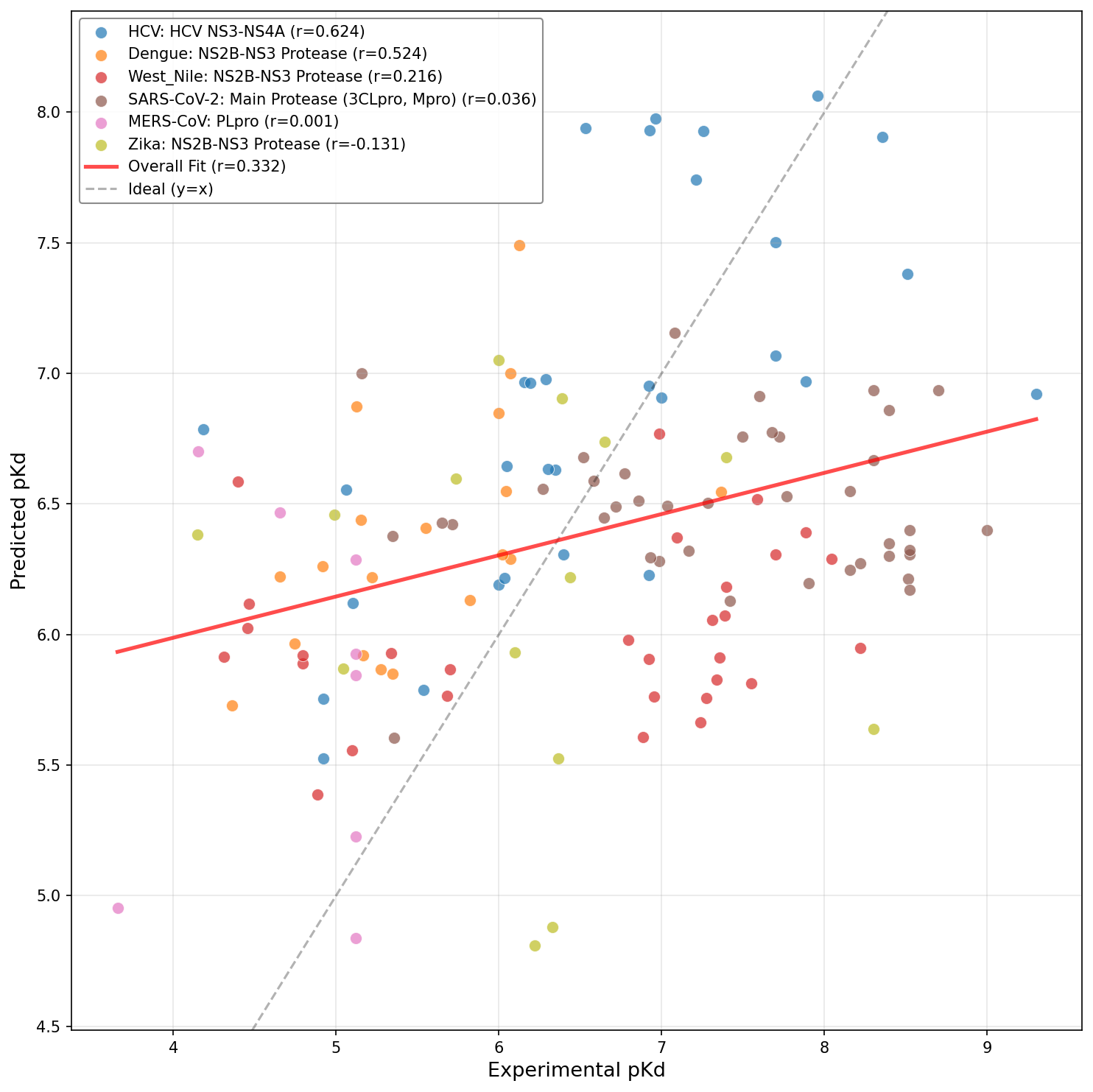}
\caption{Base model}
\end{subfigure}
\hspace{0.02\textwidth}
\begin{subfigure}[t]{0.49\textwidth}
\includegraphics[width=\textwidth]{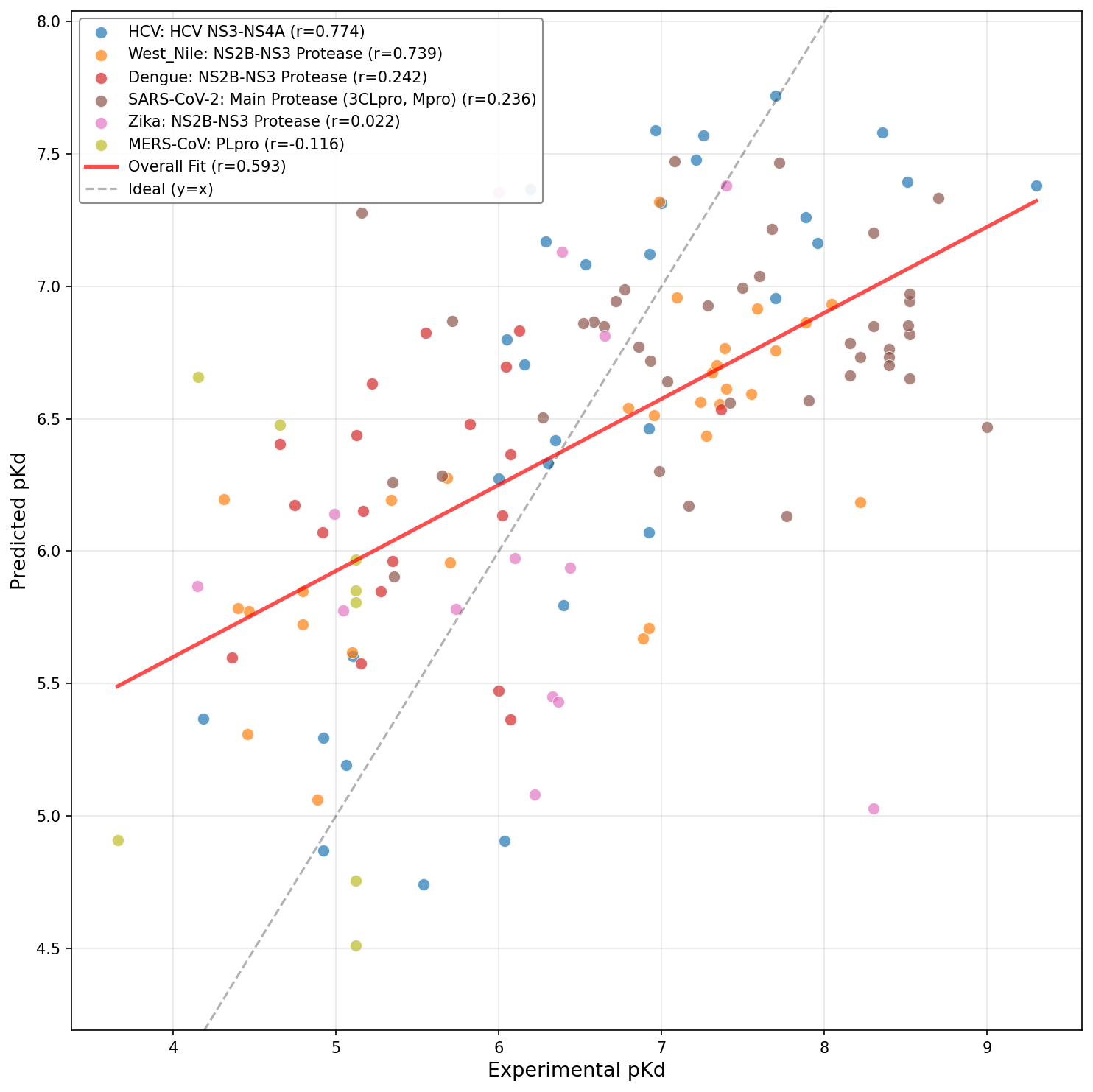}
\caption{Fine-tuned model}
\end{subfigure}
}%
\caption{Scatter plots comparing predicted versus experimental pKd values for the DrugFormDTA model - base vs fine-tuned.}
\label{fig:benchmarking_scatter}
\end{figure}

The resulting dataset was split into training and test sets for the purposes of fine-tuning the DrugFormDTA model. The final train-test split was 43,005 (99.4\%) for training and 280 (0.6\%) for testing. 2.5\% of the training set was used as a validation set during fine-tuning to monitor the training progress. The small test set used for testing the fine-tuned DrugFormDTA model is described in table \ref{tab:initial_test_set}. It consists of $K_i$/$K_d$ values only. We compared the fine-tuned model with Autodock GNINA on this test set - the results are shown in Table \ref{tab:initial-metrics}. Scatterplots for the original and fine-tuned DrugFormDTA model are shown in figure \ref{fig:benchmarking_scatter}. Fine-tuning DrugFormDTA boosted performance, as expected. We did two fine-tuning runs -- for the first fine-tuning run we used MSE loss and two stages of training with different learning rates. For the second fine-tuning run we used a more conventional setup with Huber loss and a ``reduce on plateau'' learning rate schedule. We got similar overall results for both (namely a large boost in overall performance compared to the base model). The DrugFormDTA system actually trains four models separately and ensembles them so during inference, so we replicated this by fine-tuning four separate models. We were skeptical that this actually boosts performance, so we tested fine-tuning a single model as well, finding about equivalent performance. In general ensembling leads to higher robustness and better generalization performance, so for the benchmark testing described in the next section we used the ensemble model (fine-tuned model \# 2).

\begin{table*}[ht]
\centering
\caption{Initial set of performance metrics on the test set of $K_i$ values for 141 viral protein-ligand pairs. For autodock runs, ``e'' is the ``exhaustiveness'' level.}\label{tab:initial-metrics}
\resizebox{\textwidth}{!}{%
\begin{tabular}{lcccccp{1.25cm}p{1.25cm}c}
\toprule
Model & $Q^2$ & Pearson $r$ & Spearman $\rho$ & RMSE & MAE & time/mol (RTX 3060) & time/mol (RTX 5090) & failures \\
\midrule
%Autodock Vina (e=8)      & 0.080 & 0.332 & 0.313 & 1.229 & 1.031 & 30 s & & 7 \\
Autodock GNINA (e=8)      & -0.202 & 0.226 & 0.084 & 1.426 & 1.198 & 32 s & & 8 \\
Autodock GNINA (e=32)     & -0.123 & 0.279 & 0.187 & 1.377 & 1.169 & 120 s & & 6 \\
Boltz-2                   & -0.211 & 0.504 & 0.490 & 1.416 & 1.127 & 88 s & 22 s & 1 \\
DrugFormDTA               & 0.080 & 0.332 & 0.313 & 1.229 & 1.031 & 0.8 s & 0.02 s & 0 \\
DrugFormDTA fine-tuned 1  & 0.344 & 0.593 & 0.621 & 1.038 & 0.860 & 0.8 s & 0.02 s & 0 \\
DrugFormDTA fine-tuned 2  & 0.290 & 0.660 & 0.692 & 1.072 & 0.889 & 0.8 s & 0.02 s & 0 \\
DrugFormDTA single model  & 0.407 & 0.677 & 0.675 & 0.980 & 0.810 & 0.7 s & 0.015 s & 0 \\
DrugFormDTA+Boltz2+GNINA  & 0.233 & 0.559 & 0.524 & 1.123 & 0.871 & & & 0 \\
\bottomrule
\end{tabular}%
}
\end{table*}

\FloatBarrier
% TODO: Uncomment when benchmark/plots/ images are available
%\begin{figure}[htbp]
%\centering
%\begin{subfigure}[b]{0.48\textwidth}
%    \includegraphics[width=\textwidth]{benchmark/plots/scatter_cnn_affinity.png}
%    \caption{CNN Affinity vs Experimental pK$_d$ ($r = 0.40$)}
%    \label{fig:scatter_cnn}
%\end{subfigure}
%\hfill
%\begin{subfigure}[b]{0.48\textwidth}
%    \includegraphics[width=\textwidth]{benchmark/plots/scatter_gnina_pKd.png}
%    \caption{Vina $\Delta G$-derived pK$_d$ vs Experimental pK$_d$ ($r = 0.02$)}
%    \label{fig:scatter_vina}
%\end{subfigure}
%\caption{Comparison of GNINA affinity predictions against experimental pK$_d$/pK$_i$ values for 122 antiviral compounds across nine virus-target combinations. Points are colored by virus-target pair. The CNN-based affinity score (left) shows moderate correlation with experimental values, while the physics-based Vina scoring function (right) shows no significant correlation.}
%\label{fig:gnina_benchmark}
%\end{figure}

%-------------------------------------------------------------------------------------------
%--------------------------------------------------------------------------------------------
\subsection{Testing open-source tools for binding affinity prediction on viral protein data}
We selected 15 open-source methods for testing on a larger test set. An overview is given in Table~\ref{tab:methods_comparison}. The larger test set contains 853 compounds spread across virus proteins from 10 virus species (see Table \ref{tab:newtestset} for the full breakdown). The test set includes data from the COVID Moonshot initiative, which consists of fluorescence-based IC50 data for 2,059 compounds and cellular EC50 data for 892 compounds, all against SARS-CoV-2 MPro.\cite{Boby2023} (The COVID-19 moonshot project was an initiative which ``crowdsourced'' experimental data on SARS-CoV-2 MPro binding from researchers around the world.) Data for 139 of those compounds were found not to exist in BindingDb and were added to the test set as novel ligands which are not in the DrugFormDTA training data and likely not in the Boltz-2 training data.  Overall, about 80\% of this larger test set overlaps with the training set for Boltz-2 and DrugFormDTA, but recall that polyproteins were not split for many of the viral proteins in the original training sets for both models. 

\begin{table}[h]
\centering
\caption{Comparison of protein-ligand binding affinity prediction methods evaluated in this work.}\label{tab:methods_comparison}
\small
\begin{tabular}{lp{3.2cm}p{3.2cm}p{3.2cm}l}
\toprule
\textbf{Name} & \textbf{Method} & \textbf{Inputs} & \textbf{Outputs} & \textbf{Ref.} \\
\midrule
DrugFormDTA  & ML+Chemformer+ESM                           & SMILES, AA sequence & K$_i$, pIC$_{50}$  & \cite{DrugFormDTA2025} \\
\hline
Boltz-2       & MSA+AlphaFold3 architecture+affinity module & SMILES, AA sequence & docked pose, score & \cite{Passaro2025}  \\
\hline
GNINA         & Rigid docking + ML scoring              & 3D protein structure, 3D ligand structure  & docked pose, score & \cite{mcnutt2021gnina} \\
\hline
FlowDock      & Diffusion-based generative docking + ML scoring & 3D protein structure, SMILES       & docked pose, score & 
\cite{Morehead2025} \\
\hline
DiffDock & Diffusion-based docking & 3D protein structure, SMILES & Docked pose, confidence score (no affinity) & \cite{Corso2022} \\
\hline
Protenix-Dock & Rigid docking + ML scoring             & 3D protein structure, 3D ligand structure& docked pose, $\Delta G$  & \\
\hline
GatorAffinity & ML (SE(3)-equivariant GNN)            & 3D protein pocket, SMILES   & Binding affinity (p$K_d$)  & \cite{GatorAffinity2025} \\
\hline
Interformer   & docking + transformer affinity head   & 3D protein structure, 3D ligand structure& Docked pose, binding affinity & \cite{Interformer2024} \\
\hline
Uni-Dock      & Rigid docking                         & 3D protein structure, 3D ligand structure  & docked pose, $\Delta G$ & \cite{Yu2023UniDock}\\
\hline
AutoDock-GPU  & Rigid docking                         & 3D protein structure, 3D ligand structure  & docked pose, $\Delta G$ & \cite{SantosMartins2021}\\
\hline
DrugCLIP      & ML (Contrastive learning)             & 3D binding pocket, SMILES                  & Binding affinity      & \cite{Jia2026} \\
\hline
Vina-GPU      & Rigid docking                           & 3D protein structure, 3D ligand structure  & docked pose, $\Delta G$  & \cite{Tang2024} \\
\hline
Uni{-}Mol+GNINA & ML (SE(3)-equivariant transformer) + CNN scoring & 3D protein structure, SMILES & Docked pose, binding affinity & \cite{Alcaide2025} \\
%AEV-PLIG & GNN (atomic environment vectors + PLIGs) & 3D protein-ligand complex (PDB + SDF) & Binding affinity & \cite{Valsson2025} \\
%PM7/COSMO & Semi-empirical quantum chemistry & 3D protein-ligand complex & $\Delta G$ (kcal/mol) & \cite{Klamt1993} \\
\bottomrule
\end{tabular}
\end{table}

\begin{table}[h]
\centering
\caption{Antiviral benchmark test set: 853 compounds across 8 target types, 16 crystal structures, and 10 virus species.
The threshold for distinguishing active vs inactive compounds was pKd $=$ 6.0.}\label{tab:newtestset}
\vspace{0.5em}
\small
\begin{tabular}{@{}llrrrrrr@{}}
\toprule
\textbf{Target} & \textbf{Virus / PDB} & $\boldsymbol{n}$ & \textbf{Act.} & \textbf{Inact.} & \textbf{pKd range} & $\boldsymbol{K_i/K_d}$ & \textbf{IC50} \\
\midrule
\multirow{4}{*}{Mpro}
    & SARS-CoV-2 / 7VH8  & 90  & 70 & 20  & 5.0--9.0  & 62  & 28  \\
    & SARS-CoV-1 / 2AMD  & 44  & 6  & 38  & 4.7--7.8  & 3   & 41  \\
    & HCoV-229E / 7YRZ   & 30  & 1  & 29  & 4.7--9.3  & 1   & 29  \\
    & HCoV-229E / 2ZU2   & 19  & 0  & 19  & 5.0--5.5  & 0   & 19  \\
\cmidrule(l){1-8}
\multirow{2}{*}{PLpro}
    & SARS-CoV-1 / 4OW0  & 78  & 21 & 57  & 3.1--6.9  & 63  & 15  \\
    & MERS-CoV / 4RNA    & 8   & 0  & 8   & 3.7--5.1  & 8   & 0   \\
\cmidrule(l){1-8}
HCV NS3/NS4A
    & HCV / 3P8N         & 121 & 77 & 44  & 3.4--10.7 & 98  & 23  \\
\cmidrule(l){1-8}
HCV NS5B
    & HCV / 3SKA         & 57  & 29 & 28  & 1.4--8.5  & 2   & 55  \\
\cmidrule(l){1-8}
HIV Protease
    & HIV-1 / 1HXW       & 114 & 67 & 47  & 2.0--11.0 & 88  & 26  \\
\cmidrule(l){1-8}
HIV RT
    & HIV-1 / 4G1Q       & 62  & 34 & 28  & 3.2--9.5  & 4   & 58  \\
\cmidrule(l){1-8}
\multirow{5}{*}{NS2B-NS3}
    & West Nile / 5IDK   & 45  & 33 & 12  & 4.3--8.2  & 45  & 0   \\
    & Dengue / 3U1J      & 32  & 11 & 21  & 3.6--8.1  & 1   & 31  \\
    & Dengue / 4M9K      & 28  & 3  & 25  & 2.8--7.4  & 28  & 0   \\
    & Zika / 7VLG        & 8   & 6  & 2   & 4.1--7.4  & 8   & 0   \\
    & West Nile / 2FP7   & 4   & 4  & 0   & 7.6--7.7  & 4   & 0   \\
\cmidrule(l){1-8}
Neuraminidase
    & Influenza A / 3TI5 & 113 & 63 & 50  & 2.4--10.0 & 61  & 52  \\
\midrule
\multicolumn{2}{@{}l}{\textbf{Total}} & \textbf{853} & \textbf{425} & \textbf{428} & 1.4--11.0 & \textbf{476} & \textbf{377} \\
\bottomrule
\end{tabular}
\end{table}

\begin{figure}[ht]
\centering
\begin{subfigure}[t]{0.49\textwidth}
\centering
\includegraphics[width=\textwidth]{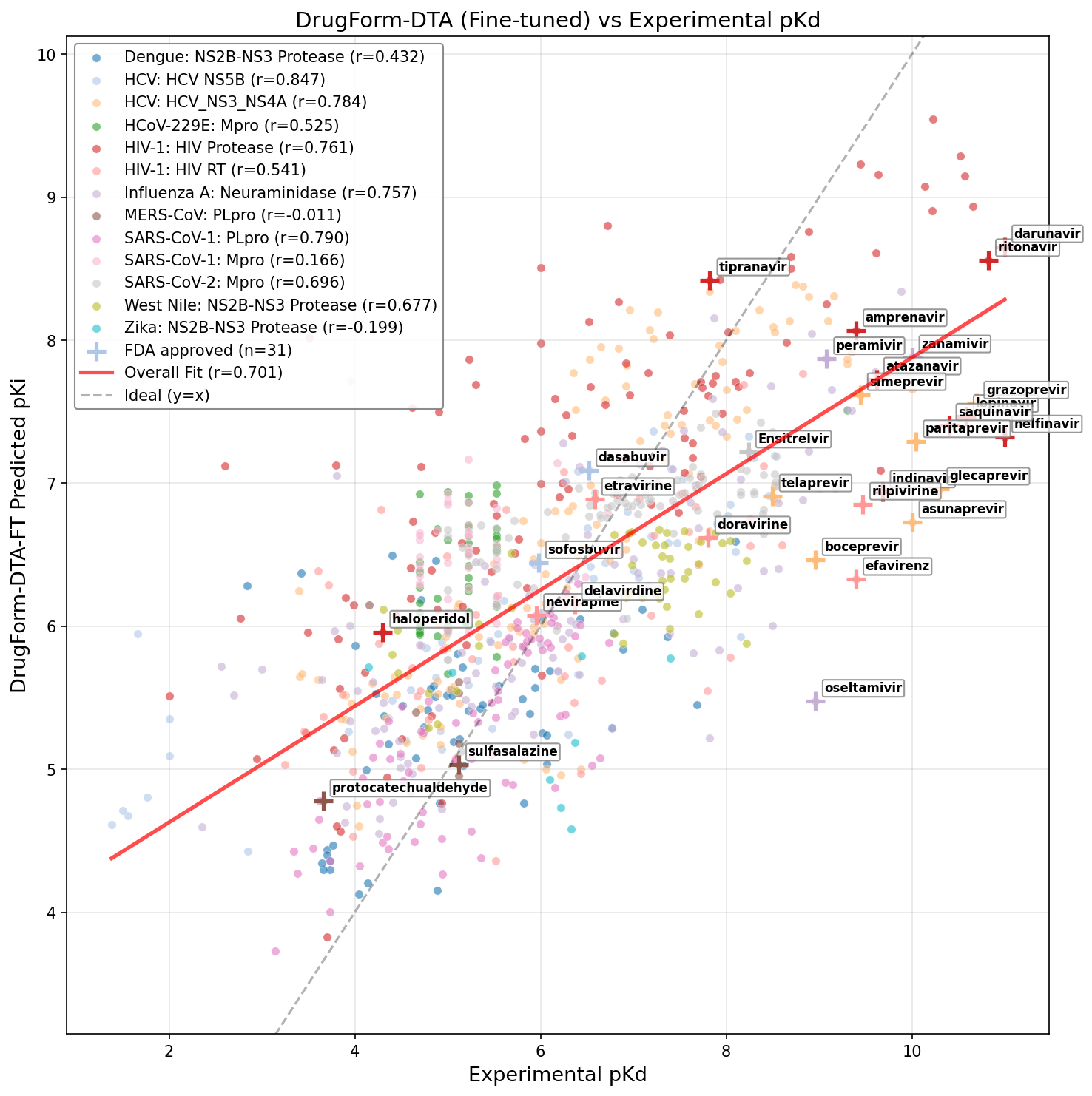}
\caption{Fine-tuned DrugFormDTA model}
\end{subfigure}
\hfill
\begin{subfigure}[t]{0.49\textwidth}
\centering
\includegraphics[width=\textwidth]{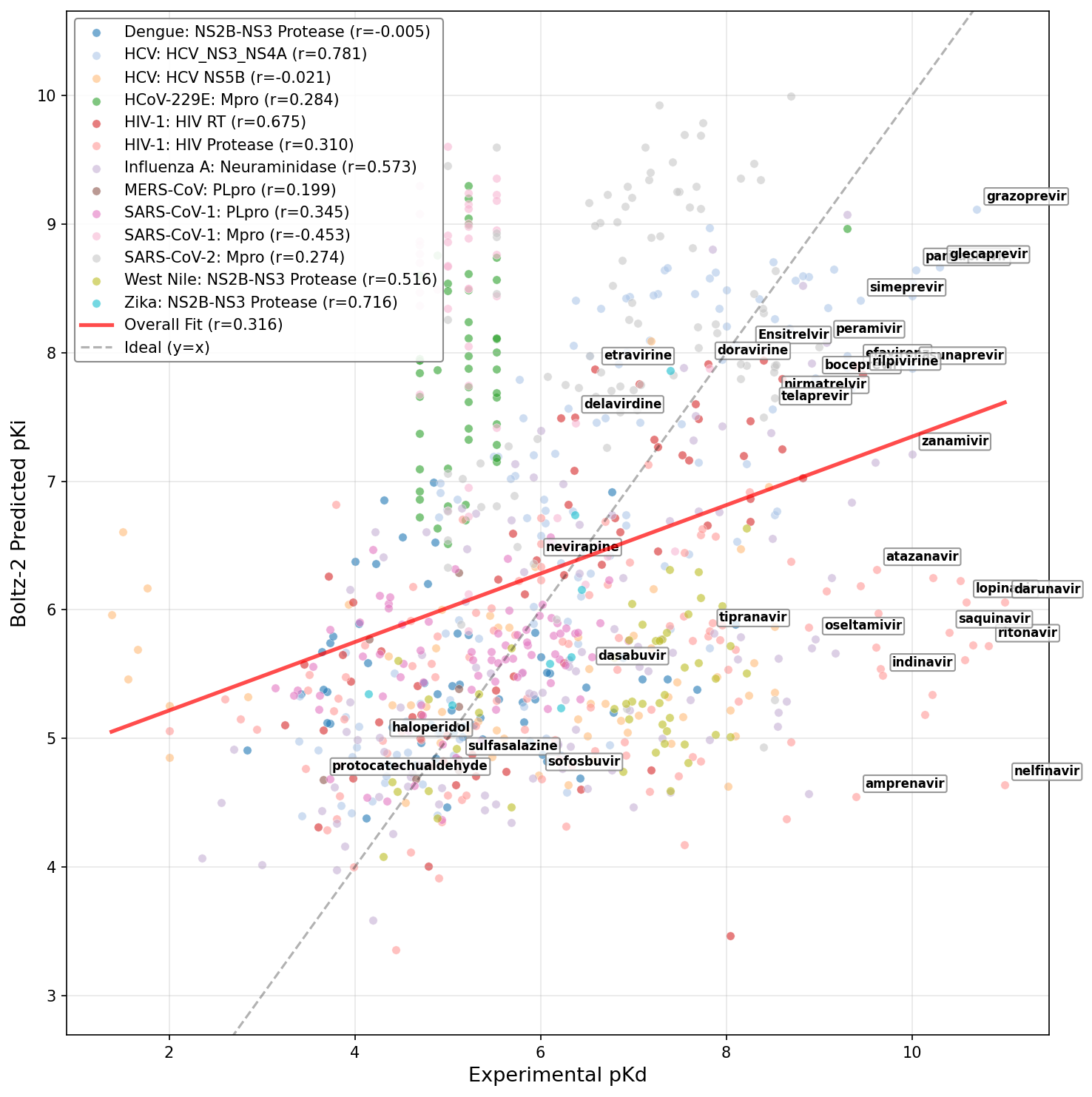}
\caption{Boltz-2}
\end{subfigure}
\caption{Performance of the fine-tuned DrugFormDTA model and Boltz-2 on the expanded test set (has overlap with the training data). FDA-approved drugs were added to the test set and are labeled.}
\end{figure}

\begin{figure}[h]
\centering
\includegraphics[width=\textwidth]{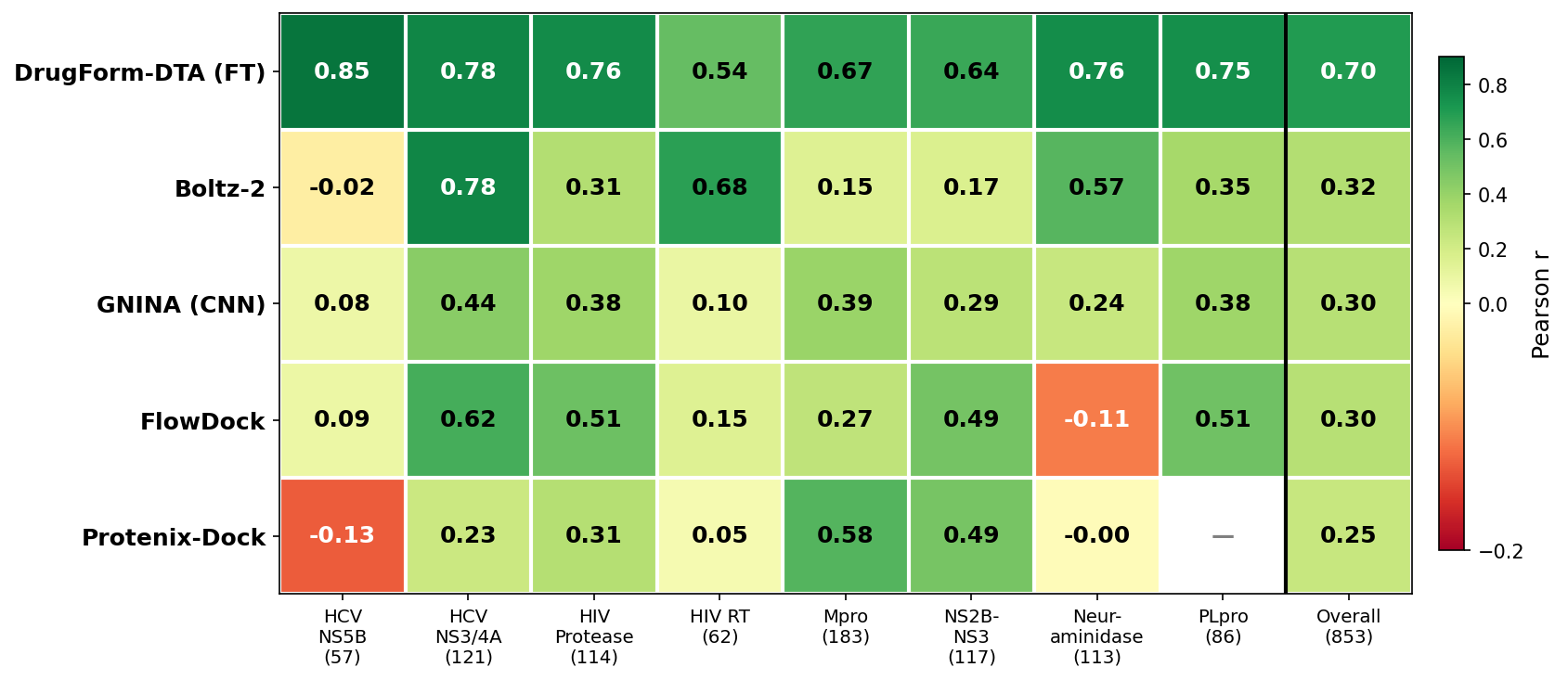}
\caption{Per-target Pearson $r$ correlation between predicted and experimental pK$_d$/pK$_i$ values for each scoring tool across eight viral targets. The number of compounds per target is shown in parentheses. The rightmost column shows the overall correlation across all 853 compounds.}
\label{fig:heatmap}
\end{figure}

The full results on the larger test set are shown in Table \ref{tab:fullresults}. The performance of these models varied greatly across viral proteins, as shown in figure \ref{fig:heatmap}. Boltz-2 did very well on HIV RT ($r=0.68$), which makes sense given the large amount of training data available for that protein. The relatively poor performance of Boltz-2 on SARS-CoV-2 MPro ($r=0.15$) is likely explained by the polyprotein issue we discussed earlier. There was a large variation in docking method performance as well across proteins, consistent with prior literature showing that docking performance varies greatly between different protein types.  
\begin{table}[h]
\centering
\resizebox{\textwidth}{!}{%
\begin{tabular}{@{}llcrrrrrrrr@{}}
\toprule
\textbf{Tool} & \textbf{Type} & \textbf{ML} & $\textbf{\textit{r}}$ & $r^2$/\textbf{CoD} & $\boldsymbol{\rho}$ & {\tiny\textbf{RMSE}} & {\tiny\textbf{AUROC}} & {\tiny\textbf{BEDROC}} & \textbf{Fail\%} & \textbf{s/mol} \\
\midrule
    DrugForm{-}DTA (FT)$^\dagger$ & Seq-only & $\checkmark$ & 0.701 & 0.492 & 0.698 & 1.19 & 0.845 & 0.962 & 0.0 & 0.03 \\
    DrugForm{-}DTA & Seq-only & $\checkmark$ & 0.498 & 0.248 & 0.472 & 1.45 & 0.725 & 0.899 & 0.0 & 0.03 \\
    Boltz{-}2 & Blind & $\checkmark$ & 0.316 & 0.100 & 0.342 & 1.59 & 0.634 & 0.615 & 0.0 & 24.18 \\
    GNINA (CNN) & Blind & $\checkmark$ & 0.302 & 0.091 & 0.312 & 1.60 & 0.652 & 0.753 & 0.0 & 4.20 \\
    FlowDock & Blind & $\checkmark$ & 0.295 & 0.087 & 0.254 & 1.60 & 0.640 & 0.832 & 0.1 & 4.00 \\
    Protenix{-}Dock & Targeted & $\times$ & 0.246 & 0.060 & 0.239 & 1.68 & 0.630 & 0.670 & 6.6 & 49.00 \\
    GatorAffinity & Targeted & $\checkmark$ & 0.199 & 0.039 & 0.200 & 1.67 & 0.587 & 0.596 & 9.3 & 0.25 \\
    Interformer & Targeted & $\checkmark$ & 0.146 & 0.021 & 0.122 & 1.66 & 0.562 & 0.672 & 0.6 & 0.35 \\
    Uni{-}Dock & Targeted & $\times$ & 0.146 & 0.021 & 0.100 & 1.66 & 0.535 & 0.587 & 1.2 & 0.52 \\
    Uni{-}Mol+GNINA{-}CNN & Blind & $\checkmark$ & 0.130 & 0.017 & 0.132 & 1.66 & 0.557 & 0.689 & 7.4 & 8.1 \\
    Interformer (full) & Targeted & $\checkmark$ & 0.126 & 0.016 & 0.071 & 1.73 & 0.547 & 0.654 & 16.2 & 22.10 \\
    AutoDock{-}GPU & Blind & $\times$ & 0.085 & 0.007 & 0.063 & 1.67 & 0.496 & 0.466 & 1.2 & 2.49 \\
    DrugCLIP & Targeted & $\checkmark$ & 0.075 & 0.006 & 0.062 & 1.67 & 0.531 & 0.474 & 0.0 & 0.22 \\
    AutoDock{-}GPU & Targeted & $\times$ & 0.047 & 0.002 & 0.032 & 1.67 & 0.485 & 0.393 & 0.0 & 2.09 \\
    GNINA (Vina, no CNN) & Blind & $\times$ & 0.007 & 0.000 & -0.034 & 1.68 & 0.476 & 0.521 & 1.9 & 3.35 \\
    Vina{-}GPU & Blind & $\times$ & -0.063 & -0.004 & -0.121 & 1.67 & 0.408 & 0.448 & 1.2 & 2.75 \\
    \bottomrule
\end{tabular}
}%
\caption{
\small
Benchmark results on 853 antiviral compounds (10 virus species, 16 crystal structures). To normalize each tool's outputs, each tool's raw score was recalibrated via linear regression to match pKd units as closely as possible before computing RMSE and the coefficient of determination (CoD). To compute the AUROC and BEDROC metrics we binarized the experimental data using an active/inactive threshold of pKd = 6.0. In the ``Type'' column, ``Seq-only'' = sequence-based (no 3D structure), ``Blind'' = input is the full protein without pocket targeting, and ``Targeted'' = input is a box around the desired binding pocket (for enzymes, this is often the active site). See section \ref{sec:eval_metrics} for detailed definitions of the different metrics.
}
\label{tab:fullresults}
\end{table}

\FloatBarrier

%-------------------------------------------------------------- 
\section{Conclusion}
In this work we first presented comprehensive overviews of the available datasets and open-source tools for antiviral drug discovery. We then created a custom dataset of 43,005 viral protein-ligand binding measurements, fixing numerous issues with polyproteins that we found in BindingDb. Next, we benchmarked fifteen binding affinity prediction methods on a diverse test set of 853 compounds spread across 16 protein targets from 10 viruses. Fine tuning the DrugFormDTA model on our custom dataset with the split polyproteins significantly boosted performance from $r=0.50$ to $r=0.70$. During the course of this work we created a library of approved drugs, GRAS compounds, and natural products. We also created a referenced list of approved and investigational antivirals which can be explored at a custom dashboard at \href{https://antivirals-database.radvac.org}{antivirals-database.radvac.org}. This work serves as a solid foundation for future work on developing tools and platforms for rapid antiviral repurposing and rapid design of antiviral combinations. 
\FloatBarrier

%---------------------------------------------------------- 
\backmatter

\bmhead{Acknowledgments}
The authors would like to thank Brian M. Delaney, Ranjan Ahuja, and Alex Hoekstra for providing proofreading, guidance, and suggestions during the writing of this whitepaper.

%\section*{Declarations}
%\begin{itemize}
%\item Conflict of interest/Competing interests (check journal-specific guidelines for which heading to use)
%\item Ethics approval and consent to participate
%\item Consent for publication
%\item Data availability 
%\end{itemize}

%---------------------------------------------------------------------------------------------------------------

\clearpage
\begin{appendices}

%---------------------------------------------------------------
\section{Building an approved, GRAS, and natural product compound library}
\begin{table}[h!]
\centering
\begin{tabular}{l r r r r r}
\toprule
\textbf{Database} & \textbf{Compounds} & \textbf{Exclusive} & \textbf{Has Species Source} & \textbf{Has Food Source} & \textbf{No Source Info} \\
\midrule
COCONUT & 686,549 & 456,285 & 33.0\% & 8.1\% & 59.4\% \\
LOTUS & 276,518 & 108,539 & 92.3\% & 2.8\% & 7.6\% \\
FooDB & 65,764 & 5,516 & 11.7\% & 100\% & 0\% \\
CMAUP & 60,215 & 4,210 & 96.8\% & 6.7\% & 1.9\% \\
ANPDB & 11,063 & 2,567 & 95.8\% & 8.8\% & 3.7\% \\
%\midrule
\textbf{Total} & & \textbf{817,252} & \textbf{41.9\%} & \textbf{8.0\%} & \textbf{51.0\%} \\
\bottomrule
\end{tabular}
\caption{Table showing the contributions of different datasets to the final natural products / GRAS database.}
\label{tab:naturalprodoverview}
\end{table}

We compiled a library of 2,096 approved small molecule drugs using two FDA datasets and extensive filtering to remove peptides and biologics.  We combined this with data from Drug Central and AntiviralDb, adding an additional 215 small molecule drugs approved by other major regulatory bodies, bringing the total to 3,311 approved drugs.  

Table \ref{tab:naturalprodoverview} summarizes the construction of our approved+GRAS+NP compound library. Our starting point was \href{https://foodb.ca/downloads}{FooDB}, a collection of food-compound associations. Most of the foods have NCBI taxonomy IDs, so one can understand precisely what species is being talked about. We then added the \href{https://zenodo.org/records/6378223}{LOTUS database}, a natural products database run on Wikidata.\cite{Rutz2022} Next, we added COCONUT (COlleCtion of Open Natural prodUcTs), an aggregation of 24 openly accessible natural product databases.\cite{Chandrasekhar2024} We then added the \href{https://african-compounds.org/anpdb/download/}{African Natural Products Database} and the Collective Molecular Activities of Useful Plants (CMAUP) dataset.\cite{Hou2023} We were careful to avoid duplicates by mapping every compound to an InChiKey. While COCONUT 2.0 (2024) contains two traditional Chinese medicine databases -- TCMDB@Taiwan and TCMID, we found it did not contain data from ``Yet another Traditional Chinese Medicine Database for Drug Discovery'' (YaTCM).\cite{Li2018} Unfortunately, we could no longer find YaTCM online. In the future we could include small peptides, although in general peptide modifications would be required to achieve greater bioavailability. From the 20 proteinogenic amino acids, there are $20^2 = 400$ dipeptides and $20^3 = 8,000$ tripeptides.

%------------------------------------------------------------------------------
\section{Quantifying combination synergy}

%\begin{figure}[ht]
%\centering
%\includegraphics[width=0.4\textwidth]{viral_load_vs_dose.png}
%\caption{Viral load as a function of drug dose.}
%\label{fig:dose}
%\end{figure}
When optimizing drug dose, it is first important to understand the non-linear nature of drug dose-response curves, which typically look a logistic curve. If the dose is too small, the virus may recover. In the worse case, a dosing regimen that is not aggressive enough can lead to illness returning in near-full force once dosing stops. To ensure a high enough dose there are a number of important factors to consider. The first is plasma protein binding, as only unbound drug will be able to cross into cells and have an effect on viral enzymes or other targets. Most drugs bind reversibly to albumin, glycoproteins, and lipoproteins in the blood. Thus, the amount of plasma binding is partly a function of what other drugs a person is taking. Next, one must understand how the drug is metabolized, which informs dosing amount and interval. Drug metabolism varies greatly between individuals depending on genetics and what other drugs a person is taking (drugs often inhibit, induce, downregulate, or upregulate the liver enzymes which are responsible for metabolizing most small molecule drugs). Finally, one should understand $C_{\text{max}}$ (peak concentration) and $C_{\text{min}}$ (trough concentration) at steady-state under repeated dosing. von Delft et al. recommend that dosing should be such that $C_{\text{min}}$ is greater than $\text{EC}_{90}$ for 90\% of the population.\cite{vonDelft2023} 

%------------------------------------------------------
\paragraph{Dose timing}
Antivirals should be given as early as possible. Ideally, the antiviral is already present in the body before infection. Modeling by Gon\c{c}alves et al.~suggests that antivirals administered prophylactically can reduce viral load by a factor of 1,000,000 compared to an antivirals given three days post-symptoms.\cite{Gonalves2020} The effects of timing are clearly seen in clinical trials on the broad spectrum antivirals Favipiravir and Remdesivir for SARS-CoV-2 -- trials that administered the drug early (within 3-5 days of symptom onset) generally show a positive effect, while those that administered later (after more than 3-5 days) generally show no effect. 

%------------------------------------------------------
\paragraph{The Hill Equation}
The Hill equation is typically a good way to model the dose-response curve. Let $E$ be a measure of drug effectiveness on a 0 to 1 scale, and $[L]$ be the concentration of the drug. The equation is: 

\begin{equation}
E = a + \frac{b - a}{1 + \left(\frac{\mbox{IC}_{50}}{[L]}\right)^m}
\end{equation}
Here, $m$ is called the Hill coefficient or Hill slope. Hill (1910) was originally studying the binding of oxygen to hemoglobin. Up to four oxygen molecules can bind to hemoglobin, and $m$ quantifies the degree of cooperativity in the binding process. If after one oxygen molecule binds it becomes easier for another to bind, the binding is cooperative and $m > 1$. If the binding of an oxygen molecule is independent of how many have already bound, $n=1$. If binding becomes harder after an oxygen has already bound, $m < 1$. Empirically, $m$ is found to be between 1.7 to 3.2. The maximum allowed $m$ theoretically is related to the number of binding sites -- since there are four in this case, $m$ must be less than 4. Shen et al. measured the Hill slope for a number of anti-HIV compounds.\cite{Shen2008} They found that for nucleoside reverse transcriptase inhibitors and integrase inhibitors, $m=1$ while for non-nucleoside reverse transcriptase inhibitors $m \approx 1.7$. For protease inhibitors they found $m$ between 1.8 and 4.5, indicating co-operative binding dynamics.\cite{Shen2011}

%------------------------------------------------------
\subsection{Quantifying synergy}\label{sec:quantifying_synergy}
The literature on drug synergy can be very confusing since several very different definitions for synergy are used. We review the main ones here. 

 %Synergistic Activity of Remdesivir–Nirmatrelvir Combination on a SARS-CoV-2 In Vitro Model and a Case Report
%https://www.mdpi.com/1999-4915/15/7/1577

%Combination of antiviral drugs inhibits SARS-CoV-2 polymerase and exonuclease and demonstrates COVID-19 therapeutic potential in viral cell culture
%https://www.nature.com/articles/s42003-022-03101-9
%"SARS-CoV-2 has an exonuclease-based proofreader, which removes nucleotide inhibitors such as Remdesivir that are incorporated into the viral RNA during replication, reducing the efficacy of these drugs for treating COVID-19. Combinations of inhibitors of both the viral RNA-dependent RNA polymerase and the exonuclease could overcome this deficiency. "

%-------------------------------------- 
\paragraph{The Bliss score}
Bliss (1939) introduced a simple mathematical framework for quantifying synergy in the context of studying the effect of toxins on insects.\cite{BLISS1939} If the replication of a virion is prevented, we say that virion is ``killed''. Suppose $p_A$ is the probability of killing a virion under treatment with drug $A$ and $p_B$ is the probability under treatment with drug $B$. If the two drugs act completely independently, then the probability the virion survives is: 
\begin{equation}
    \begin{aligned}
        p_s &= (1 - p_A)(1 - p_B) \\
        p_s &= 1 - p_A - p_B + p_A p_B
    \end{aligned}
\end{equation}
Therefore the combined probability of killing the virion is:
\begin{equation}\label{eq:ind}
    \begin{aligned}
        p_C &= 1 - p_s \\
        p_C &= p_A + p_B - p_A p_B
    \end{aligned}
\end{equation}
The ``Bliss score'' is given by: 
\begin{equation}
    S_{\mbox{Bliss}} = p^{\mbox{obs}}_C - p_C
\end{equation}
$S_{\mbox{Bliss}} > 0$ indicates synergy, $S_{\mbox{Bliss}} = 1$ indicates perfect independence, and $S_{\mbox{Bliss}} < 0$ indicates antagonism. Often, the Bliss score is analyzed over a grid of different concentrations on a percentage scale (-100 to 100). In that case all terms become concentration dependent: 
\begin{equation}
    S_{\mbox{Bliss}}(a, b) = p^{\mbox{obs}}_C(a, b) - p_C(a, b)
\end{equation}

Some numbers are instructive. If a drug that is only 50\% effective at blocking step 1 in the viral lifecycle is combined with a drug that is 50\% effective at blocking step 2, then the combination will have an improved efficacy of 75\% ($1 - 0.5 \times 0.5 = 0.75$), assuming no Bliss synergy. 

%------------------------------------------------- 
\paragraph{The Loewe synergy score}
Loewe (1926) considered synergy in terms of substitutability. Under this model, two drugs are additive (non-synergistic) if combining half the IC50 of drug A and half the IC50 of drug B also achieves a 50\% effect. This same idea can be applied to any effect level $e \in [0,1]$. Mathematically, Loewe synergy is described as follows: let $E_A(a)$ and $E_B(b)$ be the dose-response functions for drugs $A$ and $B$ at doses $a$ and $b$, with $E \in (0,1)$. The, define $D_A(e)$ as the dose of drug $A$ that produces effect $e$ and $D_B(e)$ the dose of drug $B$ that produces the same effect. 

When there is no synergy, the following holds: 
\begin{equation}\label{eqn:loeweadd}
    \frac{a}{D_A(e)} + \frac{b}{D_B(e)} = 1
\end{equation}

The first term in this equation is the fraction of the ``full required dose'' used for drug $a$. If the combination achieves effect $e$ using only half the $D_A(e)$ dose of $A$ and half the $D_B(e)$ dose of $B$, the two fractions sum to 1 and the combination is ``Loewe-additive''. If one plots all the $(a, b)$ pairs that produce a fixed effect $e$ on a 2D graph, the result is a line from $(D_A(e), 0)$ to $(0, D_B(e))$ which is called the ``isobole''. Combinations that fall below the isobole (achieving the same effect with less drug than the line predicts) are synergistic, while combinations above it are antagonistic.

The Loewe synergy score quantifies deviation from the isobole line: 
\begin{equation}
    S_{\mbox{Loewe}} = e_{\mbox{obs}} - e_{\mbox{Loewe}}
\end{equation}
where $e_{\mbox{obs}}$ is the observed combination effect and $e_{\mbox{Loewe}}$ is the effect predicted by the Loewe model when there is no synergy. To obtain $e_{\mbox{Loewe}}$, one has to plug $a$ and $b$ into equation \ref{eqn:loeweadd} and solve for $e$. If $D_A(e)$ and $D_B(e)$ are given by the Hill equation with $m=1$, then there is an analytic solution which is quite nice: 

\begin{equation}
e_{\mbox{Loewe}}([L_A], [L_B]) = a + \frac{b - a}{1 + \frac{1}{K}}
\end{equation}
where
\begin{equation}
K = \frac{[L_A]}{\mbox{IC}_{50,A}} + \frac{[L_B]}{\mbox{IC}_{50,B}}
\end{equation}

If $m\neq 1$ (common), then there is no analytic equation for $e_{\mbox{Loewe}}$, and the solution must be obtained numerically.
%$S_{\mbox{Loewe}} > 0$ indicates synergy and $S_{\mbox{Loewe}} < 0$ indicates antagonism.  

%We can also define the combination index: 
%\begin{equation}
%    CI(e) = \frac{a}{D_A(e)} + \frac{b}{D_B(e)} 
%\end{equation}

The Loewe synergy score makes the most sense when drugs share the same target, but may differ in potency.  

%------------------------------------------------------
\paragraph{The HSA score}
The Highest Single Agent (HSA) score, also known as the Gaddum score, compares the combination therapy to the highest possible effect from either drug acting alone at the same concentration as in the combination. It is thus defined as: 
\begin{equation}
    S_{\mbox{HSA}} = E^{obs}(a, b) - \max(E_A(a), E_B(b))
\end{equation}
This is the most permissive definition of synergy, presenting a lower bar for synergy than the Bliss or Loewe scores. In our view this definition is not useful and muddles the water since it doesn't align with commonsense notions of what synergy should be. For instance, if drug A achieves 40\% inhibition and drug B achieves 30\%, combining them might yield 58\% inhibition — exactly what equation~\ref{eq:ind} predicts for non-interacting drugs — yet HSA would call this synergistic since 58\% exceeds the best single agent's 40\%.

%------------------------------------------------------
%\paragraph{The ZIP score}
%The Zero Interaction Potency (ZIP) model, proposed by Yadav et al. (2015), attempts to unify the Bliss and Loewe concepts.\cite{YADAV2015}

%-----------------------------------------------------------------------
%-----------------------------------------------------------------------
%-----------------------------------------------------------------------
\section{Drug screening assay metrics}

\subsection{Protein-ligand assay metrics}\label{sec:assay_metrics}
%---------------------------- 
\paragraph{$\boldsymbol{K_d}$ and $\boldsymbol{K_a}$}
The term ``binding affinity'' usually refers to the dissociation constant $K_d$, defined as: 
\begin{equation}\label{eqn:kd}
K_d = \frac{[P][L]}{[PL]}
\end{equation}
Here $[P]$ is the concentration of free protein, $[L]$ is the concentration of free ligand, and $[PL]$ is the concentration of the protein-ligand complex. Lower $K_d$ values indicate tighter binding. When 50\% of protein is bound, $[PL] = [P]$ and eqn.~\ref{eqn:kd} simplifies to $K_d = [L]$.

$K_d$ is related to the Gibbs free energy of binding by:
\begin{equation}\label{eq:dG_kd} 
\Delta G = RT \ln K_d 
\end{equation}
where $R$ is the gas constant and $T$ is temperature. Molecular docking programs directly estimate $\Delta G$, making K$_d$ the most natural experimental metric for comparison. Unfortunately, depending where the ligand binds, a very small K$_d$ value may not mean the protein's function is inhibited, which brings us to the next metric.

Less commonly, the term ``binding affinity'' refers to the association constant $K_a$, which is defined as:
\begin{equation} 
K_a = \frac{1}{K_d}
\end{equation}

%-------------------------------------------------------
\paragraph{$\boldsymbol{K_i}$ (Inhibition Constant)}
The inhibition constant $K_i$ is intended to be an assay-independent constant defined as the concentration necessary to reduce enzyme activity by 50\%. It may be different than $K_d$, depending on whether the inhibition is competitive, non-competitive, or semi-competitive. Competitive inhibitors block an enzyme's substrate from being able to bind, which is obviously highly desirable. For competitive inhibitors that bind exclusively to the substrate binding site and follow Michaelis-Menten kinetics, $K_i \approx K_d$. However, not all compounds will be competitive binders, which is why $K_i$ may be different than $K_d$.  

%-------------------------------------------------------
\paragraph{$\boldsymbol{pK_d}$ and $\boldsymbol{pK_i}$}
It is useful to take the negative base-10 logarithm of the dissociation and inhibition constants:
\begin{equation}
    \begin{aligned}
        \text{pK}_d &= -\log_{10}(K_d)\\
        \text{pK}_i &= -\log_{10}(K_i)
    \end{aligned}
\end{equation}
There are several reasons for this. First, higher pK$_d$ and pK$_i$ values indicate tighter binding, conforming to the human intuition that ``larger is better''. Additionally, pK$_d$ is directly proportional to the binding free energy calculated by molecular dynamics, quantum chemistry, and docking software:
\begin{equation}
\text{pK}_d = \frac{-\Delta G}{2.303 \cdot RT} \approx \frac{-\Delta G}{1.364}  \text{(at 298K, in kcal/mol)}
\end{equation}
Finally, in the machine learning context, logarithmically compressing the data reduces the number of floating point figures and the range of the data, which increases training stability, resulting in faster training runs that are less likely to fail due to exploding or vanishing gradient problems. 

%-------------------------------------------------------
\paragraph{IC$_{50}$ (Half-Maximal Inhibitory Concentration)}
IC$_{50}$ is the inhibitor concentration that reduces enzyme activity by 50\% under specific assay conditions. Whereas $K_i$ is an intrinsic binding constant for a given temperature; IC$_{50}$ depends on additional experimental parameters, namely substrate concentration, enzyme concentration, and incubation time. 

The Cheng--Prusoff equation relates $K_i$ to IC$_{50}$:
\begin{equation}\label{eq:chengprusoff}
K_i = \frac{\text{IC}_{50}}{1 + \frac{[S]}{K_m}}
\end{equation}
where $[S]$ is the substrate concentration (assumed constant during the assay) and $K_m$ is the Michaelis-Menten parameter (discussed below). 

It's worth mentioning the Michaelis-Menten equation, which is simple logistic equation giving the rate of product formation $v$ of an enzyme in terms of substrate concentration: 

\begin{equation}
v = \frac{V_{\max}[S]}{K_m + [S]}
\end{equation}

Here $v$ is rate of product formation, also called the reaction velocity, $[S]$ is the substrate concentration, and $V_{\max}$ is the maximum reaction velocity that is achieved when all enzyme sites are occupied. $K_m$ is the Michaelis--Menten constant, and it corresponds to a substrate concentration at which $v = \frac{1}{2}V_{\max}$. 

For competitive inhibitors, a high substrate concentration $[S]$ means the inhibitor has trouble competing with the substrate for access to the enzyme's active site. If $[S]$ is very high or very low, difficulties arise for experimentalists trying to distinguish strong and weak inhibitors. It is common for assays to operate at $[S] = K_m$, which again is the concentration where the rate of product formation is half its max. In that case:  

\begin{equation}\label{eqn:ic50}
    K_i \approx \frac{\text{IC}_{50}}{2}
\end{equation}

In reality, this equation is quite approximate and $K_i$ may differ from IC$_{50}$ by a factor of 2 to 5x. In cheminformatics workflows, IC$_{50} < 10~\mu$M is sometimes used as a threshold to classify compounds as ``active''. 

%-------------------------------------------------------
\paragraph{Note on covalent inhibition}
A small percentage of approved antivirals form covalent bonds with their substrate (a notable example is nirmatrelvir, the active ingredient in Paxlovid.) The above metrics are based on an assumption of reversible kinetics driven by classical statistical mechanics and they break down if the ligand forms a covalent bond with the substrate. Covalent binding occurs in two stages - a fast initial non-covalent binding followed by a slower covalent bond formation stage. Once formed, the covalent bond may be broken later, usually through hydrolysis. During hydrolysis, a protein's histidine catalyzes the splitting of water into a proton (\(H^{+}\)) and a highly reactive hydroxide-like ion (\(OH^{-}\)) which breaks the bond.  

Potency for covalent binders is typically characterized by the following ratio:
\begin{equation}
\mbox{inactivation efficiency} = k_{\text{inact}}/K_I
\end{equation}
Here $k_{\text{inact}}$ is the rate of covalent bond formation (units = s$^{-1}$) after the non-covalent complex forms, and $K_I$ is the equilibrium inhibition/dissociation constant for the initial, non-covalent  complex. This metric focuses on achieving speed in the covalent bond formation. K$_d$ doesn't make sense for a covalent binder. However, IC$_{50}$ may be reported for covalent inhibitors, but it becomes time dependent and an incubation time must be given (IC$_{50}(t_I)$). For irreversible covalent inhibitors, the IC$_{50}$(t$_I$) tends towards 0 as the incubation time goes to infinity. For reversible covalent inhibitors, IC$_{50}$(t$_I$) tends to a finite value in a protocol-dependent manner. ML models trained on the K$_i$ and IC$_{50}$ data available in public datasets are generally considered unreliable when it comes to predicting the efficacy of covalent inhibitors. 

%------------------------------------------------------------------------------
\subsection{Cell-Based Efficacy Metrics}

%--------------------------------- 
\paragraph{IC$_{50}$, IC$_{90}$, EC$_{50}$, EC$_{90}$, and AC$_{50}$}
IC$_{50}$, IC$_{90}$, EC$_{50}$, and EC$_{90}$ measure the concentration required to achieve 50\% or 90\% of a maximal Effect or maximal Inhibition in a cell-based assay. EC$_{90}$ and IC$_{90}$, while not as commonly used, may be more relevant when therapeutic suppression of viral replication is desired. Another synonymous term is AC$_{50}$, which refers to the concentration achieving a 50\% reduction in some cellular Activity. Unlike biochemical assays, these metrics are affected by multiple factors beyond target binding including the degree to which the compound binds to other proteins, the compound's cell membrane permeability, the stability of the compound in the cellular environment, and cellular efflux pump activity. 

%-------------------------------------- 
\paragraph{CC$_{50}$ (Cytotoxic Concentration)}
CC$_{50}$ is the compound concentration that reduces cell viability by 50\% in the absence of virus. This metric quantifies compound toxicity to the host cell and is essential for evaluating therapeutic windows. A compound with potent antiviral activity (low EC$_{50}$) but high cytotoxicity (low CC$_{50}$) may be unsuitable for therapeutic use.

%-------------------------------------------------------
\paragraph{SI (Selectivity Index)}
SI, which can stand for ``Selectivity Index'' or ``Safety Index'', is defined as SI $=$ CC$_{50}$/EC$_{50}$. In the context of cells it is called the cytotoxic index, and in the context of animals it is called the therapeutic index. It quantifies the trade-off between antiviral efficacy and cytotoxicity (all compounds become 50\% toxic at some dose). During screening, thresholds of SI $> 10$ or SI $> 100$ are commonly used to select compounds that are good enough to proceed to the next stage. To give an interesting example, when tested in cell culture against SARS-CoV-2, hydroxychloroquine had SI = 22, which Xiao et al.~note was a less-favorable safety profile compared to six ``hit'' compounds they found which had SI > 600.\cite{Xiao2020}

%------------------------------------------------------------ 
%------------------------------------------------------------
\section{Model evaluation metrics}\label{sec:eval_metrics}

Let $y_i$ denote the experimental pKd value and $\hat{y}_i$ denote the predicted pKd value for compound $i$, where $i = 1, \ldots, n$. In the following section we define several common metrics for accessing performance: 

%------------------------------------------------------
\paragraph{Pearson Correlation Coefficient ($r$)}
The Pearson correlation coefficient measures the linear correlation between predicted and experimental values:
\begin{equation}
r = \frac{\sum_{i=1}^{n}(y_i - \bar{y})(\hat{y}_i - \bar{\hat{y}})}{\sqrt{\sum_{i=1}^{n}(y_i - \bar{y})^2} \sqrt{\sum_{i=1}^{n}(\hat{y}_i - \bar{\hat{y}})^2}}
\end{equation}
where $\bar{y}$ and $\bar{\hat{y}}$ are the means of the experimental and predicted values, respectively. Values range from $-1$ to $+1$.

%------------------------------------------------------
\paragraph{Coefficient of Determination ($R^2$)}
The coefficient of determination is calculated as :
\begin{equation}
Q^2 = 1 - \frac{SS_{res}}{SS_{tot}} = 1 - \frac{\sum_{i=1}^{n}(y_i - \hat{y}_i)^2}{\sum_{i=1}^{n}(y_i - \bar{y})^2}
\end{equation}
where $SS_{res}$ is the residual sum of squares and $SS_{tot}$ is the total sum of squares. Unlike $r^2$ from a fitted regression, this formulation compares predictions directly against experimental values (not against a fitted line), and can therefore be negative when predictions perform worse than simply predicting the mean. While this metric is often designated by $R^2$, we follow the binding affinity prediction literature and use $Q^2$, which also helps distinguish this metric from $r^2$.

%------------------------------------------------------
\paragraph{Modified Coefficient of Determination ($r_m^2$)}
The modified $r^2$ metric, proposed by Roy and Roy \cite{Roy2008}, accounts for the difference between the standard $r^2$ and $r_0^2$ (the coefficient of determination with intercept forced through zero) to better assess external predictability:
\begin{equation}
r_m^2 = r^2 \left(1 - \sqrt{|r^2 - r_0^2|}\right)
\end{equation}
where $r_0^2$ is calculated as:
\begin{equation}
r_0^2 = 1 - \frac{\sum_{i=1}^{n}(y_i - \hat{y}_i)^2}{\sum_{i=1}^{n}y_i^2}
\end{equation}
When predictions systematically over- or under-predict, $r_0^2$ will differ significantly from $r^2$, and $r_m^2$ will be substantially lower than $r^2$, indicating poor external predictability. The metric penalizes models that achieve good correlation but poor calibration.

%------------------------------------------------------
\paragraph{Spearman Rank Correlation Coefficient ($\rho$)}
The Spearman correlation coefficient measures monotonic relationships between variables using ranks rather than raw values:
\begin{equation}
\rho = \frac{\sum_{i=1}^{n}(R_i - \bar{R})(\hat{R}_i - \bar{\hat{R}})}{\sqrt{\sum_{i=1}^{n}(R_i - \bar{R})^2} \sqrt{\sum_{i=1}^{n}(\hat{R}_i - \bar{\hat{R}})^2}}
\end{equation}
where $R_i$ and $\hat{R}_i$ are the ranks of $y_i$ and $\hat{y}_i$, respectively. Spearman $\rho$ is robust to outliers and captures non-linear monotonic relationships, making it particularly useful for evaluating compound ranking ability.

%------------------------------------------------------
\paragraph{Root Mean Square Error (RMSE)}
RMSE measures the average magnitude of prediction errors:
\begin{equation}
\text{RMSE} = \sqrt{\frac{1}{n}\sum_{i=1}^{n}(y_i - \hat{y}_i)^2}
\end{equation}
RMSE is expressed in the same units as the target variable (pKd units) and penalizes larger errors more heavily due to the squaring operation. 

%------------------------------------------------------
\paragraph{Mean Absolute Error (MAE)}
MAE measures the average absolute deviation between predictions and experimental values:
\begin{equation}
\text{MAE} = \frac{1}{n}\sum_{i=1}^{n}|y_i - \hat{y}_i|
\end{equation}
Like RMSE, MAE is expressed in pKd units. MAE is more robust to outliers than RMSE since it does not square the errors.

%------------------------------------------------------
\paragraph{Concordance Index (CI)}
The concordance index measures whether the predicted binding affinity values of two random drug-target pairs are predicted in the same order as their true values \cite{Gonen2005}. For a set of predictions, the CI is defined as:
\begin{equation}
\text{CI} = \frac{1}{Z}\sum_{i,j: \delta_i > \delta_j} h(b_i - b_j)
\end{equation}
where $b_i$ is the prediction value for the larger affinity $\delta_i$, $b_j$ is the prediction value for the smaller affinity $\delta_j$, $Z$ is a normalization constant equal to the number of pairs, and $h(x)$ is the step function \cite{Pahikkala2014}:
\begin{equation}
h(x) = \begin{cases}
1 & \text{if } x > 0 \\
0.5 & \text{if } x = 0 \\
0 & \text{if } x < 0
\end{cases}
\end{equation}
The CI ranges from 0 to 1, where 0.5 indicates random predictions and 1 indicates perfect concordance. We used paired t-test for statistical significance tests with 95\% confidence interval.

%------------------------------------------------------
\paragraph{Enrichment Factor (EF)}
The enrichment factor measures how many more active compounds are found in a specific top fraction of a ranked list (the ``early recognition subset'') compared what would be expected in a randomly-ordered list. It is defined as: 
\begin{equation}
EF_{\chi} = \frac{n_a / n_{total}}{N_a / N_{total}}
\end{equation}

Here $\chi$ is the fraction in the top subset (e.g., 0.01 for 1\%) and 
\begin{itemize}
    \item $n_a$: Number of active compounds found in the top subset.
    \item $n_{total}$: Total number of compounds in the top subset (calculated as $\chi \times N_{total}$).
    \item $N_a$: Total number of active compounds in the entire dataset.
    \item $N_{total}$: Total number of compounds in the entire dataset.
\end{itemize}

%------------------------------------------------------
\paragraph{Area Under the Receiver Operating Characteristic Curve (AUROC)}
AUROC evaluates the overall ranking performance across the entire dataset. It is a well known metric for binary classification performance. It can interpreted as giving the probability that a randomly chosen active compound is ranked higher than a randomly chosen inactive compound. In the context of ranked lists, AUROC is calculated using ranks:
\begin{equation}
\mbox{AUROC} = 1 - \frac{1}{N_a N_i} \sum_{j=1}^{N_a} (r_j - j)
\end{equation}

\textbf{Where:}
\begin{itemize}
    \item $N_a$: Total number of active compounds.
    \item $N_i$: Total number of inactive compounds.
    \item $r_j$: The rank of the $j$-th active compound in the sorted list.
    \item $j$: The ideal rank position for that active compound (1, 2, 3...).
\end{itemize}

%------------------------------------------------------
\paragraph{Boltzmann-Enhanced Discrimination of ROC (BEDROC)}
BEDROC is a custom metric which heavily weights the discriminatory power of the top of a ranked list using an exponential decay parameter $\alpha$. BEDROC is derived from the Robust Initial Enhancement (RIE):
\begin{equation}
\mbox{RIE} = \frac{\frac{1}{N_a} \sum_{j=1}^{N_a} e^{-\alpha r_j / N}}{\frac{1}{N} \left( \frac{1 - e^{-\alpha}}{e^{\alpha / N} - 1} \right)}
\end{equation}

The BEDROC score is then normalized between 0 and 1:
\begin{equation}
\mbox{BEDROC} = \frac{\mbox{RIE} - \mbox{RIE}_{min}}{\mbox{RIE}_{max} - \mbox{RIE}_{min}}
\end{equation}

\textbf{Where:}
\begin{itemize}
    \item $\alpha$: The ``early recognition'' parameter (typically $\alpha=20$ for the top 8\% of the list).
    \item $r_j$: The rank of the $j$-th active compound.
    \item $N$: Total number of compounds in the dataset.
    \item $N_a$: Total number of active compounds.
\end{itemize}

\end{appendices}


\bibliography{sn-bibliography}

\end{document}